\documentclass[prx,aps,twocolumn,footinbib,superscriptaddress,floatfix]{revtex4-1}

\usepackage{CJK}
\usepackage[colorlinks,bookmarks=false,citecolor=blue,linkcolor=red,urlcolor=blue]{hyperref}
\usepackage{color} 
\usepackage{graphicx}
\usepackage{float}
\usepackage{multirow}
\usepackage{amsmath}
\usepackage{bm}
\usepackage{amsfonts}
\usepackage{braket}
\usepackage{subfigure}
\usepackage{cleveref}

\usepackage{dcolumn}
\usepackage{amssymb}
\usepackage{microtype}
\usepackage{xfrac}
\usepackage{array}
\usepackage[dvipsnames]{xcolor}

\newcolumntype{P}[1]{>{\centering\arraybackslash}p{#1}}

\begin{document}

\title{Exploiting polarization dependence in two
dimensional coherent spectroscopy: examples of Ce$_2$Zr$_2$O$_7$ and Nd$_2$Zr$_2$O$_7$}

\author{Mark Potts}
\affiliation{Max Planck Institute for the Physics of Complex Systems, N\"{o}thnitzer Str. 38, Dresden 01187, Germany}
\author{Roderich Moessner} 
\affiliation{Max Planck Institute for the Physics of Complex Systems, N\"{o}thnitzer Str. 38, Dresden 01187, Germany}
\author{Owen Benton} 
\affiliation{Max Planck Institute for the Physics of Complex Systems, N\"{o}thnitzer Str. 38, Dresden 01187, Germany}
\affiliation{School of Physical and Chemical Sciences, Queen Mary University of London, London, E1 4NS, United Kingdom}

\date{\today}

\begin{abstract}
Two dimensional coherent spectroscopy (2DCS) probes the nonlinear optical response of correlated systems. An interesting application is the study of fractionalized excitations, which are challenging to distinguish unambiguously in linear response. 
Here we demonstrate how the sensitivity of optical matrix elements to variations in the photon polarization allows one to probe different aspects of low lying excitations in models of the candidate fractionalized materials Ce$_2$Zr$_2$O$_7$ and Nd$_2$Zr$_2$O$_7$, which host effective one-dimensional spin chains when subjected to a [110] magnetic field.
We show how both fractionalized spinon excitations or conventional magnons can be picked out in the 2DCS response, and how the response from polarized spin chains can be used to probe the dipolar-octupolar mixing angle $\theta$ through the relative intensity of one- and two-magnon signals. Further, we find that a $[001]$ polarization of the probe field is particularly sensitive to lower band edge of the spinon continuum and can be used as a measure of the proximity of a quantum critical point in Ce$_2$Zr$_2$O$_7$. 2DCS can thus be employed to provide invaluable and detailed information both on the constituent degrees of freedom of a quantum material and on their collective behaviour. 
\end{abstract}

\maketitle

\section{Introduction}
\label{sec:intro}

A defining feature of quantum spin liquids
is the presence of fractionalized excitations, which cannot be created on their own.
When probed with linear response
methods, such as inelastic
neutron scattering, this leads to
signal which is broad in wave-vector and frequency, corresponding to the many ways of distributing the scattered momentum and energy amongst the multiple fractional excitations 
\cite{tennant95, helton07, balents12, han12}.
The lack of a sharp signal leads to
a challenge in the experimental study of spin liquids: how can continua of fractionalized excitations be distinguished from broad scattering due to the simple absence of long-lived excitations?

Two dimensional coherent spectroscopy (2DCS) offers a 
resolution to this conundrum. 2DCS probes the nonlinear response of the system to a pair of radiation pulses, at different times. By varying the temporal separation of the pulses and the measurement, one can measure the response as a function of two time coordinates, or, when Fourier transformed, two independent frequencies: $\omega_t$ and $\omega_{\tau}$. 
Fractionalized excitations show up as streaks of 
intensity in the $\omega_t - \omega_{\tau}$ plane.
Of particular importance is 
the rephasing signal:
a streak along the 
$\omega_t = - \omega_{\tau}$ direction \cite{wan19}.
This signal is important because it is robust to broadening effects, and allows such broadening effects to be distinguished from extended signals from fractionalized continua: the broadening due to finite lifetimes appears transverse to the streak, while the intrinsic continuum width is given by the length of the streak.
It thus becomes possible to directly separate out fractionalized quasiparticle lifetimes, and identify them as long-lived excitations.
Since the original proposal to use
2DCS as a probe of fractional excitations \cite{wan19}, several theoretical calculations have demonstrated its usefulness in principle for various models in 1 and 2 dimensions \cite{parameswaran20, li21, choi20, hart23, nandkishore21}. 

\begin{figure*}
\subfigure[]{\includegraphics[width=0.35\textwidth]{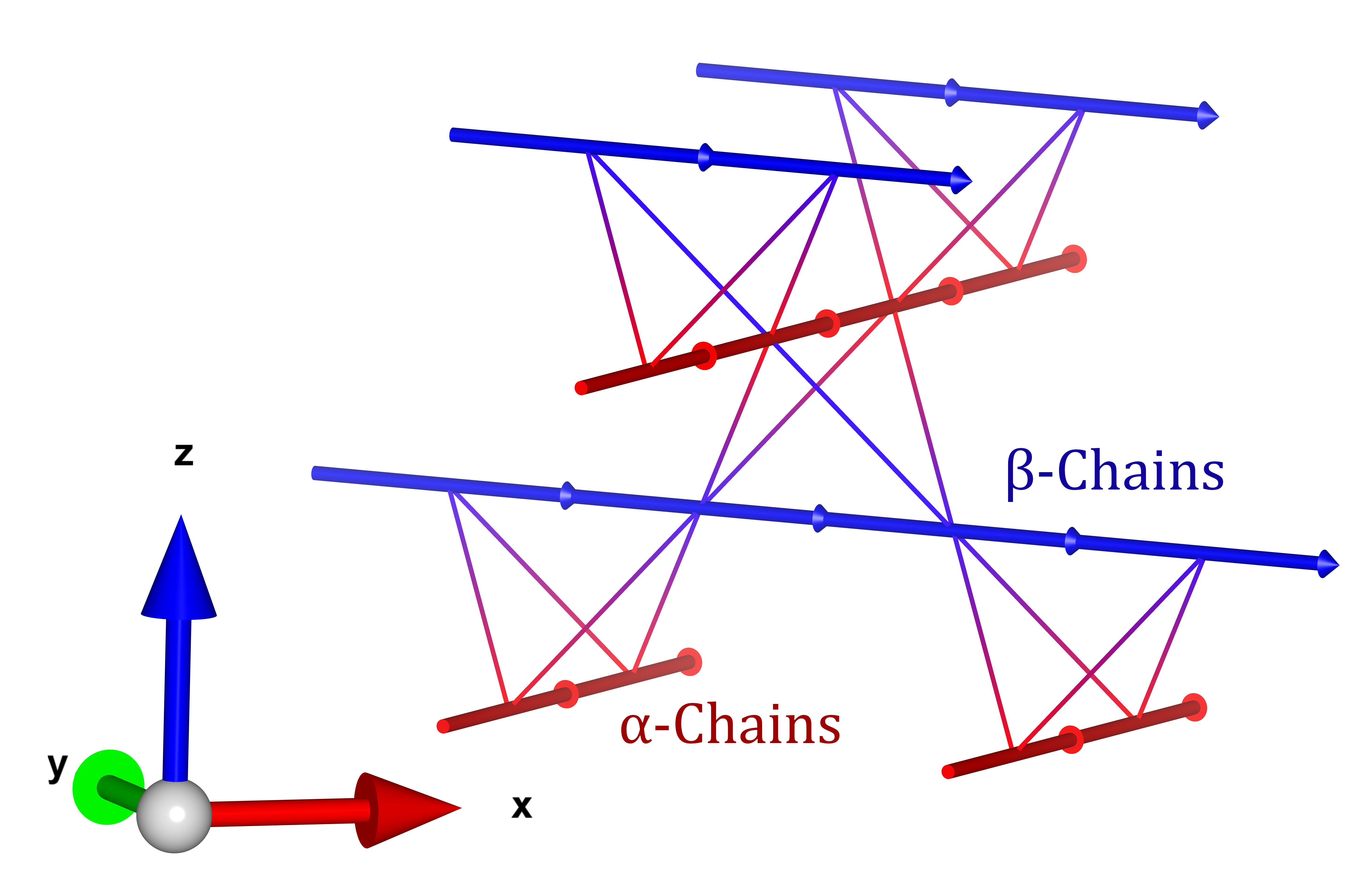}}
\subfigure[]{\includegraphics[width=0.35\textwidth]{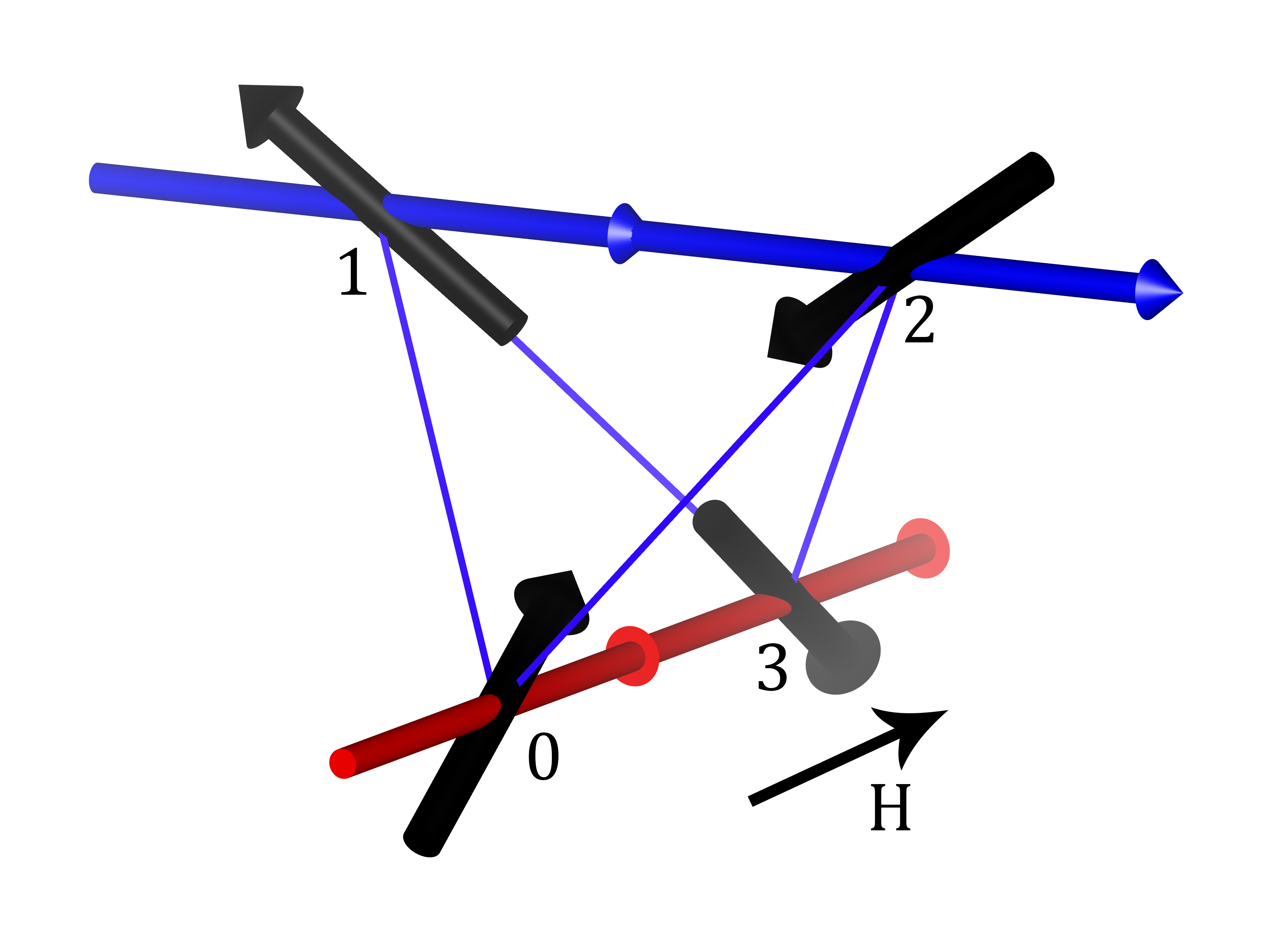}}
\caption{
(a) The pyrochlore lattice and its decomposition
into $\alpha$ and $\beta$ chains in a [110] magnetic field.
$\alpha$ chains run parallel to the field, and are coloured red in the left panel. In a sufficiently large applied field, spins on the $\alpha$ chain become polarized, whilst the perpendicular $\beta$ chains do not feel the field.
(b) A single tetrahedron
demonstrating one possible alignment of spins that satisfy the `2-in-2-out' rule while also polarising the $\alpha$ chain spins with respect to the applied field. The $\beta$ spin chains can then be in one of two configurations that preserve the ice-rule.
The numerical labels 0-3 indicate the labelling convention for the four fcc sublattices 
$\mathcal{L}_{\mu}$ used throughout this Article.
}
\label{fig:chains}
\end{figure*}

Some of the most promising spin liquid candidates are found in the rare-earth pyrochlore oxides 
R$_2$M$_2$O$_7$ \cite{rau19}.
In recent years, Ce-based pyrochlores
Ce$_2$M$_2$O$_7$ (M=Zr, Sn, Hf) have emerged as particularly likely candidates \cite{sibille20, poree23-sn, gao19, poree22}, and much effort has been devoted to understanding their correlations and microscopic coupling parameters 
\cite{smith22, gao22, yahne22, bhardwaj22, poree23}.
As with other spin liquid candidates, the question of how to 
interpret inelastic scattering data, in the absence of sharply defined excitations, is a key issue.

These materials also exhibit interesting responses to externally 
applied magnetic fields, due 
 to the highly anisotropic nature
of the crystal electric field (CEF) environments at the rare-earth sites.
In particular a magnetic field aligned with the $[110]$ crystal axis splits the system into two sets
of spin chains, $\alpha$ and $\beta$, with the $\alpha$ chains being polarized by the applied field while the $\beta$ chains feel neither the effect of the field, nor the neighbouring $\alpha$ chains.
This emergence of effectively one-dimensional physics out of a three-dimensional correlated system has been observed in the quantum pyrochlores Ce$_2$Zr$_2$O$_7$ \cite{smith23} and 
Nd$_2$Zr$_2$O$_7$ \cite{xu18}, as well as in the canonical classical spin ices Dy$_2$Ti$_2$O$_7$ \cite{fennell02, fennell05} and 
Ho$_2$Ti$_2$O$_7$ \cite{fennell05, clancy09}.

In this work we investigate how variations in the polarization of a probe field can be used to extract different information from the 2DCS response of a system, focusing on the quantum pyrochlores Ce$_2$Zr$_2$O$_7$ and Nd$_2$Zr$_2$O$_7$ with an applied [110] magnetic field.
Their highly anisotropic microscopic Hamiltonians provide a way to explore how the dependence of optical matrix elements on probe field polarization allows one to tune the 2DCS response. The emergence of effectively one-dimensional spin chains in an applied field allows us to study this response in a controlled way while remaining in a setting where fractional excitations appear, in this case on the unpolarized $\beta$ chains.

Further, we focus on providing predictions for 2DCS experiments in the materials
Ce$_2$Zr$_2$O$_7$ and Nd$_2$Zr$_2$O$_7$ as estimates of the interaction parameters are available \cite{smith22, bhardwaj22, xu19, lhotel18, leger2021}, and the emergence of chains has been established \cite{smith23, xu18}. We show how polarization variations allow separate access to
the dynamical correlations of both $\alpha$ and $\beta$ chains, and calculate the dependence of the signal on the nature of the in-field ground state and on the microscopic interaction parameters.

For Ce$_2$Zr$_2$O$_7$ we find that a $[001]$ polarization of the probe field is particularly powerful for probing the proximity to a nearby quantum critical point, predicted from the parameter values in \cite{smith22, smith23}.
This is because the $[001]$ polarization suppresses the matrix elements to excite all states bar those at the lower band edge of the $\beta$-chain spinon continuum.
On the other hand, a $[1\bar{1}0]$ polarization probes the spinon density of states across the full spectrum.
We demonstrate that a streak-like response is still observed  on the $\alpha$ chains are probed using a $[110]$ field, despite the fact that excitations are not fractionalized on these chains, due to the probe field having no matrix element to excite single magnons.

For Nd$_2$Zr$_2$O$_7$, we find that the 2DCS response is profoundly affected by a strong mixing between dipolar and octupolar degrees of freedom found in the microscopic Hamiltonian \cite{xu19}, expressed through a mixing angle $\theta$.  
In particular, we find that $\theta$ controls the relative contribution of one- and two- magnon states to the 2DCS response of the $\alpha$ chains in a $[110]$ polarization set-up.
For the $[1\bar{1}0]$ polarization, where the experiment probes spinons on the $\beta$ chains, we find that the relative amplitude of the rephasing signal along $\omega_t = -\omega_{\tau}$ is suppressed at values of $\theta$ far from a multiple of $\pi/2$. 

The Article will proceed as follows:
in Section \ref{sec:background} we review some necessary background 
theory regarding both the materials and the experimental method, as well as presenting some new 
results regarding the effect of quantum fluctuations on the ground state phase diagram;
in Section \ref{sec:calculations} we give details of our analysis, which makes use of several methods in various limits of the problem (Jordan-Wigner fermionisation, Hartree-Fock mean field theory and perturbation theory);
in Section \ref{sec:results} we show the results of our calculations as they apply to Ce$_2$Zr$_2$O$_7$ and Nd$_2$Zr$_2$O$_7$, before concluding in Section \ref{sec:summary}.

\section{Background}
\label{sec:background}

\subsection{Chain Phases in Dipolar-Octupolar Pyrochlores in a 110 magnetic field}
\label{subsec:chains}

Both Ce$_2$Zr$_2$O$_7$ and Nd$_2$Zr$_2$O$_7$ are rare-earth pyrochlore magnets of a dipolar-octupolar character. In either material, the magnetic ions (Ce$^{3+}$ or Nd$^{3+}$) lie on a pyrochlore lattice, consisting of corner sharing tetrahedra. The geometry of the pyrochlore lattice is displayed in Fig. \ref{fig:chains}. 
There are two kinds of tetrahedra in the lattice, related by inversion symmetry, conventionally referred to as `A' and `B' or `up' and `down' tetrahedra. Each site
is shared between one `up' and one `down' tetrahedron.
The lattice can be split up into four fcc sublattices $\mathcal{L}_i$ ($i=0,1,2,3)$, each generated by taking one of the vertices of a single `up' tetrahedron and acting on it with the translational symmetries of the lattice. 
We define vectors $\hat{\mathbf{z}}_i$ to be unit vectors pointing from the centre of the host `up' tetrahedron to the given lattice site.

Each rare-earth ion
has a fixed total angular momentum $J$, determined by Hund's rules.
The crystal electric field at each site
breaks the degeneracy of the $2J+1$ different angular momentum states $m_J$ of each ion. Both Ce$^{3+}$ and Nd$^{3+}$ are Kramers ions, and so the lowest energy states arising from this splitting form a degenerate time-reversal paired doublet. For rare-earth pyrochlore magnets with Kramers ions, this lowest energy doublet can be of two types depending on the transformation properties of the doublet under the double group of point symmetries and time-reversal \cite{huang14, rau19}. One finds that the doublet states either transform in the same way as spin-$1/2$ states under the action of the symmetry group, or in the more exotic `dipolar-octupolar' \cite{huang14} fashion seen in our materials of study. 

In dipolar-octupolar compounds, one can construct a basis of pseudo-spin operators $S^x,S^y,S^z$ acting on the lowest energy doublet such that the magnetisation operator $\mathbf{\mu}$ projected into this subspace is diagonal, and proportional to $S^z$. 
\begin{equation}
    \mathbf{\mu}_i = \mu_B g_z \hat{\mathbf{z}}_i S^z_i .
    \label{eq:magnetisation definition}
\end{equation}
One finds that the $S^{\pm}$ in this basis correspond not to physical dipole operators, but to magnetic octupoles. Nevertheless, $S^x$ in this pseudospin basis transforms identically to $S^z$ under the symmetries of the crystal, and transforms differently from $S^y$ \cite{huang14}. Thus the $S^x$ pseudospin operator, despite physically being a magnetic octupole operator, is often referred to as `dipolar'.

The degeneracy in the set of crystal field doublets is lifted by interactions between ions on the pyrochlore lattice. Taking into account all symmetry allowed interactions between nearest neighbour pseudospins, one arrives at the following form for the Hamiltonian:
\begin{multline}
    H=\sum_{\langle ij \rangle} J_{xx} S^x_iS^x_j + J_{yy} S^y_iS^y_j + J_{zz} S^z_iS^z_j \\ +J_{xz}\left(S^z_iS^x_j+S^z_jS^x_i\right)- \mu_B g_z \sum_i \mathbf{H}\cdot \hat{\mathbf{z}}_i S^z_i .
    \label{eq:DO Hamiltonian pseudospin basis}
\end{multline}
Note that we are allowed a coupling between $S^x$ and $S^z$ due to their identical transformation properties under the symmetries of the crystal. We have also allowed for a uniform magnetic field $\mathbf{H}$, which couples only to the magnetic dipole operator $S^z$.
The $S^z$-$S^x$ coupling terms are often difficult to work with directly, so we elect to perform the following rotation of our pseudospin basis::
\begin{align}
S^x_i &= \cos\theta {S}^{\tilde{x}}_i-\sin\theta {S}^{\tilde{z}}_i , \label{eq:Pseudo x rotation}\\ 
S^z_i &=  \cos\theta {S}^{\tilde{z}}_i+\sin\theta {S}^{\tilde{x}}_i , \label{eq:Pseudo z rotation} \\ 
\tan(2\theta) &= \frac{2J_{xz}}{J_{xx}-J_{zz}} . \label{eq:Pseudo rotation angle}
\end{align}
Crucially, in this new basis, both ${S}^{\tilde{x}}$ and ${S}^{\tilde{z}}$ have some dipolar character, and so both couple with varying strengths to the external field. ${S}^{\tilde{y}}$ naturally remains purely octupolar. In the new rotated basis, the Hamiltonian now reads:
\begin{multline}
H=\sum_{\langle ij \rangle} J_{x} {S}^{\tilde{x}}_i {S}^{\tilde{x}}_j 
+ J_{y} {S}^{\tilde{y}}_i {S}^{\tilde{y}}_j 
+ J_{z} {S}^{\tilde{z}}_i {S}^{\tilde{z}}_j \\ - \mu_B g_z \sum_i \mathbf{H}\cdot \hat{\mathbf{z}}_i(\cos\theta {S}^{\tilde{z}}_i+\sin\theta {S}^{\tilde{x}}_i) . \label{eq:DO Hamiltonian rotated basis}
\end{multline}
The angle $\theta$ describes the symmetry-allowed mixing of dipolar and octupolar degrees of freedom, and plays an important role in the response of the system to external probes.

In this paper we are interested in phases that exist within Ce$_2$Zr$_2$O$_7$ and Nd$_2$Zr$_2$O$_7$ when they are placed within a [110] magnetic field. Such a field is orthogonal to $\mathbf{\hat{z}}_1$ and $\mathbf{\hat{z}}_2$, giving rise to chains of pseudospins within the pyrochlore lattice that couple to the external field (sites on sublattices 0 and 3 which are termed $\alpha$ chains), and chains of spins that do not ($\beta$ chains). Fig. \ref{fig:chains} highlights these chains on the pyrochlore lattice.

We shall here briefly review the types of phases that are predicted by classical mean field theory for the Hamiltonian Eq. (\ref{eq:DO Hamiltonian rotated basis}) in the [110] field \cite{placke20}. At zero field, the classical ground state can be in one of six possible phases. Three of these are spin-ice like regimes SI$_{\lambda}$ ($\lambda=x,y,z$), where pseudospins obey a `2-in-2-out' rule with respect to the dominant pseudospin component. For two of these regimes, SI$_x$ and SI$_z$, this ordering is of mixed dipole-octupole character, whilst whilst (SI$_y$) hosts purely octupolar ordering. The system is in one of these regimes if one of the coupling constants is sufficiently large and positive:
\begin{equation}
    J_{\lambda} > \text{max}(-3J_{\lambda'},J_{\lambda'}) \forall \lambda' \neq \lambda .\label{eq:SI condition}
\end{equation}

The remaining three phases host `all-in-all-out' type order, and again there are three flavours of this phase depending on the dominant exchange coupling, with AIAO$_y$ hosting purely octupolar order. These phases exist when:
\begin{equation}
    J_{\lambda} < \text{min}(-J_{\lambda'}/3,J_{\lambda'}) \forall \lambda' \neq \lambda .\label{eq:AIAO condition}
\end{equation}

\begin{figure*}
\subfigure[]
{\includegraphics[scale=0.38]{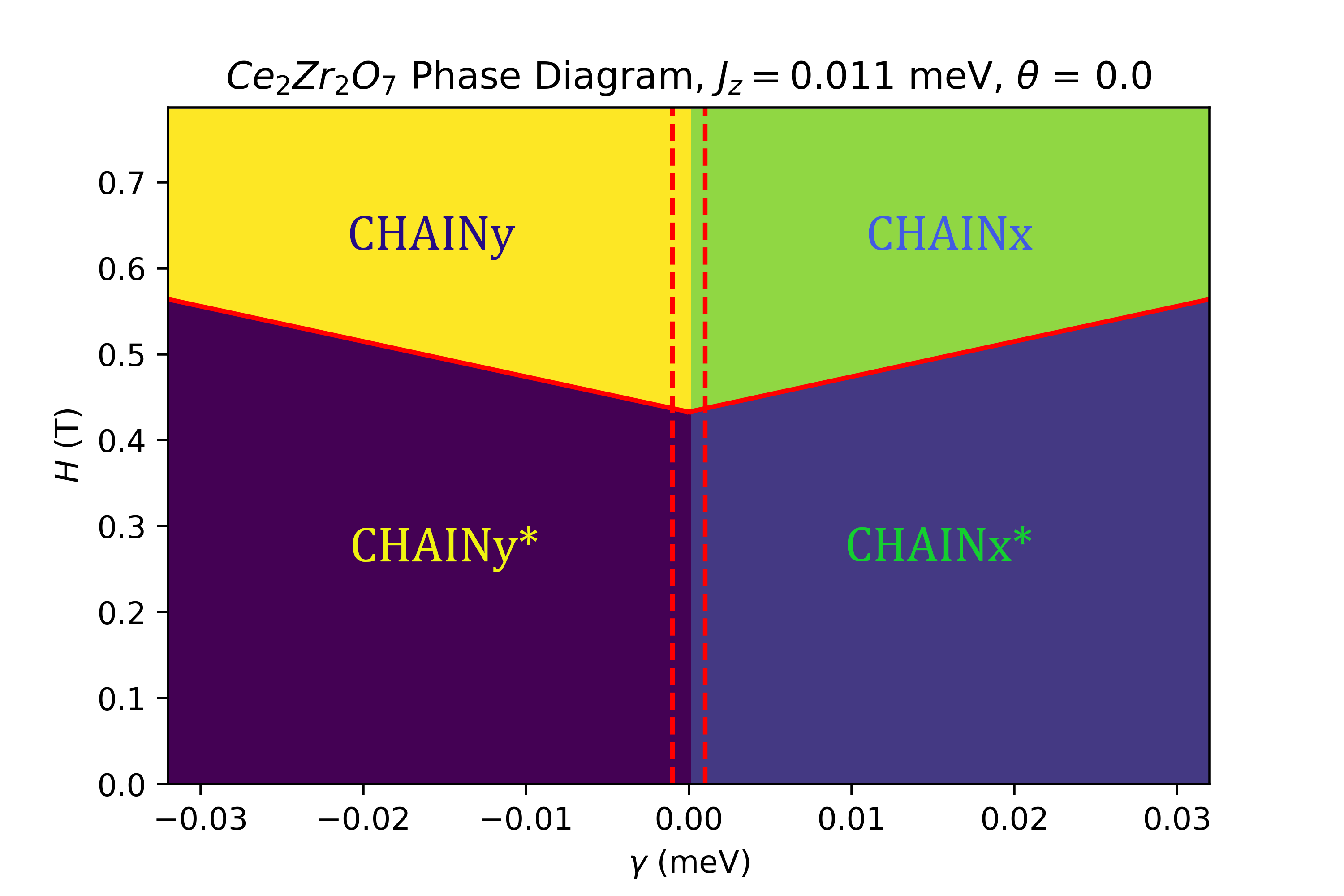}}
\subfigure[]
{\includegraphics[scale=0.38]{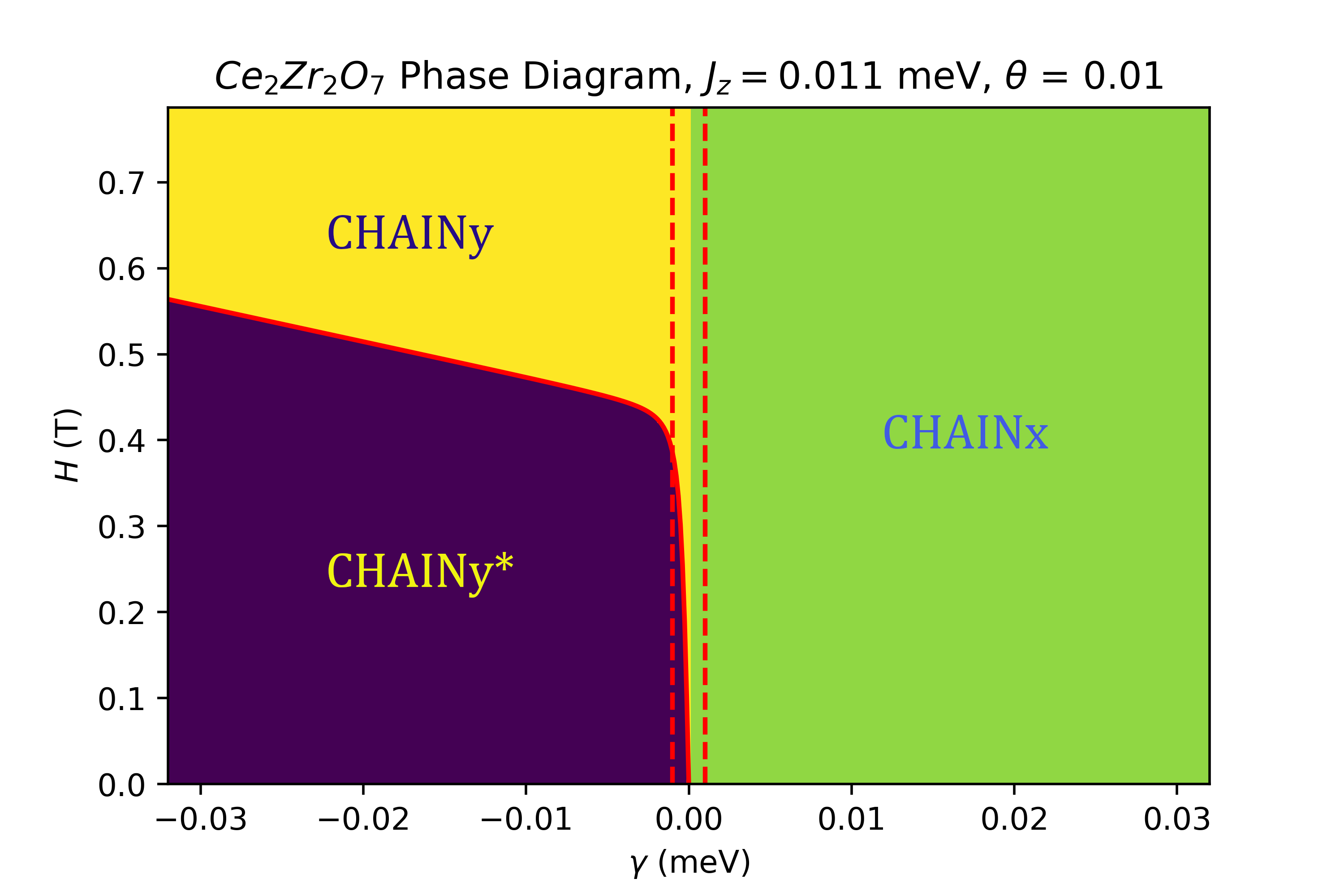}}
\subfigure[]
{\includegraphics[scale=0.38]{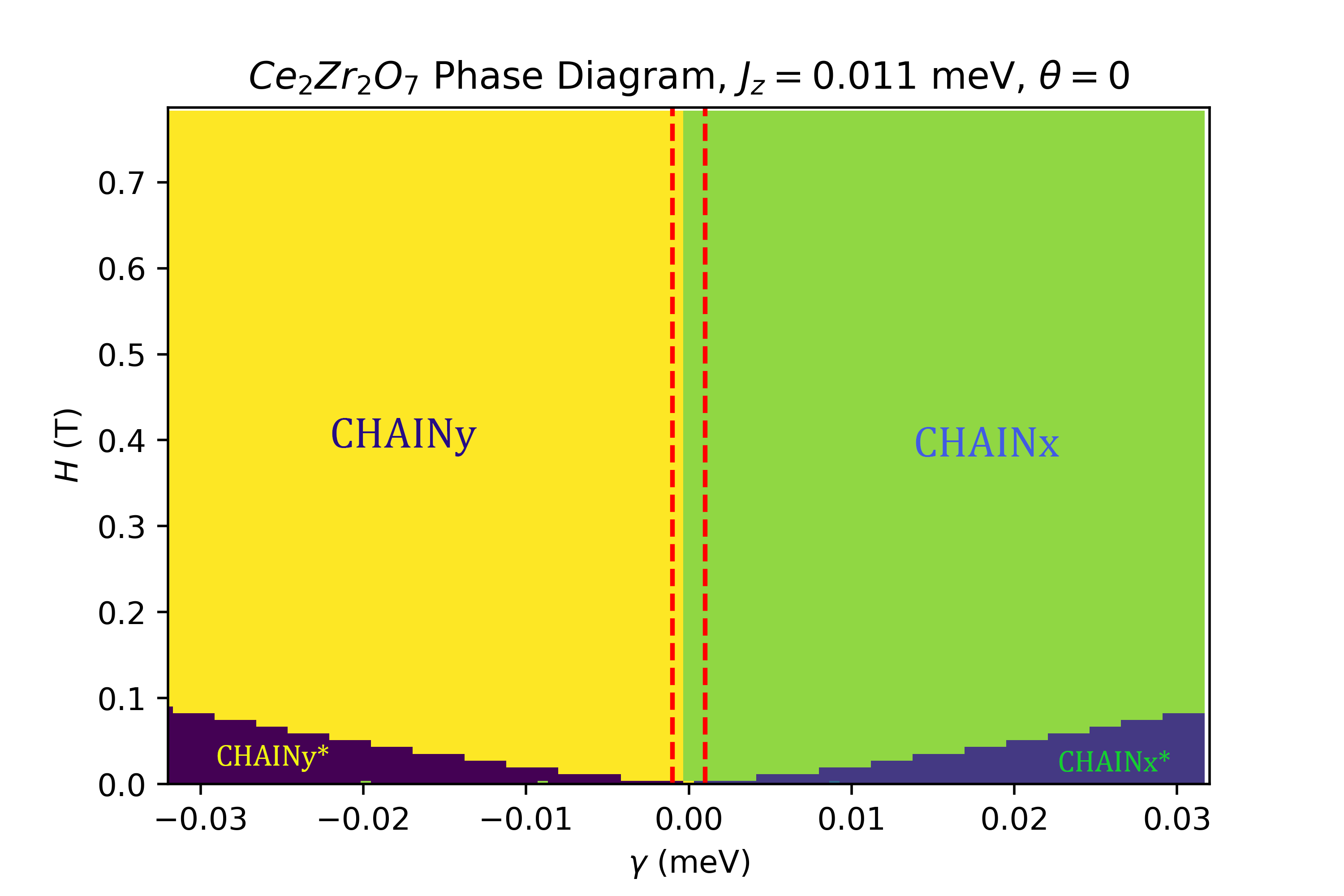}}
\caption{Ground state phase diagrams for Ce$_2$Zr$_2$O$_7$ for a finite [110] field, for parameters in the vicinity of the values determined in \cite{smith22}.
Phase diagrams are shown as a function of field strength and $\gamma=J_x-J_y$.
In all figures, the two vertical dotted red lines represent the predicted model parameters of Ce$_2$Zr$_2$O$_7$ with increasing magnetic field $H$, which place the material close to a transition between states with dominant correlations along the $\tilde{x}$ and $\tilde{y}$ pseudospin axes.
(a) Classical mean field phase diagram with $\theta=0$ [Eq. (\ref{eq:Pseudo x rotation})-(\ref{eq:Pseudo rotation angle})].
In this case, with $\theta$  precisely vanishing, the system transitions to a CHAIN$_{\lambda}^*$ regime at finite field ($\lambda=\tilde{x}, \tilde{y}$), and then through a second order transition to a CHAIN$_{\lambda}$ phase for a critical value of $H$.
(b) Classical mean field phase diagram with $\theta=0.01$.
Here the transition at finite field vanishes for $J_x>J_y (\gamma>0)$, due to the small but finite value of $\theta$ \cite{placke20}.
(c) Phase diagram including the effects of 
 intra-chain quantum fluctuations with $\theta=0$. It is observed that the CHAIN$_{\lambda}$ phases, within which the $\alpha$ chains are polarized parallel to the [110] field, are greatly stabilised, with the CHAIN$_{\lambda}^{\ast}$ regimes vanishing in the isotropic limit $J_x=J_y$.}
\label{fig:pd_czo}
\end{figure*}

\begin{figure}
\includegraphics[scale=0.5]{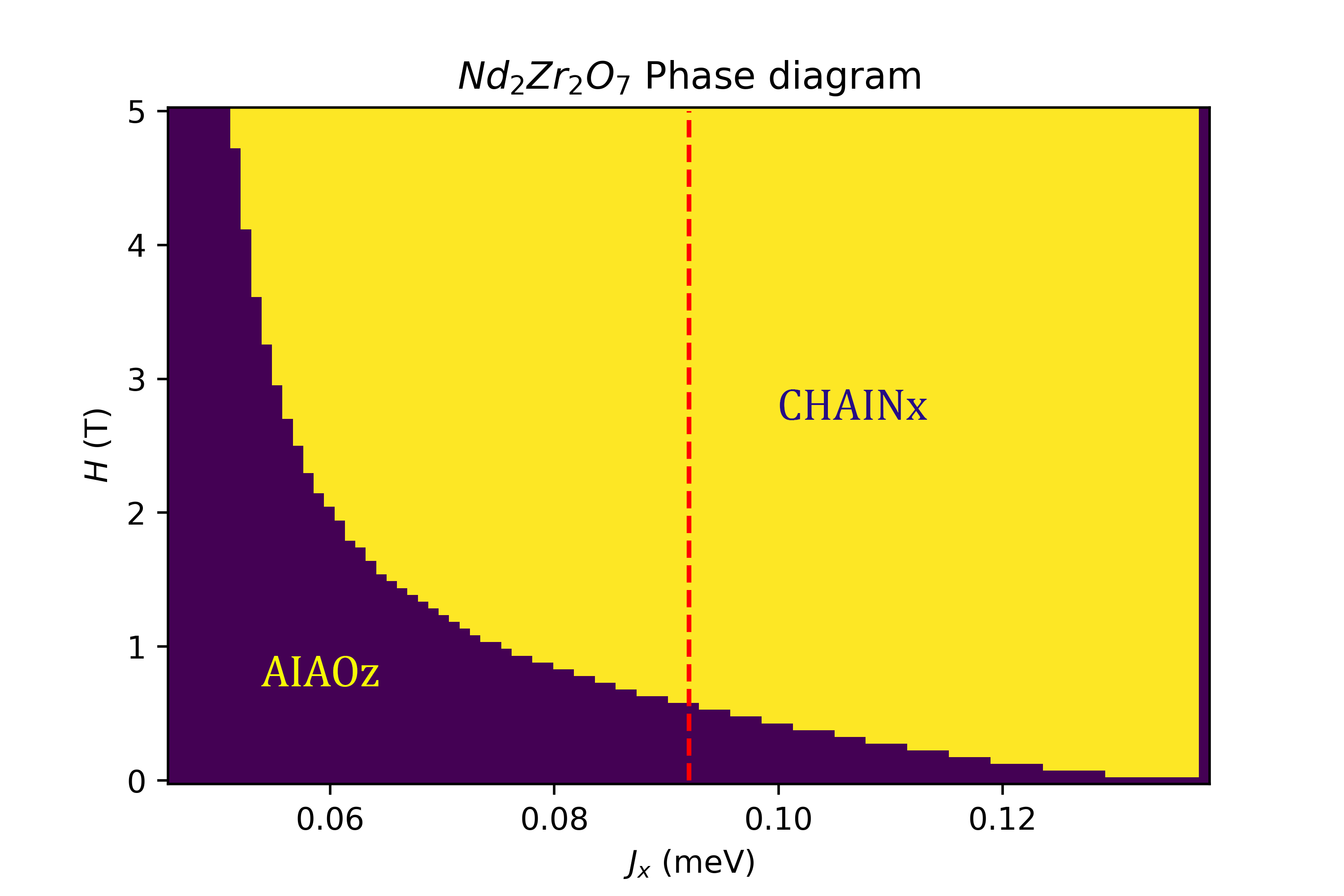}
\caption{Classical mean field ground state phase diagram for parameters close to those of Nd$_2$Zr$_2$O$_7$, as determined in \cite{xu19}, for a finite [110] field. For small to moderate fields, the system retains AIAO$_z$ ordering, however a phase transition is found for fields of approximately 0.6 Tesla to the CHAIN$_x$ phase. We find that, in the regime for which the spin chain model is applicable, the coupling to the field is already comparable in size to the largest coupling constant, $J_x$.}
\label{fig:pd_nzo}
\end{figure}

Fig. \ref{fig:pd_czo} and Fig. \ref{fig:pd_nzo} present phase diagrams for the Hamiltonian of Eq. (\ref{eq:DO Hamiltonian rotated basis}) with a [110] field, for model parameters close to those of Ce$_2$Zr$_2$O$_7$ and Nd$_2$Zr$_2$O$_7$. It is found within classical mean field theory that the AIAO$_{\lambda}$ phases remain stable to a finite field, whilst the SI$_{\lambda}$ regimes are replaced by CHAIN$_{\lambda}$ phases, wherein the system orders along the $\alpha$ and $\beta$ chains. Within the CHAIN phases, the classical ground states still obey the `2-in-2-out' rule, but the ground state degeneracy becomes subextensive due to spins on the $\alpha$ chains preferentially orienting themselves parallel to the applied field. With the $\alpha$ chain spins pinned by the external field there are two spin configurations for each $\beta$ chain that preserve the ice rules on the tetrahedra they intersect. This leaves an Ising degree of freedom on each $\beta$ chain in the ground state.

This is at least the picture for the CHAIN$_x$ and CHAIN$_z$ phases. The ${S}^{\tilde{y}}$ pseudospins do not however couple to the external field, and so one finds for low fields that the $\alpha$ chains also retain an Ising degree of freedom in the ground states \cite{placke20}. When the field strength becomes sufficiently large, the coupling between the field and the small $x$ and $z$ components of the pseudospin dominates, and the $\alpha$ chains are again pinned. The first phase at low fields is denoted CHAIN$_y^*$, which has a second order phase transition at some critical field to the CHAIN$_y$ phase. It is also worth noting that at special values of theta, $\theta=0,\pi/2,\pi,3\pi/2$, the coupling between the field and one of ${S}^{\tilde{x}}$ or ${S}^{\tilde{z}}$ vanishes, allowing for the possibility of CHAIN$_x^*$ or CHAIN$_z^*$ regimes at these points.

It is important to note that, at the mean field level, interactions between spins on differing chains are frustrated, and cancel out within each of the CHAIN ground states. One can also demonstrate that the lifting of the ground state degeneracy in the CHAIN phases through quantum fluctuations occurs at sixth order in perturbation theory, resulting in a very small energy splitting \cite{placke20} much less important than the effects of quantum fluctuations within each chain, which will consider in detail in this paper.

Let us specialise now to Ce$_2$Zr$_2$O$_7$ and Nd$_2$Zr$_2$O$_7$. These materials have the following experimentally estimated model parameters:
\begin{align}
(J_x,J_y) = (0.064,0.063)& \ \text{meV} \ \ J_z = 0.011 \ \text{meV}; \nonumber \\ \ \  \theta \approx 0 \ \text{rad};& \ \ g_z = 2.57, \nonumber
\end{align}
for Ce$_2$Zr$_2$O$_7$ \cite{smith22, smith23}, with $(...)$ indicating that the values of $J_x$ and $J_y$ can be swapped without affecting the agreement with experiment , and 
\begin{align}
J_x = 0.091(9) \ \text{meV};& \ \ J_y = 0.014(6) \ \text{meV}; \nonumber \\ \ \ J_z = -0.046(2) \ \text{meV}; \ \  \theta =& 0.98(3) \ \text{rad}; \ \ g_z = 5.0(1), \nonumber
\end{align}
for Nd$_2$Zr$_2$O$_7$ \cite{xu19}.

Looking first at Ce$_2$Zr$_2$O$_7$, we find that the experimentally estimated parameters lie close to the boundary between the SI$_x$ and SI$_y$ regimes in zero field. If we enforce strictly that $\theta=0$, then for finite field the $T=0$ classical phase diagram is as presented in  Fig. \ref{fig:pd_czo}(a). The two red lines present the evolution of the two experimental possibilities (permutations of $J_x$ and $J_y$) for the location in parameter space with increasing $H$. In either case, we evolve through a CHAIN$_{\lambda}^*$ regime until a critical field is reached, at which point there is a second order phase transition into a CHAIN phase with fully polarized $\alpha$ chains. The CHAIN$_x^*$ regime is however unstable to any finite change in $\theta$, as shown in the second panel of Fig. \ref{fig:pd_czo}. For $\theta=0.01$, the system transitions to a CHAIN$_x$ phase immediately with increasing $H$ if $J_x>J_y$. If $J_y$ is instead the dominant positive exchange parameter, then as the dominant interaction is octupolar, we still observe a finite range of $h$ for which the CHAIN$_y^*$ phase is found, before a transition to the pinned $\alpha$ chain phase.

In the third panel of Fig. \ref{fig:pd_czo} we present the phase diagram predicted for $\theta=0$ if quantum fluctuations on each chain are considered. Further details on these calculations will be presented in Section \ref{subsec:hartreefock}. Using the result that interactions between spins on neighbouring chains cancel out for classical ground states, and that energy splittings between these ground states is introduced at high orders in perturbation theory, we take each chain to be an independent quantum spin chain. The phase of the ground state on the $\beta$ chains is obviously independent of the applied field, and so the quantum phase transitions observed occur, as a function of $h$, on the $\alpha$ chains.

We find that the effect of these quantum fluctuations is to drastically lower the critical field at which the transition between the CHAIN$_{\lambda}^{\ast}$ and CHAIN$_{\lambda}$ phases occurs. At $J_x=J_y$ this line passes through $H=0$, indicating that along this phase boundary any finite field will remove the ground state degeneracy of the $\alpha$ chains.

We also see that $J_x=J_y$ marks a transition between distinct chain phases.
This transition is a $T=0$ critical point, and is associated to the emergence of gapless excitations on the $\beta$ chains.
The proximity of Ce$_2$Zr$_2$O$_7$ to this critical point, and the possible application of 
2DCS in probing this is discussed further in Section \ref{sec:results}.

Experimental model parameters for Nd$_2$Zr$_2$O$_7$ place it within the AIAO$_z$ phase, which is stable to finite magnetic fields. For a sufficiently large field however, it has been observed \cite{xu19}, and is predicted in classical mean field theory, that a transition to a CHAIN$_x$ phase occurs at finite fields $H$. Fig. \ref{fig:pd_nzo} presents the classical mean field phase diagram for Nd$_2$Zr$_2$O$_7$ with increasing field strength. Classical mean-field theory predicts a transition at a field strength of around 0.6 $\rm{T}$, which is rather higher than found in experiments \cite{xu18}, suggesting that mean field theory overestimates the stability of the AIAO order. We will focus on a high field limit for Nd$_2$Zr$_2$O$_7$, where we are certainly in the CHAIN$_x$ phase, and within which we can treat the pinned $\alpha$ chains perturbatively.

For both Ce$_2$Zr$_2$O$_7$ and Nd$_2$Zr$_2$O$_7$, placed within a 
sufficiently strong [110] magnetic field, we then expect to observe chain-like structures, which behave which behave approximately independently from one another on energy scales greater than $\sim 10^{-4} J$ \cite{placke20}, where $J$ is of the scale of the coupling parameters. Within this picture, the leading order quantum fluctuations are excitations which propagate along individual $\alpha$ and $\beta$ chains, and are either domain walls (spinons), or spin flips (magnons) depending on the quantum phase of that chain.

\subsection{Two dimensional coherent spectroscopy (2DCS)}
\label{subsec:2DCS}

In systems with fractionalized excitations, or in other situations where particles are created in groups, any zero momentum probe will produce a continuum of different excitations due to the many different ways of satisfying momentum conservation. For the example of pairs of spinons on a one-dimensional spin chain, we find absorption peaks at frequencies $\epsilon=\lambda_{-k}+\lambda_{k}$, where $\lambda_k$ is the energy of a spinon with momentum $k$. In linear response, we thus find a broad continuum, potentially difficult to distinguish from a single peak broadened by life-time or disorder effects. The key application of two-dimensional coherent spectroscopy is to produce sharp signatures of such excitation continua and remove this ambiguity.

2DCS is a non-linear technique which produces response functions of two measured frequencies $\omega_t$ and $\omega_{\tau}$. The desired sharp signatures of continuum excitations are diagonal streaks within this $\omega_t-\omega_{\tau}$ plane.
This includes a ``rephasing'' or ``spinon echo'' streak along  the line $\omega_t =- \omega_{\tau}$,
for which lifetime effects broaden the response transverse to the peak, while the continuum bandwidth
is measured by the length of the streak \cite{wan19}.
This allows a clean separation between fractionalisation and lifetime effects.

The procedure we consider is as set out in \cite{wan19}. We couple to the magnetic dipole moments in the magnet through weak, uniform, pulses of magnetic field, and then measure the response of the total magnetisation response to these probe fields along a certain axis. Two pulses are applied, a pulse $A$ at time zero, and a pulse $B$ at time $\tau$. The total magnetisation is then measured at a later time $t+\tau$, denoted $M_{AB}(t+\tau)$. The signature we are interested lies in the non-linear part of the response, thus the linear part of the response must be removed. The magnetisation response to the pulses $A$ and $B$ alone must then be measured and the results subtracted off the full response. If the time integrated magnetic field strengths, in units of $\hbar$, for the two pulses are $A_0$ and $A_{\tau}$ and we assume that each pulse is sufficiently brief that they can be approximated as delta functions, one finds that the non-linear response of the magnetisation is given by \cite{wan19}:
\begin{multline}
    M_{NL}(t,\tau)=A_0A_{\tau}\chi^{(2)}(t,t+\tau)+A_0A_{\tau}^2\chi^{(3)}(t,t,t+\tau) \\ +A_0^2A_{\tau}\chi^{(3)}(t,t+\tau,t+\tau) + ... \ \label{eq:Non_linear_response}.
\end{multline}
If the probe fields are assumed to be perturbatively weak, we can evaluate the higher order magnetisation-magnetisation susceptibilities $\chi^{(n)}$ using Kubo formulae. If we measure the magnetisation in the same direction as the probe fields, then we can express these susceptibilities in terms of a single magnetisation $M$ parallel to an as yet unspecified axis. We then define the second and third order susceptibilities as:
\begin{equation}
    \chi^{(2)}(t,t+\tau)=\frac{i\Theta(t)\Theta(\tau)}{V}\langle[[M(t+\tau),M(\tau)],M(0)]\rangle, \label{eq:Kubo_second_order}
\end{equation}
and
\begin{multline}
    \chi^{(3)}(t_3,t_3+t_2,t_3+t_2+t_1)= \\
\frac{i\Theta(t_1)\Theta(t_2)\Theta(t_3)}{V}\langle[[[M(t_3+t_2+t_1),M(t_2+t_1)],M(t_1)],M(0)]\rangle. \\\label{eq:Kubo_third_order}
\end{multline}
Here $V$ is the total number of sites, $\Theta(t)$ is a step function ensuring causality and the expectation value is taken with respect to the ground state. Finite temperature effects can be included by replacing this expectation value with a trace of the commutator with a thermal density matrix, which can be evaluated in certain cases using Keldysh contour techniques \cite{hart23}. We find that for the analytically tractable spin-chain models we consider, finite temperatures generically lead to a suppression of the non-linear response, with the contribution from absorption at a given energy $\epsilon$ falling away with $T$  as $\tanh\left(\frac{\epsilon}{2T}\right)$. As we are chiefly interested in polarization effects on the shape of the response in the two dimensional frequency plane, we elect to work at $T=0$ throughout this work.

The choice of polarization for the probe fields and the measurement axis determines the operator $M$ entering into the Kubo formulae. In the magnetic dipole limit of coupling to an external magnetic field, this will in general be a linear combination of spin operators, and so altering the polarization axis will have highly non-trivial effects on the commutators in the Kubo formulae. 
In the particular example of dipolar-octupolar pyrochlores, the coupling of magnetic dipoles to a probe field field $\mathbf{m}$ is relatively simple; as shown in Eq. (\ref{eq:magnetisation definition}) it is just proportional to $\mathbf{m} \cdot \mathbf{\hat{z}}_i S^z$, where $\mathbf{\hat{z}}_i$ is the local $z$ axis for each of the four fcc sublattices. The effect of the polarization on the optical matrix elements thus only enters as a numerical factor and the operator content of $M$ is unchanged, which greatly simplifies our analysis. 

Even with this simplification, the variation of $\mathbf{m} \cdot \mathbf{\hat{z}}_i$ has a significant impact on how the 2DCS response probes the excitations of dipolar-octupolar pyrochlores. If we align both of the probe fields, and the axis along which we measure the magnetisation response, with the [1$\bar{1}$0] direction then we can couple exclusively to the $\beta$ chains, whilst a [110] polarization singles out the $\alpha$ chains.

For both of these polarizations, the sign of $\mathbf{m} \cdot \mathbf{\hat{z}}_i$ varies from one site to the next along the relevant chains, however if we choose to perform the experiment with a [001] polarization, we find uniform strength coupling to all sites with constant sign along each $\alpha$ and $\beta$ chain (but a relative sign between types of chain). Comparing the magnetisation operator in the [001] case to the [110] and [1$\bar{1}$0] cases, there is an effective momentum shift of $\pi$, and so one can use the [001] polarization to probe different excitation processes within the system.

Throughout this work, our analysis of the 2DCS response of our chosen materials will be within an independent chain approximation, and we will focus on the [001], [110], and [1$\bar{1}$0] polarization set ups. We will then need the second and third order magnetic susceptibilities of both the $\alpha$ and $\beta$ chains to a probe field which alternates in sign from site to site: $\chi^{(n)}_{\alpha,\text{alt}}$ and $\chi^{(n)}_{\beta,\text{alt}}$, and the susceptibilities for a field with uniform sign along each chain: $\chi^{(n)}_{\alpha,\text{direct}}$ and $\chi^{(n)}_{\beta,\text{direct}}$.

To be precise, for the [110] polarization, the relevant susceptibilities $\chi_{\parallel}^{(n)}$, and the required magnetisation operator are defined as:
\begin{align}
    \chi_{\parallel}^{(n)} =& \left(\frac{\mu_B g_z}{\sqrt{3}}\right)^n 2 \ \chi^{(n)}_{\alpha,\text{alt}} ,\\
    M_{\alpha,\text{alt}} =& \sum_{j\in \alpha} (-1)^j \left(\cos\theta{S}^{\tilde{z}}_j+\sin\theta{S}^{\tilde{x}}_j\right).
\end{align}
The `parallel' subscript is used to denote this polarization mode as it is parallel to the constant, uniform applied field $\mathbf{H}$.

For the [1$\bar{1}$0] polarization, we similarly define $\chi^{(n)}_{\perp}$ and their relevant magnetisation operator to be:
\begin{align}
    \chi_{\perp}^{(n)} =& \left(\frac{\mu_B g_z}{\sqrt{3}}\right)^n 2 \ \chi^{(n)}_{\beta,\text{alt}}, \\
    M_{\beta, \text{alt}} =& \sum_{j\in \beta} (-1)^j \left(\cos\theta{S}^{\tilde{z}}_j+\sin\theta {S}^{\tilde{x}}_j\right).
\end{align}

Finally, for the [001] polarization, the susceptibilities $\chi^{(n)}_z$ involve contribution from both $\alpha$ and $\beta$ chains. Both chains have equal magnetisation operators, up to a sign which we factor out of the $\beta$ susceptibilities:
\begin{align}
    \chi_{z}^{(n)} =& \left(\frac{\mu_B g_z}{\sqrt{3}}\right)^n ( \ \chi^{(n)}_{\alpha,\text{direct}} - (-1)^n \chi^{(n)}_{\beta,\text{direct}}), \\
    M_{\nu, \text{direct}} =& \sum_{j\in \nu} \left(\cos\theta {S}^{\tilde{z}}_j+\sin\theta {S}^{\tilde{x}}_j\right), \ \ \nu=\alpha,\beta .
\end{align}

In the following sections, we  examine how the 2DCS response varies between these three polarization modes, and how this can be used to probe the properties of our example materials Ce$_2$Zr$_2$O$_7$ and Nd$_2$Zr$_2$O$_7$. In the following sections, we shall analyse the functions $\chi^{(n)}_{\alpha,\text{alt}},\chi^{(n)}_{\beta,\text{alt}}$,$\chi^{(n)}_{\alpha,\text{direct}}$, and $\chi^{(n)}_{\beta,\text{direct}}$.This analysis will primarily be through a mean field Hartree-Fock treatment, however we make use of a high field perturbation expansion for the $\alpha$ chains of Nd$_2$Zr$_2$O$_7$, as the non-zero value of the dipolar-octupolar mixing angle $\theta$ introduces interaction terms to the Hamiltonian that are otherwise difficult to handle analytically.

\section{Analysis}
\label{sec:calculations}

\subsection{Non-interacting fermions $(J_{{z}}=0, \theta=0)$}
\label{subsec:noninteract}

The Hamiltonian of Eq. (\ref{eq:DO Hamiltonian rotated basis}) can be mapped to a system of non-interacting fermions for $J_z=0, \theta=0$ \cite{lieb61}, and fortunately this limit is very close to the model parameters predicted for Ce$_2$Zr$_2$O$_7$. Further, the most troublesome non-integrable term for general parameters is the effectively twisted field acting on the pseudospins for non-zero values of dipolar-octupolar mixing angle $\theta$, which does not appear in the Hamiltonian of the $\beta$ chains. We thus expect the behaviour on $\beta$ chains at least to also be well described by the exactly solvable limit more generally, provided $J_z$ is not too large.

We thus begin our calculation of the required susceptibilities by evaluating them within this exactly soluble limit. This will entail finding the fermionic modes that diagonalize Eq. (\ref{eq:DO Hamiltonian rotated basis}) in this limit, and much of this work follows that done elsewhere \cite{wan19}.

We choose always to work in a local spin basis for which the total magnetisation operator $M_{\mu,\text{alt}}$ or $M_{\mu,\text{direct}}$ is uniform. In the case of $M_{\mu,\text{alt}}$, this requires a rotation of $\pi$ about the local $y$ axis on every alternate spin site, which both removes the alternating $(-1)^j$ factor from the Hamiltonian, and introduces a minus sign in front of $J_x$ ($J_z=0$ for now, but this coupling also has its sign flipped). For the `alternating' case then, the exactly solvable Hamiltonian expressed in this new basis reads:
\begin{equation}
    H_{0} = \frac{1}{4}\sum_{j} -J_x \sigma^x_j \sigma^x_{j+1}+ J_y \sigma^y_j \sigma^y_{j+1} -\frac{h}{2}\sum_j \sigma^z_j. \label{eq:Integrable_Hamiltonian}
\end{equation}
We have taken out a factor of $1/2$ from each pseudospin operator, and have retained the definitions of the parameters in Eq. (\ref{eq:DO Hamiltonian rotated basis}) for clarity. The magnetisation operator in this new basis is:
\begin{equation}
    M=\frac{1}{2}\sum_j \sigma^z_j.\label{eq:Integrable_magnetisation}
\end{equation}
To evaluate $\chi^{(n)}_{\alpha,\text{alt}}$ or $\chi^{(n)}_{\beta,\text{alt}}$, we need to time evolve Eq. (\ref{eq:Integrable_magnetisation}) and evaluate the nested commutators in the Kubo formulae. For Eq. (\ref{eq:Integrable_Hamiltonian}), the fermionization procedure follows that given in \cite{wan19}, which we will only sketch here. One can map Eq. (\ref{eq:Integrable_Hamiltonian}) to a quadratic fermion Hamiltonian via the Jordan-Wigner transformation. This is then diagonalised using a suitable Bogoliubov transformation, with Bogoliubov fermions defined in terms of the Jordan-Wigner fermions as $c_k=\cos\frac{\phi_k}{2}\psi_k+i\sin\frac{\phi_k}{2}\psi^{\dagger}_{-k}$. The Hamiltonian is reduced to:
\begin{equation}
    H_{0}= \sum_{k>0} \lambda_k \bigl(c^{\dagger}_kc_k-c_{-k}c^{\dagger}_{-k}\bigr), \label{eq:bog_Hamiltonian}
\end{equation}
with the functions $\sin\phi_k$, $\cos\phi_k$, and $\lambda_k$ given by
\begin{footnotesize}
\begin{align}
\sin\phi_k =& \frac{-(J_x+J_y)\sin(k)}{\sqrt{J^2_x+J^2_y+4h^2-4h(-J_x+J_y)\cos(k)-2J_xJ_y\cos(2k)}}, \label{eq:XYh_sin_theta}\\
\cos\phi_k =& \frac{(-J_x+J_y)\cos(k)-2h}{\sqrt{J^2_x+J^2_y+4h^2-4h(-J_x+J_y)\cos(k)-2J_xJ_y\cos(2k)}}, \label{eq:XYh_cos_theta}\\
\lambda_k =& \frac{1}{2}\sqrt{4h^2+J_x^2+J^2_y-4h(-J_x+J_y)\cos(k)-2J_xJ_y\cos(2k)}. \label{eq:XYh_lambda}
\end{align}
\end{footnotesize}
The time dependent probe field operator $M(t)$ can be expressed in terms of the $c_k$ fermions in the same way as outlined in \cite{wan19}, and then inserted into either Eq. (\ref{eq:Kubo_second_order}) or (\ref{eq:Kubo_third_order}) to determine the non-linear susceptibilities. Dephasing/depolarising and lifetime effects can also be included phenomenologically by using a quantum channel model for time evolution, as described in the supplementary material for \cite{wan19}. Including these effects, one arrives at the following expressions for the second and third order susceptibilities in the limit $J_z=0$, $\theta=0$:

\begin{multline}
    \chi^{(2)}_{zzz}(t,t+\tau) = \frac{4\Theta(t)\Theta(\tau)}{L} \sum_{k>0} \sin^2\phi_k\cos\phi_k \\ \times \left(\cos(2\lambda_k\tau)e^{-t/T_1}-\cos(2\lambda_k(t+\tau))e^{-(t+\tau)/T_2}\right) \label{eq:Chi2_exact},
\end{multline}
and
\begin{multline}
\chi^{(3)}_{zzzz}(t_3,t_2+t_3,t_1+t_2+t_3) =  \\ -\frac{\Theta(t_1)\Theta(t_2)\Theta(t_3)}{L} \sum_k A^{(1)}_k +A^{(2)}_k + A^{(3)}_k + A^{(4)}_k, \label{eq:Chi3_exact}
\end{multline}
with
\begin{align}
A^{(1)}_k =& 8\sin^2 \phi_k \cos^2 \phi_k \sin(2\lambda_k(t_1+t_2+t_3))\nonumber\\&e^{-(t_1+t_2+t_3)/T_2},
\label{eq:A1}
\\ A^{(2)}_k =& -8\sin^2 \phi_k \cos^2 \phi_k \sin(2\lambda_k(t_1+t_2))\nonumber
\\&e^{-(t_1+t_2)/T_2}e^{-t_1/T_1},
\label{eq:A2}
\\  A^{(3)}_k =& 4\sin^{4}\phi_k \sin(2\lambda_k(t_3+t_1))e^{-(t_1+t_3)/T_2}e^{-t_2/T_1},
\label{eq:A3}
\\ 
A^{(4)}_k =& 4\sin^{4}\phi_k \sin(2\lambda_k(t_3-t_1))e^{-(t_1+t_3)/T_2}e^{-t_2/T_1}.
\label{eq:A4}
\end{align}
Here $T_1$ and $T_2$ are phenomenological time scales characterising depolarization and dephasing respectively, and are as defined in \cite{wan19}.

At this exactly soluble point, we note that the second order susceptibilities vanish if $h=0$, and so $\chi^{(2)}_{\beta,\text{alt}}$ always vanishes, since the beta chains don't feel the field.

We note here that the Jordan-Wigner fermion construction is valid in both low and high field phases, and so equally well describes spinons for small $h$ and magnons for large $h$. However, only spinons are fractionalized excitations, whilst continuum responses are predicted in all limits. The existence of the continuum only indicates the creation of multiple particles in the interaction with the probe field, and does not necessarily rule out singly created particle excitations. As discussed in \cite{wan19}, excitations of single particles can be created in high field limits of quantum spin chains by probing the system using a transverse field ($S^x$ or $S^y$). For dipolar-octupolar pyrochlores, such processes would occur through magnetic octupole coupling, or magnetic dipole coupling for non-zero mixing angle $\theta$, as discussed in Section \ref{subsec:lswt_and_ed}. 

Also note that in the high field limit the function $\sin\phi_k$ falls away as $J_{\mu}/h$, and so all parts of the response diminish at least as fast as $J_{\mu}^2/h^2$, with $A^{(3)}_k$ and $A^{(4)}_k$ falling off with $J_{\mu}^4/h^4$. The entire response formally vanishes in the infinite field limit, where excitations are non-dynamic magnons (single spin flips).

We now move to consider $\chi^{(n)}_{\alpha,\text{direct}}$ and $\chi^{(n)}_{\beta,\text{direct}}$, for which we effectively have an alternating external field $h$ but a uniform probe field along each chain. $\chi^{(n)}_{\beta,\text{direct}}$ involves no external field, so can be calculated directly using the above results with the sign of $J_x$ reversed. Accordingly, $\chi^{(2)}_{\beta,\text{direct}}$ also vanishes. For $\chi^{(n)}_{\alpha,\text{direct}}$ however, we can either work with an alternating field in the Hamiltonian, or put the alternating sign in the magnetisation operator. We choose to do the former, which will result in a two band formulation of the dispersion. The fermionized Hamiltonian in $k$-space is found to be:
\begin{footnotesize}
\begin{equation}
H = \sum_{0<k<\frac{\pi}{2}}
\Psi^{\dagger}_k
\begin{pmatrix}
\epsilon_k & i\Delta_k & 0 & -h \\
-i\Delta_k & -\epsilon_k & h & 0 \\
0 & h & \epsilon_k & i\Delta_k \\
-h & 0 & -i\Delta_k & -\epsilon_k 
\end{pmatrix}
\Psi_k;
\ \ \Psi_k=\begin{pmatrix}
\psi_k \\
\psi^{\dagger}_{-k} \\
\psi^{\dagger}_{\pi-k} \\
\psi_{k-\pi}
\end{pmatrix},
\end{equation}
\end{footnotesize}
where $\epsilon_k=(J_x+J_y)/2\cos(k)$ and $\Delta_k=(J_x-J_y)/2\sin(k)$. One finds that the alternating field term in the Hamiltonian introduces couplings between $k$ and $k-\pi$ momentum fermion modes.

To diagonalize this Hamiltonian, one must  perform a $4\times4$ Bogoliubov transformation. The required unitary transformation is of the form:
\begin{equation}
\begin{pmatrix}
\psi_k \\
\psi^{\dagger}_{-k} \\
\psi^{\dagger}_{\pi-k} \\
\psi_{k-\pi}
\end{pmatrix}
=\frac{1}{\sqrt{2}}\begin{pmatrix}
U_{+} & iU_{-} \\
iU_{+} & U_{-} 
\end{pmatrix}
\begin{pmatrix}
c_{k,+} \\
c^{\dagger}_{-k,+} \\
c_{k,-} \\
c^{\dagger}_{-k,-}
\end{pmatrix},
\end{equation}
with
\begin{equation}
U_{\pm} =
\begin{pmatrix}
\cos\frac{\phi_{k,\pm}}{2} & -i\sin\frac{\phi_{k,\pm}}{2} \\
-i\sin\frac{\phi_{k,\pm}}{2} & \cos\frac{\phi_{k,\pm}}{2}
\end{pmatrix}.
\end{equation}
The final Hamiltonian is then:
\begin{multline}
H = \sum_{0<k<\frac{\pi}{2}} \lambda_{k,+} \Bigl\{c_{k,+}^{\dagger}c_{k,+}-c_{-k,+}c^{\dagger}_{-k,+}\Bigr\} + \\ \lambda_{k,-} \Bigl\{c^{\dagger}_{k,-}c_{k,-}-c_{-k,-}c^{\dagger}_{-k,-}\Bigr\}.
\end{multline}
The new fermions $c_k$ are only defined for momenta in the range $k\in [-\pi/2,\pi/2]$. The Bogoliubov rotation angle, and fermion dispersion relations are found to be:
\begin{footnotesize}
    \begin{align}
    \cos\phi_{k,\pm} =& \frac{(J_x+J_y)\cos(k)}{\sqrt{4h^2+J_x^2+J^2_y+2J_xJ_y\cos(2k)\mp 4h(J_x-J_y)\sin(k)}},  \\
    \sin\phi_{k,\pm} =& \frac{(J_x-J_y)\sin(k)\mp 2h}{\sqrt{4h^2+J_x^2+J^2_y+2J_xJ_y\cos(2k)\mp 4h(J_x-J_y)\sin(k)}}, \\
    \lambda_{k,\pm} =& \frac{1}{2}\sqrt{4h^2+J_x^2+J_y^2+2J_xJ_y\cos(2k)\mp 4h(J_x-J_y)\sin(k)}. 
\end{align}
\end{footnotesize}
As this is ultimately the same $\alpha$ chain Hamiltonian diagonalized in a different basis, the two band spectrum is entirely equivalent to the single band case.

Translating the magnetisation operator $M_{\alpha,\text{direct}}$ into this fermion basis, one obtains:
\begin{footnotesize}
    \begin{equation}
    M_{\alpha,\text{direct}}=C^{\dagger}_k
        \begin{pmatrix}
0 & 0 & i\cos\bar{\phi}_k & \sin\bar{\phi}_k \\
0 & 0 & -\sin\bar{\phi}_k & -i\cos\bar{\phi}_k \\
-i\cos\bar{\phi}_k & -\sin\bar{\phi}_k & 0 & 0 \\
\sin\bar{\phi}_k & i\cos\bar{\phi}_k & 0 & 0 
\end{pmatrix}C_k,
\end{equation}
\end{footnotesize}
where we have introduced the notation $\bar{\phi}_k=(\phi_{k,+}+\phi_{k,-})/2$, and $C_k=(c_{k,+},c^{\dagger}_{-k,+},c_{k,-},
c^{\dagger}_{-k,-})^{T}$. The magnetisation operator now creates fermions with equal and opposite momenta from different energy bands, for a total excitation energy of $2\bar{\lambda}_k=\lambda_{k,+}+\lambda_{k,-}$,and can also facilitate transitions between bands, with and associated energy change $\Delta\lambda_k=\lambda_{k,+}-\lambda_{k,-}$.

Thus, when the second and third order susceptibilities are recalculated, we pick up new time dependencies arising from inter-band transitions. We also find that the second order response vanishes, just as it does in the $h=0$ limit, leaving one only with the third order contribution:
\begin{footnotesize}
    \begin{align}
\chi^{(3)}_{zzzz}&(t_3,t_3+t_2,t_1+t_2+t_3) = \nonumber\\ &\frac{\Theta(t_1)\Theta(t_2)\Theta(t_3)}{L} \sum_{0<k<\frac{\pi}{2}} A^{(1)}+  A^{(2)}+  A^{(3)}+  A^{(4)}, \label{eq:Chi3_exact_alternating}\\
A^{(1)} =& -16\sin^2\bar{\phi}_k\cos^2\bar{\phi}_k\sin(\bar{2\lambda}_k(t_3+t_2+t_1))\cos(\Delta\lambda_kt_2)\nonumber\\&e^{-(t_1+t_2+t_3)/T_2}, \\
A^{(2)} =& 16\sin^2\bar{\phi}_k\cos^2\bar{\phi}_k\sin(\bar{2\lambda}_k(t_2+t_1))\cos(\Delta\lambda_k(t_3+t_2))\nonumber\\&e^{-(t_1+t_2)/T_2}e^{-t_1/T_1}, \\
A^{(3)} =& -8\sin^4\bar{\phi}_k\sin(\bar{2\lambda}_k(t_3+t_1))e^{-(t_1+t_3)/T_2}e^{-t_2/T_1}, \\
A^{(4)} =& -8\sin^4\bar{\phi}_k\sin(\bar{2\lambda}_k(t_3-t_1))e^{-(t_1+t_3)/T_2}e^{-t_2/T_1}.
\end{align}
\end{footnotesize}
Here we introduce the notation $\bar{\phi}_k=(\phi_{k,+}+\phi_{k,-})/2$. When $h=0$, the two bands are exactly degenerate (equivalent to the symmetry of the positive momentum spectrum about $k=\pi/2$ in the rotated basis), and so we recover the zero field result. 
Taking $T_1$ and $T_2$ to be phenomenological constants independent of other model parameters, we then insert the same exponential in time factors as in Eq. (\ref{eq:Chi2_exact}) and Eq. (\ref{eq:Chi3_exact}), so as to be consistent with the zero field limit.

Whilst the form of the response to a direct rather than an alternating probe field is somewhat altered, it retains the crucial features that make the 2DCS response useful in identifying fractionalized excitations, in particular the rephasing part of the response, $A^{(4)}$ is broadly unchanged, save for the values of the excitation energies. We also note that the additional minus sign in front of $J_x$ in $\chi^{(n)}_{\mu,\text{alt}}$ compare to $\chi^{(n)}_{\mu,\text{direct}}$ greatly alters the resulting 2DCS response, a fact that makes the [001] polarization channel particularly useful in analysing Ce$_2$Zr$_2$O$_7$ and the proximity of its model parameters to the critical point.
 
\subsection{Effect of $J_{\tilde{z}}\neq0$: Hartree-Fock Approximation}
\label{subsec:hartreefock}

Thus far we have only calculated the required non-linear magnetic susceptibilities, $\chi^{(n)}_{\mu,\text{alt}}$ and $\chi^{(n)}_{\mu,\text{direct}}$ at an exactly solvable limit of the model, with $\theta=0$ and $J_z=0$. The latter of these can be relaxed for sufficiently small values of $J_z$, at the expense of a density-density interaction between Jordan-Wigner fermions. If indeed $J_z$ is not too large, this interaction can be treated in the Hartree-Fock approximation.
This is the method that was used to calculate the phase diagram in Fig. \ref{fig:pd_czo}(c), and is also used to calculate the 2DCS response of Ce$_2$Zr$_2$O$_7$ presented in Section \ref{sec:results}.

We choose to work with a basis such that the Hamiltonian takes the form of Eq. (\ref{eq:Integrable_Hamiltonian}). Including finite $J_z$, the modified Hamiltonian now reads:
\begin{multline}
    H=\frac{1}{4}\sum_j (-J_x+J_y)(\psi^{\dagger}_j\psi_{j+1}-\psi_j\psi^{\dagger}_{j+1}) \\ -(J_x+J_y)(\psi^{\dagger}_j\psi^{\dagger}_{j+1}-\psi_j\psi_{j+1}) \\-J_z(4\psi^{\dagger}_j\psi_j\psi^{\dagger}_{j+1}\psi_{j+1}-4\psi^{\dagger}_j\psi_j+1) \\ -\frac{h}{2}\sum_j(2\psi^{\dagger}_j\psi_j-1). \label{eq:H_Jz_fermions}
\end{multline}
We then take the following mean field approximation for the density-density interaction term:
\begin{multline}
\psi^{\dagger}_j\psi_j\psi^{\dagger}_{j+1}\psi_{j+1} \approx
\langle \psi^{\dagger}_j\psi_j \rangle \psi^{\dagger}_{j+1}\psi_{j+1} + \langle \psi^{\dagger}_{j+1}\psi_{j+1} \rangle \psi^{\dagger}_{j}\psi_{j} + \\\langle \psi_j\psi^{\dagger}_{j+1} \rangle  \psi^{\dagger}_{j}\psi_{j+1} + \langle \psi^{\dagger}_j\psi_{j+1} \rangle \psi_{j}\psi^{\dagger}_{j+1} \\ -\langle \psi^{\dagger}_j\psi^{\dagger}_{j+1} \rangle \psi_{j}\psi_{j+1} - \langle \psi_j\psi_{j+1} \rangle \psi^{\dagger}_{j}\psi^{\dagger}_{j+1}.
\end{multline}
Note that one cannot set the anomalous expectation values to zero, as the Bogoliubov rotation angle is non-zero.

The above expectation values are then evaluated using the ground state of the bare $J_z=0$ model, which has the effect of re-scaling the parameters $J_x$, $J_y$, and $h$. A new ground state can then be determined, and a new estimate of the expectation values calculated. This procedure then defines a set of iterative equations for our model parameters, and the functions $\cos\phi_k$ and $\sin\phi_k$ which depend on them:
\begin{align}
    J_x^{(i+1)} =& J_x^{(0)}+\frac{J_z}{L}\sum_k\cos(\phi_k^{(i)}+k), \\
    J_y^{(i+1)} =& J_y^{(0)}-\frac{J_z}{L}\sum_k\cos(\phi_k^{(i)}-k), \\
    h^{(i+1)} =& h^{(0)} -\frac{J_z}{L}\sum_k \cos\phi_k^{(i)}.
\end{align}
If these sequences converge, then the new model parameters are their limiting values. 
We refer to the renormalized parameters generated from the Hartree-Fock procedure as $J_{\alpha}'$, $h'$. 
We importantly find that $\sum_k \cos\phi_k^{(0)}$ vanishes when $h^{(0)}=0$ (equivalent to a half filling of Jordan-Wigner fermion modes) and as a result $h=0$ is a fixed point of the iterative equations. No effective field is generated by interactions, however they can re-scale an already existing field. 

Thus at this order of approximation, the effect of finite $J_z$ is only to re-scale the model parameters. No new processes are introduced, and the 2DCS response is not fundamentally altered.

Further corrections beyond Hartree-Fock can be considered perturbatively, and as shown in \cite{hart23} these lead to finite fermion lifetimes, and other anomalous broadening effects. For our purposes, finite lifetime effects and other sources of broadening will continue to be considered phenomenologically.

\begin{figure*}
\subfigure[]{\includegraphics[scale=0.58]{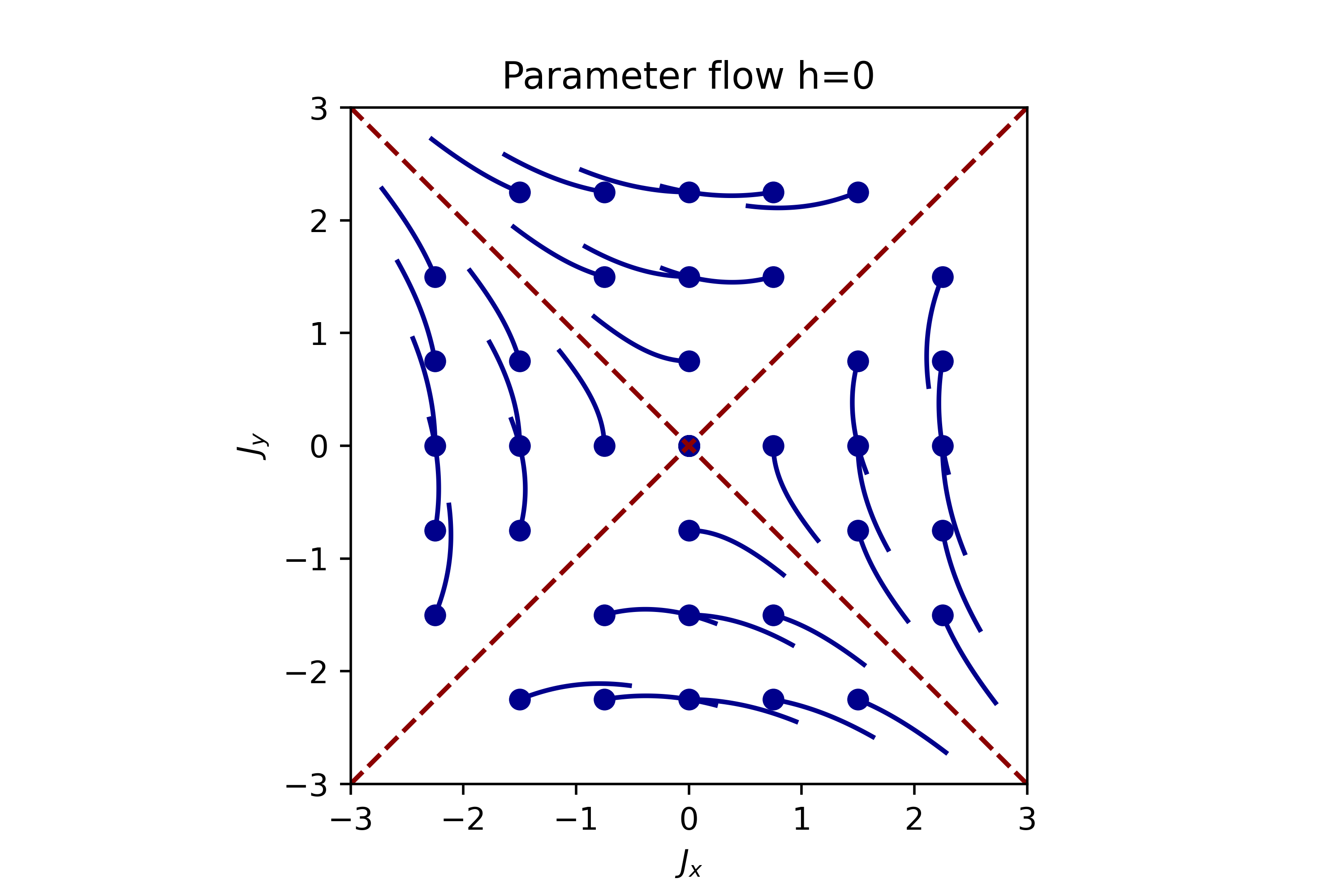}}
\subfigure[]{\includegraphics[scale=0.58]{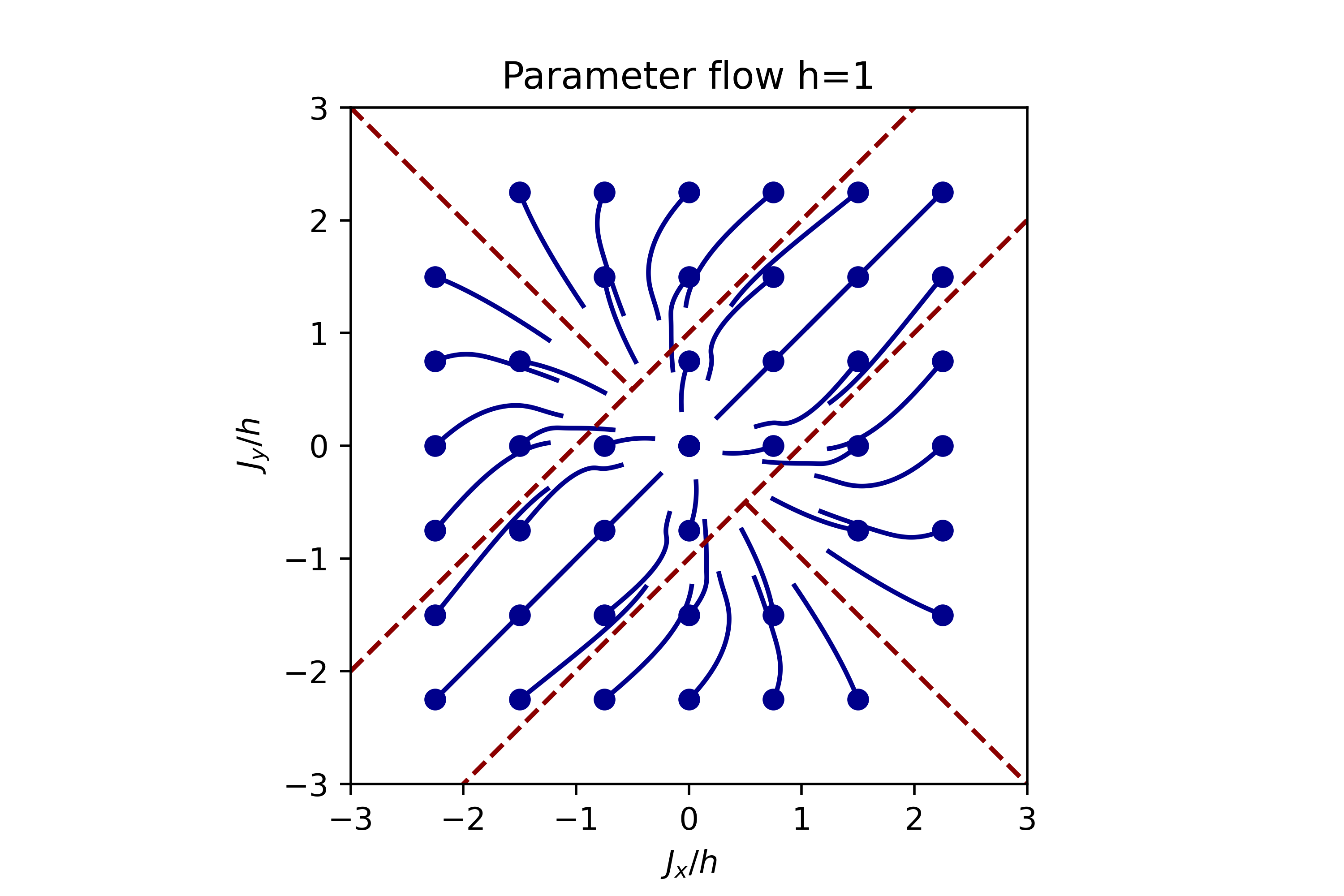}}
\caption{Parameter flows under the Hartree-Fock approximation as $J_z$ is increased for the zero field case ((a)), and for an initial field $h=1$ ((b)). The blue dots show starting parameter values, and the curves show the evolution of the re-scaled parameters as $J_z$ is increased from $0$ to $1$. The red dashed lines indicate phase boundaries. The central region of (b) is the field polarized paramagnetic regime, the four other phases are ferro- or antiferromagnetic phases with dominant correlations along either $\tilde{x}$ or $\tilde{y}$ spin axes. In (b), note that the axes are re-scaled to $J_x/h$ and $J_y/h$, such that the phase diagram can be presented in two dimensions. We observe that no phase boundary crossings occur with increasing $J_z$ in the zero field case, whilst finite $J_z$ can generate transitions into the central paramagnetic phase in the presence of a finite field.}
\label{fig:HF_parameter_flow}
\end{figure*}

Fig. \ref{fig:HF_parameter_flow} shows how the model parameters flow for increasing values of $J_z$, both in the absence of an applied field (relevant for $\beta$ chains) and for $h=1$. The red dotted lines show points where a quantum phase transition occurs in the chain model, accompanied with a gap closing in the excitation spectrum. The phase diagrams shown pertain to the individual spin chains, and a much wider range of parameters is presented than those that correspond to the CHAIN phases of dipolar-octupolar pyrochlores.

If $h=0$, the quantum spin chains host four distinct phases, corresponding to antiferromagnetic, or ferromagnetic ordering parallel to either the $x$ or $y$ pseudo-spins. For example, for large positive $J_y$, the chains are in an antiferromagnetic-$y$ phase. Note that this notion of ferromagnetic vs. antiferromagnetic ordering is basis dependent. For finite fields, a fifth field polarized phase exists across the $J_x=J_y$ diagonal, with ordering parallel to the applied field.

We observe that finite $J_z$ does not lead to any phase changes if $h=0$, provided $J_z$ remains small enough for Hartree-Fock to converge. If $h\neq 0$, then the field polarized phase is stabilised for positive $J_z$, and destabilised for negative $J_z$. Further, unlike in the $h=0$ limit, finite $J_z$ can lead to gap closings, and a transition into or out of the field polarized phase.

For completeness, we also present a formulation of the Hartree-Fock procedure in the basis with an alternating applied field:
\begin{align}
    J_x^{(i+1)} =& J_x^{(0)}+\frac{4J_z}{L}\sum_{0<k<\frac{\pi} {2}}\cos(\bar{\phi}_k^{(i)}+k)\cos\left(\frac{\Delta\phi_k^{(i)}}{2}\right), \\
    J_y^{(i+1)} =& J_y^{(0)}+\frac{4J_z}{L}\sum_{0<k<\frac{\pi} {2}}\cos(\bar{\phi}_k^{(i)}-k)\cos\left(\frac{\Delta\phi_k^{(i)}}{2}\right), \\
    h^{(i+1)} =& h^{(0)} -\frac{2J_z}{L}\sum_{0<k<\frac{\pi} {2}} (\sin\bar{\phi}_{k,+}^{(i)}-\sin\bar{\phi}_{k,-}^{(i)}).
\end{align}
As the Hamiltonian in this basis is just a rewriting of the constant $h$ field Hamiltonian, the resulting phase diagram and parameter flows are identical to that in the right panel of Fig. \ref{fig:HF_parameter_flow}. Note that whilst we have moved between bases where the Zeeman field is alternating and constant, we have not at any point re-defined our parameters, and so Fig. \ref{fig:HF_parameter_flow} is the same for both calculations without the need for any further flips in the signs of $J_x$ and $J_z$.

\begin{figure}
\includegraphics[scale=0.5]{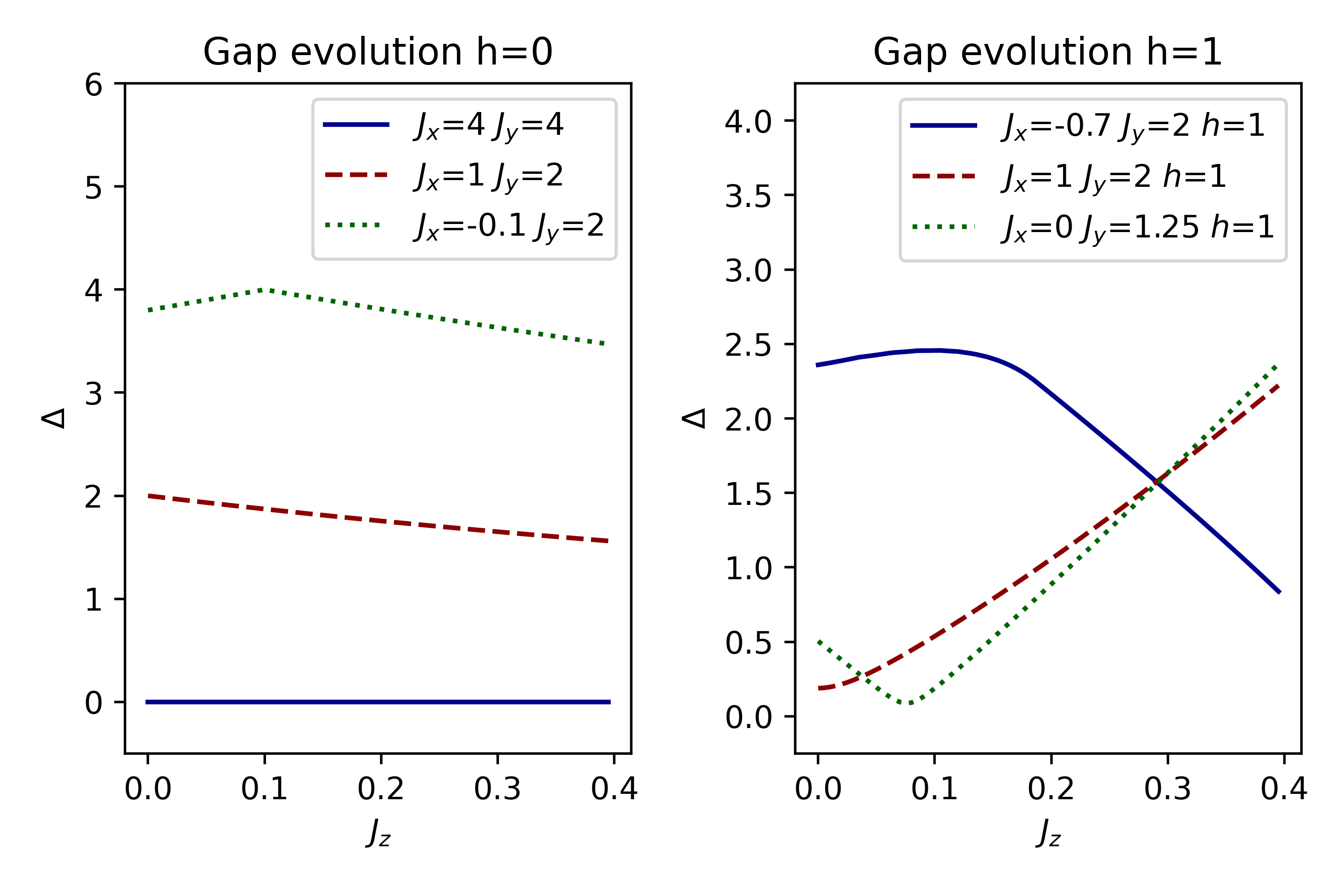}
\caption{
Evolution of the gap in the fermion dispersion, $\Delta$, with
increasing $J_z$, within the Hartree-Fock approximation,
for $h=0$ (left panel) and $h=1$ (right panel).
We observe that moderate values of $J_z$ do not alter the
gapped or gapless character of the dispersion if $h$ is initially zero, however in a finite field $J_z$ can drive the Hamiltonain through a gapless point, as observed in the dotted green line in the right panel.
}
\label{fig:gap_HF}
\end{figure}

\subsection{Effect of $\theta\neq0$: Matrix Pfaffians and perturbation theory}
\label{subsec:lswt_and_ed}

The approximation $\theta=0$ is suitable for Ce$_2$Zr$_2$O$_7$, but Nd$_2$Zr$_2$O$_7$ has a significant dipolar-octupolar mixing angle, which cannot be ignored. For the calculation of the $\beta$ chain responses $\chi^{(n)}_{\beta,\text{alt}}$ and $\chi^{(n)}_{\beta,\text{direct}}$, the effects of finite theta will only be to modify the magnetisation operator $M$, such that it acts on each site as a mixture of $S^z_j$ and $S^x_j$. We have given an analytic treatment of the susceptibility for purely $S^z_j$ polarized probe fields, but more general susceptibilities can be constructed in terms of Majorana operators and matrix Pfaffians \cite{wan19}. In this section we shall briefly review this modification to the calculation, before discussing the effects of finite $\theta$ on the $\alpha$ chains through perturbation theory in the high field limit.

\subsubsection{$\theta \neq 0$ calculation for $\beta$ chains}
The magnetisation operator acting on the $\beta$ chains, with a finite mixing angle $\theta$, can be written in a basis such that it takes the form:
\begin{equation}
    M=\sum_{j\in \beta} \left(\cos\theta {S}^{\tilde{z}}_j+\sin\theta {S}^{\tilde{x}}_j\right).
\end{equation}
In general we will find expectation values involving both ${S}^{\tilde{z}}_j$ and ${S}^{\tilde{x}}_j$ operators contributing to the second and third order susceptibilities. However, as the Hamiltonian itself is unchanged, we still retain rotational symmetry by $\pi$ radians about any of the three principal pseudospin axes, which forces all expectation values appearing in the second order susceptibility to vanish. Using the notation $\chi^{(3)}_{zzzz}$ to represent the contribution to the full susceptibility from terms with four ${S}^{\tilde{z}}_j$ operators, and similar for other combinations of $z$ and $x$ operators, we find that $\chi^{(3)}$ is now given by:
\begin{multline}
    \chi^{(3)}= \cos^4\theta\chi^{(3)}_{zzzz} + \sin^4\theta \chi^{(3)}_{xxxx} + \\ \sin^2\theta\cos^2\theta \left\{\chi^{(3)}_{zzxx}+\chi^{(3)}_{zxzx}+\chi^{(3)}_{zxxz} \right.\\ \left.+\chi^{(3)}_{xzzx}+\chi^{(3)}_{xzxz}+\chi^{(3)}_{xxzz}\right\}.
\end{multline}
From symmetry considerations, only contributions with an even number of $x$ and $z$ operators can be non-zero. From symmetry considerations, only contributions with an even number of $x$ and $z$ operators can be non-zero. Following the calculation given in the supplementary material in \cite{wan19}, one can express $\sigma^x_j$ and $\sigma^y_j$ in terms of the Majorana fermion modes $\alpha_j$ and $\beta_j$, defined as:
\begin{equation}
    \alpha_j =\psi^{\dagger}_j+\psi_j, \ \
    i\beta_j = \psi^{\dagger}_j-\psi_j.
\end{equation}
These Majorana fermions obey the following anticommutation relations: $\{\alpha_i,\alpha_j\}=2\delta_{i,j}$,$\{\beta_i,\beta_j\}=2\delta_{i,j}$, and $\{\alpha_i,\beta_j\}=0$. In terms of these operators, $\sigma^x_j$ becomes:
\begin{equation}
    \alpha_1i\beta_1\alpha_2i\beta_2...\alpha_{j-1}i\beta_{j-1}\alpha_j .
\end{equation}
As detailed in appendix \ref{sec:Pfaffian_Appendix} and in \cite{wan19}, the Hamiltonian can be recast in terms of these majorana modes, and the $S^x$ and $S^y$ expectation values can be evaluated using a Wick expansion. This results in a series of matrix pfaffians involving bilinear majorana expectation values such as $\langle \alpha_i(t) i\beta_j(0)\rangle$. Evaluating each pfaffian required to exactly calculate non-linear susceptibilities quickly becomes computationally expensive, and so certain approximations have been made when evaluating susceptibilities using this method in section \ref{sec:results}, as further discussed in appendix \ref{sec:Pfaffian_Appendix}.

\subsubsection{$\theta \neq 0$ calculation for $\alpha$ chains: Perturbation Theory}
Moving now to the situation on the $\alpha$ chains, things are somewhat more complicated. The term $-h\sin\theta\sum_j {S}^{\tilde{x}}_j$ is highly non-local in Jordan-Wigner fermions, precluding an interpretation in terms of a simple interaction between particles. Instead, we focus on the high field limit in the original pseudospin basis defined in terms of $J_{xx}$, $J_{yy}$, $J_{zz}$, and $J_{xz}$ [Eq. (\ref{eq:DO Hamiltonian pseudospin basis})], wherein the unperturbed part of the Hamiltonian takes the simple form:
\begin{equation}
    H_0=-\frac{h}{2} \sum_j \sigma^z_j. 
\end{equation}
The ground state is then simply the state with all pseudospins in the  $\sigma^z=1$ eigenstate. Periodic boundary conditions are imposed. Excitations from the fully polarized ground state are spin-waves, or magnons, with the magnon excitation energy being $h$. To account for small but non-zero values for the other coupling constants, we turn to leading order degenerate perturbation theory.

First examining the single magnon sector, the effective Hamiltonian has components
\begin{equation}
    H'_{ij} = \frac{(J_{xx}+J_{yy})}{4}(\delta_{i,j-1}+\delta_{i,j+1}) +\frac{J_{zz}}{4}(L-4) \delta_{i,j},\label{eq:H_eff_PT}
\end{equation}
where the roman indices $i,j$ denote the location of the single flipped spin (magnon). The eigenstates of the Hamiltonian are simply momentum eigenstates $\ket{k}$. To first order, the energy of the ground state is $E_{\O}=-\frac{hL}{2}+\frac{J_{zz}L}{4}$, and we find that the magnon excitation energies above the ground state are given by:
\begin{equation}
    \lambda_k= h+ J_{zz} +\frac{J_{xx}+J_{yy}}{2}\cos(k)
\end{equation}

We can similarly write down an effective Hamiltonian for the two magnon sector. Even in this perturbative limit the magnons are interacting, both through a hard-core constraint, and a non-zero $J_{zz}$, which gives an energy cost or benefit to domain walls depending on its sign. We find that $J_{zz}$ promotes the existence of eigenstates where the two magnon are primarily on neighbouring sites, and propagate as a single free particle, i.e. as bound magnon pairs.

Speaking qualitatively, the first order correction to the ground state $\ket{\delta \O}$ includes both unbound zero momentum magnon pairs, $\ket{k,-k}$, and the zero momentum bound state $\ket{0}_{2,\text{bound}}$. Due to the effect of interactions, the exact first order correction is a little more complex, but the above description is approximately correct. In the limit where $J_{zz}=0$, we find that the ground state couples only to the unbound magnons, and that this coupling is proportional to $2\sin(k)$. The sine function here reflects the fact that the hard-core magnons must avoid being on top of one-another, and so behave in this way like fermions.

Writing a general two-magnon sector eigenstate as $\ket{n}_{2}$, whose energy is $\epsilon_{n;2}$, and introducing coupling parameters $f_n= \sum_j\braket{j,j+1|n}_{2}$ ($\ket{j,j+1}$ is the state with both spins $j$ and $j+1$ flipped relative to the ground state), we find that the first order correction to the ground state is:
\begin{equation}
    \ket{\delta \O} = -\frac{J_{xz}\sqrt{L}}{2h} \ket{k=0} -\frac{J_{xx}-J_{yy}}{8h} \sum_n f_n \ket{n}_{2}.
\end{equation}

We now proceed in calculating the required non-linear susceptibilities via a Lehmann expansion of the required time-dependent expectation values that arise out of the nested commutators in the Kubo formulae. This entails inserting several resolutions of the identity in terms of energy into Eqs. (\ref{eq:Kubo_second_order}) and (\ref{eq:Kubo_third_order}), such that all time evolution operators act on energy eigenstates. We can then take a time-independent matrix element out of the time-dependent commutator. Let the remaining time and energy dependent factor be written as a function $G(\{t_i\},\{E_i\})$, then the third order susceptibility, for example, becomes:
\begin{multline}
   \chi^{(3)}= \frac{\Theta(t_1)\Theta(t_2)\Theta(t_3)}{8L}\sum_{pqr} \bra{\O}\Sigma^z\ket{E_p}\bra{E_p}\Sigma^z\ket{E_q} \\ \times \bra{E_q}\Sigma^z\ket{E_r}\bra{E_r}\Sigma^z\ket{\O} \text{Im}(G(\{t_i\},\{E_i\})),\label{eq:PT_Lehmann}
\end{multline}
where $\Sigma^z=\sum_j \sigma^z_j$. The calculation then proceeds by expanding each of the energy eigenstates appearing in Eq. (\ref{eq:PT_Lehmann}) in powers of $J_{\mu}/h$. At zeroth order, the $\Sigma^z$ operators act trivially on every eigenstate, and each energy $E_p$, $E_q$, $E_r$ introduced in the Lehmann expansion is set equal to the ground state energy. As a result, $G$ is found to be time-independent and real, and the whole response vanishes. We find at first order in $J_{\mu}/h$ that the matrix element is proportional to $\braket{\O|\delta \O}$, which is zero, and so again the response vanishes. The first non-zero contribution thus comes at second order. We find that two types of terms are generated, those which involve excitations of $k=0$ magnons, and those that excite two magnon states with zero total momentum. Both of these have identical forms for their time dependencies, and we find no rephasing behaviour to leading order in perturbation theory. The full third order susceptibility for constant probe and external fields is found to be:
\begin{footnotesize}
    \begin{multline}
\chi^{(3)}\approx \frac{2\Theta(t_1)\Theta(t_2)\Theta(t_3)}{L} \left\{\frac{J_{xz}^2L}{4h^2}\Bigl(\sin(\lambda_0(t_1+t_2)) \right.\\ -\sin(\lambda_0(t_1+t_2+t_3))\Bigr)  \\ + \frac{(J_{xx}-J_{yy})^2}{4h^2} \sum_n f_n^2 \Bigl(\sin(\epsilon_{2;n}(t_1+t_2)) \\ \left.-\sin(\epsilon_{2;n}(t_1+t_2+t_3))\Bigr)\right\}.
\label{eq:Nd_high_h_Chi3}
\end{multline}
\end{footnotesize}

In the limit $J_{zz}=0$, $f_n=2\sin(k)$ for all non vanishing $f_n$, and the corresponding $\epsilon_{2,n}$ are approximately equal to $2\lambda_k$. This matches precisely the large $h$ limit of Eqs. [(\ref{eq:XYh_sin_theta})-(\ref{eq:XYh_lambda})] to second order in $J_{\mu}/h$.

We can perform a similar analysis for the case of second order susceptibilities. Again we find that the first non-vanishing terms arise at second order in $J/h$. In this case, we obtain:
\begin{multline}
    \chi^{(2)} \approx \frac{\Theta(t)\Theta(\tau)}{L} \left\{ \frac{J_{xz}^2L}{2h^2} \Bigl(\cos(\lambda_0(t+\tau))-\cos(\lambda_0 \tau)\Bigr) \right.\\  \left.+\frac{(J_{xx}-J_{yy})^2}{4h^2} \sum_n f_n^2 \Bigl(\cos(\epsilon_{2;n}(t+\tau))-\cos(\epsilon_{2;n} \tau)\Bigr) \right\}.
    \label{eq:Nd_high_h_Chi2}
\end{multline}

The second order response then also has a part due to the excitation of $k=0$ magnons, and a part due to the two magnon sector.

Thus the key result for non-zero $\theta$ is the possibility of exciting single magnons, rather than pairs of excited particles, demonstrating that the polarized regime does not possess fractionalized excitations. As mentioned previously, single magnon excitations should be observable even with $\theta=0$ if one couples directly to the pseudospins ${S}^{\tilde{x}}$ or ${S}^{\tilde{y}}$ directly. Such processes should be seen for our magnetic probe field if we extend our treatment to include octupolar couplings, however the resulting responses will be much weaker than the dipolar contribution.

We are also interested in the response of Nd$_2$Zr$_2$O$_7$ to a [001] probe field. The full response will have contributions from $\chi^{(3)}_{\alpha,\text{direct}}$, $\chi^{(2)}_{\alpha,\text{direct}}$, and $\chi^{(3)}_{\beta,\text{direct}}$. The $\beta$ chain response can be treated within the Hartree-Fock approximation using the Pfaffian expansion as detailed above and in appendix \ref{sec:Pfaffian_Appendix}. For the $\alpha$ chain responses in the limit of large $h$, we can make use of a similar perturbative expansion to as above.

The only change to the analysis is to replace the operator $\Sigma^z$ with $\tilde{\Sigma}^z=\sum_j (-1)^j\sigma^z_j$, as we choose to remain in a basis in which $\mathbf{z}_i\cdot\mathbf{H}$ is constant, which causes the direct probe field to acquire this alternating sign. $\tilde{\Sigma}^z$ is no longer proportional to the unperturbed Hamiltonian $H_0$, and so its action on the unperturbed states is in principle non-trivial. However, if we consider an system with an even number of sites, $\tilde{\Sigma}^z$ annihilates the ground state $\ket{\O}$, and the states with two neighbouring spin flips $\ket{j,j+1}$. As a result, the only surviving contribution to the non-linear response of the $\alpha$ chains in this limit, to second order in $(J_{\mu}/h)$, is the single magnon contribution to $\chi^{(3)}_{\alpha,\text{direct}}$.

Acting on the $\ket{k=0}$ magnon state with $\tilde{\Sigma}^z$ returns $-2\ket{k=\pi}$, and similarly $\tilde{\Sigma}^z\ket{k=\pi}=-2\ket{k=0}$. There is then only one non-vanishing term in the Lehman expansion of $\chi^{(3)}$, and we find that the full $\alpha$ chain response to second order in perturbation theory is given by:

\begin{footnotesize}
    \begin{multline}
\chi^{(3)}\approx -\frac{\Theta(t_1)\Theta(t_2)\Theta(t_3)}{2} \frac{J_{xz}^2}{h^2}\Bigl\{\sin(\lambda_{\pi}(t_1+t_3)+\lambda_0 t_2) \\ +\sin(\lambda_{\pi}(t_1+2_3)-\lambda_0 t_3)  +\sin(\lambda_{\pi}(t_3-t_1)-\lambda_0t_3) \\ +\sin(\lambda_{\pi}(t_3-t_1)-\lambda(t_3+t_2))\Bigr\} .
\label{eq:Nd_mz_high_h_Chi3}
\end{multline}
\end{footnotesize}

Here $\lambda_0$ and $\lambda_{\pi}$ are the excitation energies of $k=0$ and $k=\pi$ magnons respectively. The time dependence of Eq. (\ref{eq:Nd_mz_high_h_Chi3}) does not depend on only one energy. This is similar to the predicted [001] response of $\alpha$ chains for zero mixing angle given in Eq. (\ref{eq:Chi3_exact_alternating}), and again we interpret the modified time dependence as stemming from transitions between magnon modes induced by the effective $\pi$ momentum probe field. 

One thing to note is that we have no contribution at second order to the response from the two magnon sector. Comparing this result to Eq. (\ref{eq:Chi3_exact_alternating}), we find that if the high field limit is taken for the $\theta=0$ exact expressions, the leading order contribution occurs at fourth order in $J_{\mu}/h$, which is consistent with the absence of such terms in this perturbative calculation.

\section{Results for $\text{Ce}_2 \text{Zr}_2 \text{O}_7$ and  $\text{Nd}_2 \text{Zr}_2 \text{O}_7$}
\label{sec:results}

In this section, we present predictions for the 2DCS responses detailed in Section \ref{sec:calculations} for the two rare earth pyrochlore magnets Ce$_2$Zr$_2$O$_7$ and Nd$_2$Zr$_2$O$_7$ in the presence of a sufficiently strong [110] field such that mean field theory predicts a CHAIN ground state. We examine the responses of both materials for the [110], [1$\bar{1}$0], and [001] polarization channels, and discuss the qualitatively different information provided by these different experiments about excitations in both systems.

Where the Hamiltonian of each $\alpha$ or $\beta$ chain is tractable in the Hartree-Fock approximation, and $\theta=0$, analytical results are presented. Otherwise, a mixture of numerical results calculated using the matrix Pfaffian or perturbative methods discussed in Section \ref{subsec:lswt_and_ed} are employed. 
Note that while the matrix Pfaffian method is in principle numerically exact, we employ several approximations to the exact calculation to improve computation time, as described in Appendix \ref{sec:Pfaffian_Appendix}. These approximations give rise to a discrepancy in the global scale of the response of a factor of order one when compared to the exact calculation, but the functional form of the response in $\omega_t$ and $\omega_{\tau}$ is unchanged. Thus the arbitrary units scales of response functions presented in this section may differ between plots where different computation methods have been used. The scale of different panels within the same figure are however always directly comparable.

\subsection{Ce$_2$Zr$_2$O$_7$}
\label{subsec:czo}

\begin{figure}
\includegraphics[scale=0.5]{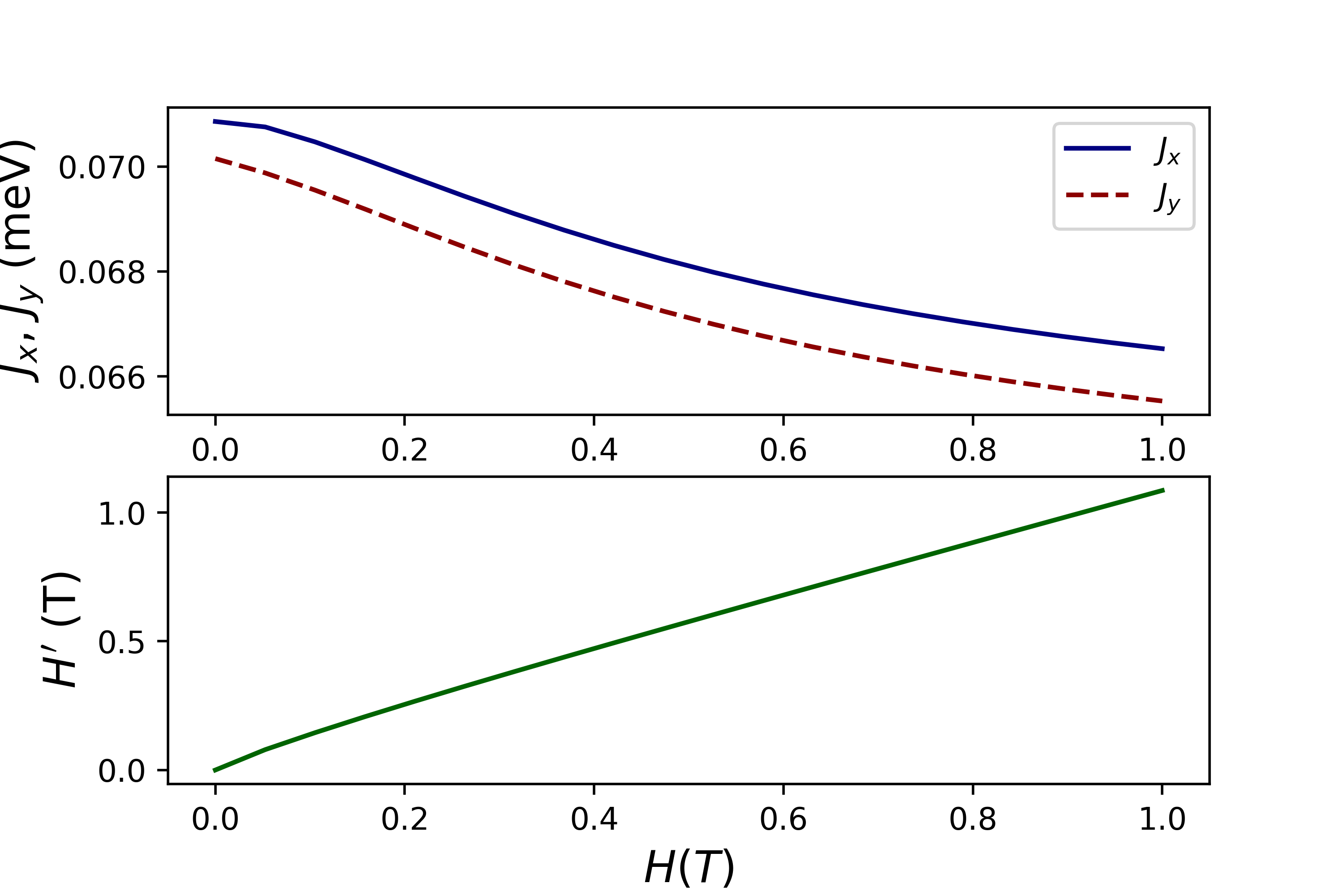}
\caption{
Evolution of re-scaled Ce$_2$Zr$_2$O$_7$ model parameters in the Hartree-Fock approximation with varying strength of the applied field $H$. We observe only small modifications to all parameters, with $H'$ scaling approximately linearly with $H$. It is also noted that the anisotropy $J_x-J_y$ remains approximately constant with increasing $H$.}
\label{fig:Ce_parameter_H_vary_HF}
\end{figure}

The model parameters predicted for Ce$_2$Zr$_2$O$_7$  \cite{smith22} are as previously stated:
\begin{align}
(J_x,J_y) = (0.064,0.063)& \ \text{meV} \ \ J_z = 0.011 \ \text{meV}; \nonumber \\ \ \  \theta \approx 0 \ \text{rad};& \ \ g_z = 2.57. \nonumber
\end{align}
That we have $\theta\approx 0$ means that the most troubling non-integrable term in the Hamiltonian of the $\alpha$ chains, the coupling to ${S}^{\tilde{x}}$, is removed and the response from both chains, for direct and alternating probe fields, can be treated within the Hartree-Fock approximation. {The re-scaled $\beta$ parameters are independent of the applied field}, and are given by:
\begin{equation}
({J}^{\prime}_x,{J}^{\prime}_y) = (0.071,0.070) \ \text{meV}. \nonumber
\end{equation}
The re-scaling of the $\alpha$ chain parameters is field dependent, as shown in Fig. \ref{fig:Ce_parameter_H_vary_HF}.

For both $\alpha$ and $\beta$ chains, both the rescaled $J_x$ and $J_y$ are moderately larger than their bare values, whilst their difference remains approximately constant. As $H$ is increased, the values of $J_x$ and $J_y$ smoothly decrease, but remain higher than their bare values. The re-scaling of the effective magnetic field acting on each $\alpha$ chain is the most significant modification, with an increase in the effective field strength of approximately $0.06$ Tesla observed for an initial field strength of $0.2$ Tesla. The Hartree-Fock rescalings are independent of which of $J_x$ or $J_y$ is assigned the larger value, and so too is the 2DCS response. For definiteness, we take $J_x=0.064$meV, and $J_y=0.063$meV throughout.

Before discussing the results in detail, it is worth describing the types of signatures observed in the Fourier transforms of the magnetic susceptibilities $\chi^{(3)}(t,t,t+\tau)$ and $\chi^{(3)}(t,t+\tau,t+\tau)$. Their Fourier transforms are functions of two variables, $\omega_t$ and $\omega_{\tau}$. The signature we are interested in lies in the absorptive part of the response, so we take the imaginary part of $\chi^{(3)}(\omega_t,\omega_{\tau})$. Looking at one term in particular, consider the $A^{(1)}$ term [Eq. (\ref{eq:A1})] of $\chi^{(3)}(t,t,t+\tau)$ in the limit $1/T_1=1/T_2=0$. This contribution has the following form as a function of time:
\begin{equation}
    \frac{\Theta(t)\Theta(\tau)}{L}\sum_{k>0}8\sin^2\phi_k \cos^2\phi_k \sin(2\lambda_k(\tau+t)).
\end{equation}
Fourier transforming this and taking the imaginary part, one obtains peaks in the response at $\omega_t=\omega_{\tau}=\pm 2\lambda_k$. However, these peaks are not pure delta functions, as one is taking the imaginary part of a product of two complex functions. Generically, the full Fourier transform at a given peak has the functional form:
\begin{equation}
    i\frac{1}{\omega_t-\epsilon}\cdot\frac{1}{\omega_{\tau}-\epsilon} .
\end{equation}
Making use of the identity $\frac{1}{x-a}=-i\pi\delta(x-a)+\mathcal{P}\frac{1}{x-a}$, where $\mathcal{P}$ denotes the principal value of a function, we find that the imaginary part of a given peak function has the form:
\begin{equation}
    -\pi^2\delta(\omega_t-\epsilon)\delta(\omega_{\tau}-\epsilon)+\mathcal{P}\frac{1}{\omega_t-\epsilon}\mathcal{P}\frac{1}{\omega_{\tau}-\epsilon} .\label{eq:peak_functional_form}
\end{equation}
The delta function part is formally diverging, and negative, whilst the principal value part of the response peak changes in sign as $(\omega_t-\epsilon)$ or $(\omega_{\tau}-\epsilon)$ swap sign.

The $A^{(1)}$ term gives rise to many such peaks along the line $\omega_t=\omega_{\tau}$ in the $\omega_t-\omega_{\tau}$ plane. For small chains and only small amounts of broadening, each set of peak remains distinct, however with greater broadening and larger system size, the regions of different sign begin to overlap.

\begin{figure}
\includegraphics[scale=0.55]{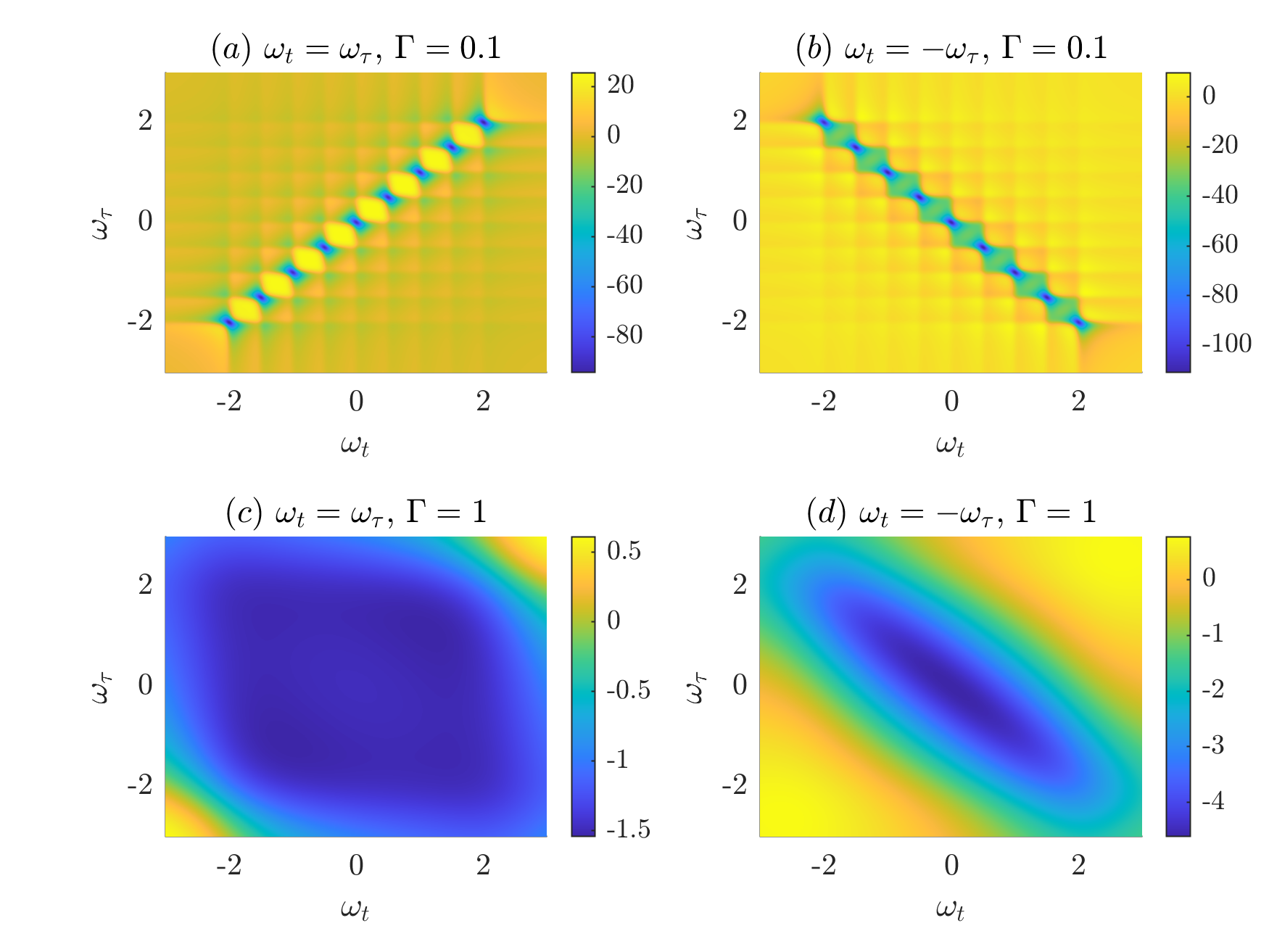}
\caption{Demonstration of the robustness of the rephasing signal ((b) and (d)) to broadening effects, when compared to the $\omega_t=\omega_{\tau}$ signature ((a) and (c)). (a) A set of nine peaks along the $\omega_t=\omega_{\tau}$ with small broadening $\Gamma=0.1$. The function varies rapidly from positive to negative between each peak. (b) A similar response from nine closely spaced peaks, but this time organised along the $\omega_t=-\omega_{\tau}$ line. Here the function is always negative along the length of the response, and is flanked by positive regions. When broadening is increased to $\Gamma=1$, we see in (c) that a broad signal is obtained, whilst in (d) a diagonal streak is still clearly observed along the $\omega_t=-\omega_{\tau}$ line.}
\label{fig:diag_streak_comparison}
\end{figure}

Referring to Eq. (\ref{eq:peak_functional_form}) and panel (a) in Fig. \ref{fig:diag_streak_comparison}, if we follow the $\omega_t=\omega_{\tau}$ line, we see that as we cross through each peak generated by $A^{(1)}$, that the function changes from large and positive, to large and negative at the Lorentzian (broadened delta function) part of the peak, to large and positive again. If the broadening is further increased, we find that this sign variation tends to wash out the signal leaving a featureless broad response, as shown in Fig. \ref{fig:diag_streak_comparison} (c).

On the other hand, the $A^{(4)}$ term of $\chi^{(3)}(t,t,t+\tau)$ [Eq. \ref{eq:A4}] produces identical peaks, but along the $\omega_t=-\omega_{\tau}$ line. Moving along this line, the second term of Eq. (\ref{eq:peak_functional_form}) is negative, the same sign as the delta functions, and so as one moves along $\omega_t=-\omega_{\tau}$, the sign of the function does not vary, as shown in Fig. \ref{fig:diag_streak_comparison} (b). The signal resembles a negative `trench' flanked by positive ridges. It is found that when the broadening on the response peaks is increased, this orientation of the streak is resistant to being washed out in the same way as the $A^{(1)}$ signal, as shown in Fig. \ref{fig:diag_streak_comparison} (d). This result underlies the focus on this  rephasing part of 2DCS responses. 

\subsubsection{$[1{\bar1}0]$ polarization: probing $\beta$-chain spinon density of states}
\label{subsubsec:czo_1-10}

A particularly interesting property of the Ce$_2$Zr$_2$O$_7$ model parameters is how close they lie to the predicted phase transition between the CHAIN$_x$ and CHAIN$_y$ phases. In terms of the behaviour on the $\alpha$ and $\beta$ chains, for all but the very smallest fields, the $\alpha$ chains are in a field polarized phase. The $\beta$ chains are however close to a critical phase transition, with a predicted gap to excitations of approximately $1 \mu\text{eV}$, corresponding to the excitation of two $k=0$ spinons. 

Being so close to criticality means that a [1$\bar{1}$0] polarized probe field ($m_{\perp}$), will provide a particularly clear picture of the full spectrum of spinon pair excitations, as $\sin\phi_k$ is very close to one for all $k$. The response to $m_{\perp}$ is determined entirely by $\chi^{(3)}_{\beta,\text{alt}}$ because the second order response vanishes on the $\beta$ chains. Referring to Eq. (\ref{eq:Non_linear_response}), two different time orderings contribute to the full non-linear response, $\chi^{(3)}_{\beta,\text{alt}}(t,t,t+\tau)$ and $\chi^{(3)}_{\beta,\text{alt}}(t,t+\tau,t+\tau)$, both of which are given by the expression in Eq. (\ref{eq:Chi3_exact}), with the  functions $\cos\phi_k$, $\sin\phi_k$, and $\lambda_k$ as given in Eq. (\ref{eq:XYh_sin_theta}-\ref{eq:XYh_lambda}) with $h$ set to zero.

\begin{figure}
\includegraphics[scale=0.5]{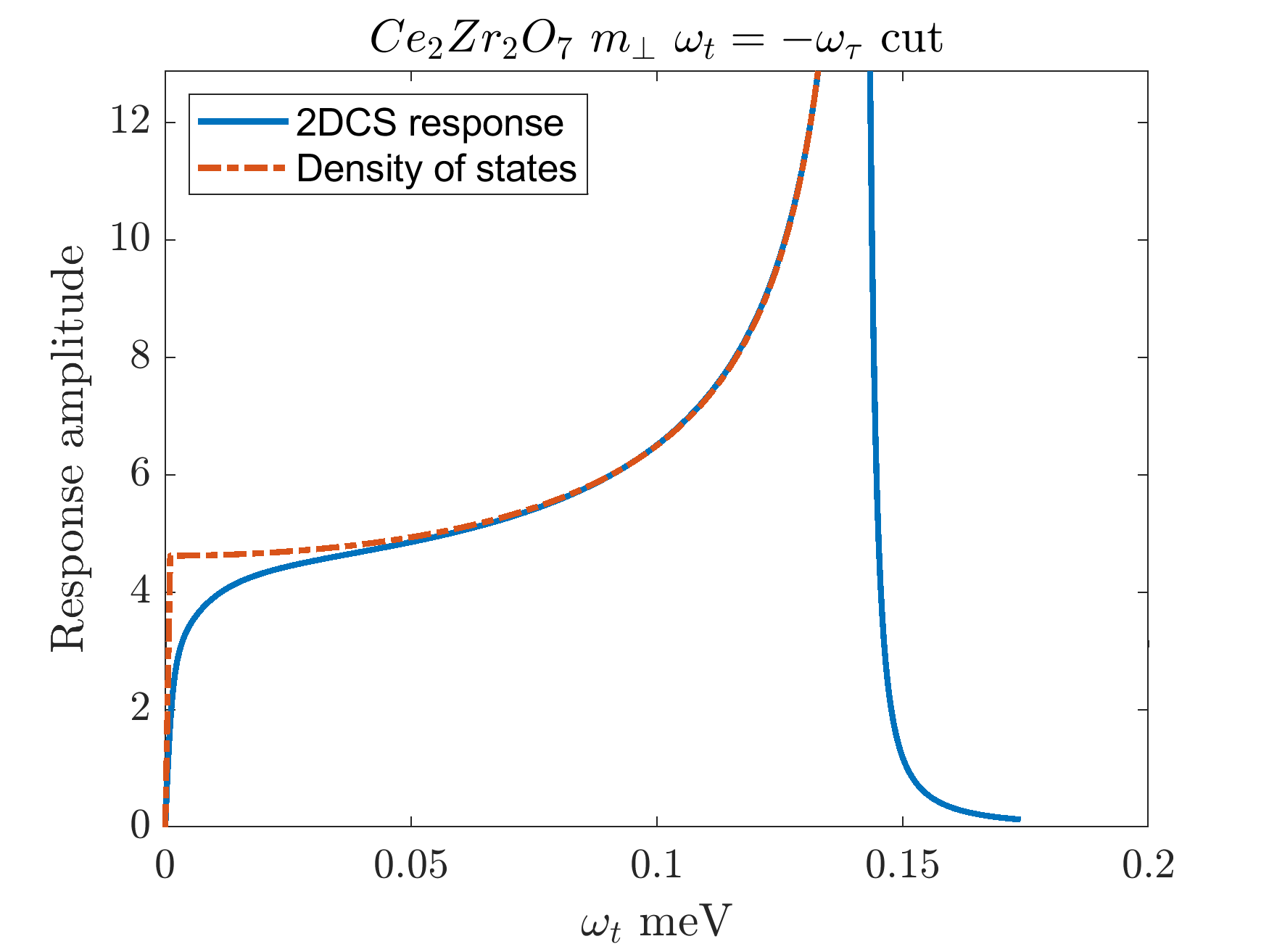}
\caption{A cut along $\omega_t=-\omega_{\tau}$ in the 2DCS response of Ce$_2$Zr$_2$O$_7$. The solid blue curve is the calculated non-linear response, whilst the red dot-dashed line is the density of states for the reported model parameters. A large dephasing time of $T_2=750\text{meV}^{-1}$ is used to minimise the effects of broadening on the shape of the response, whilst maintaining a smooth curve. We set $A_0:A_{\tau}=1:100$ (see Eq. (\ref{eq:Non_linear_response})) so as to single out the rephasing signature. We observe close agreement between the two curves, but note some deviations at the upper and lower band edges arising from the still finite broadening.
}
\label{fig:Ce_mperp_DOS_cut}
\end{figure}

One finds that precisely at the limit $J_x=J_y$ that $\cos\phi_k$=0 and $\sin^2\phi_k=1$ (see Eqs. [(\ref{eq:XYh_sin_theta})-(\ref{eq:XYh_cos_theta})]). This means that the rephasing signature would have uniform strength across the full spectrum. In the continuum limit, and for minimal broadening, the variation in the height of the response is approximately equal to the density of states in energy, as shown in Fig. \ref{fig:Ce_mperp_DOS_cut}.
Thus the 2DCS response with the $[1\bar{1}0]$ polarization provides a close approximation of the spinon density of states on the $\beta$ chains.

\begin{figure*}
\subfigure[]{\includegraphics[scale=0.39]{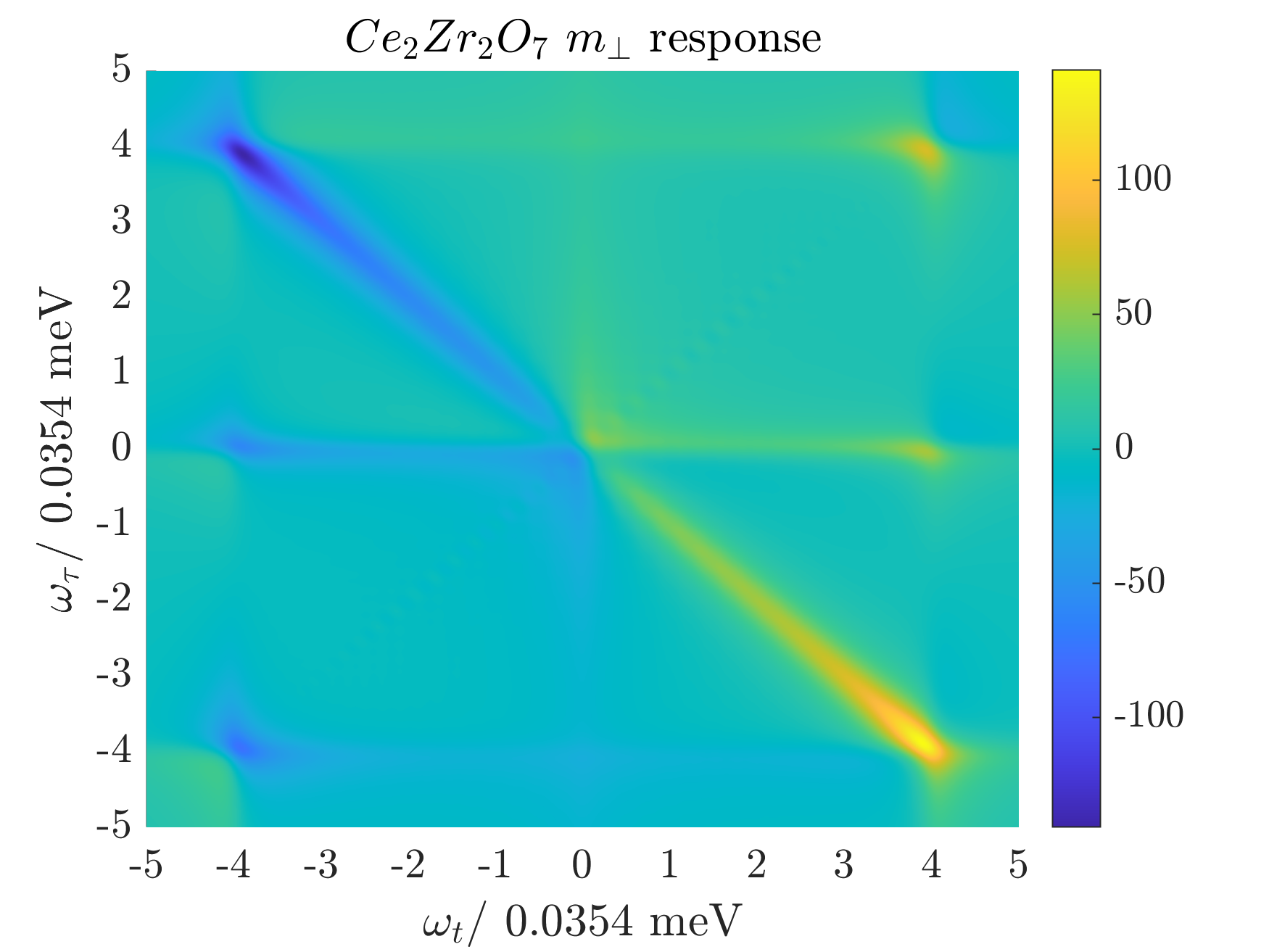}}
\subfigure[]{\includegraphics[scale=0.39]{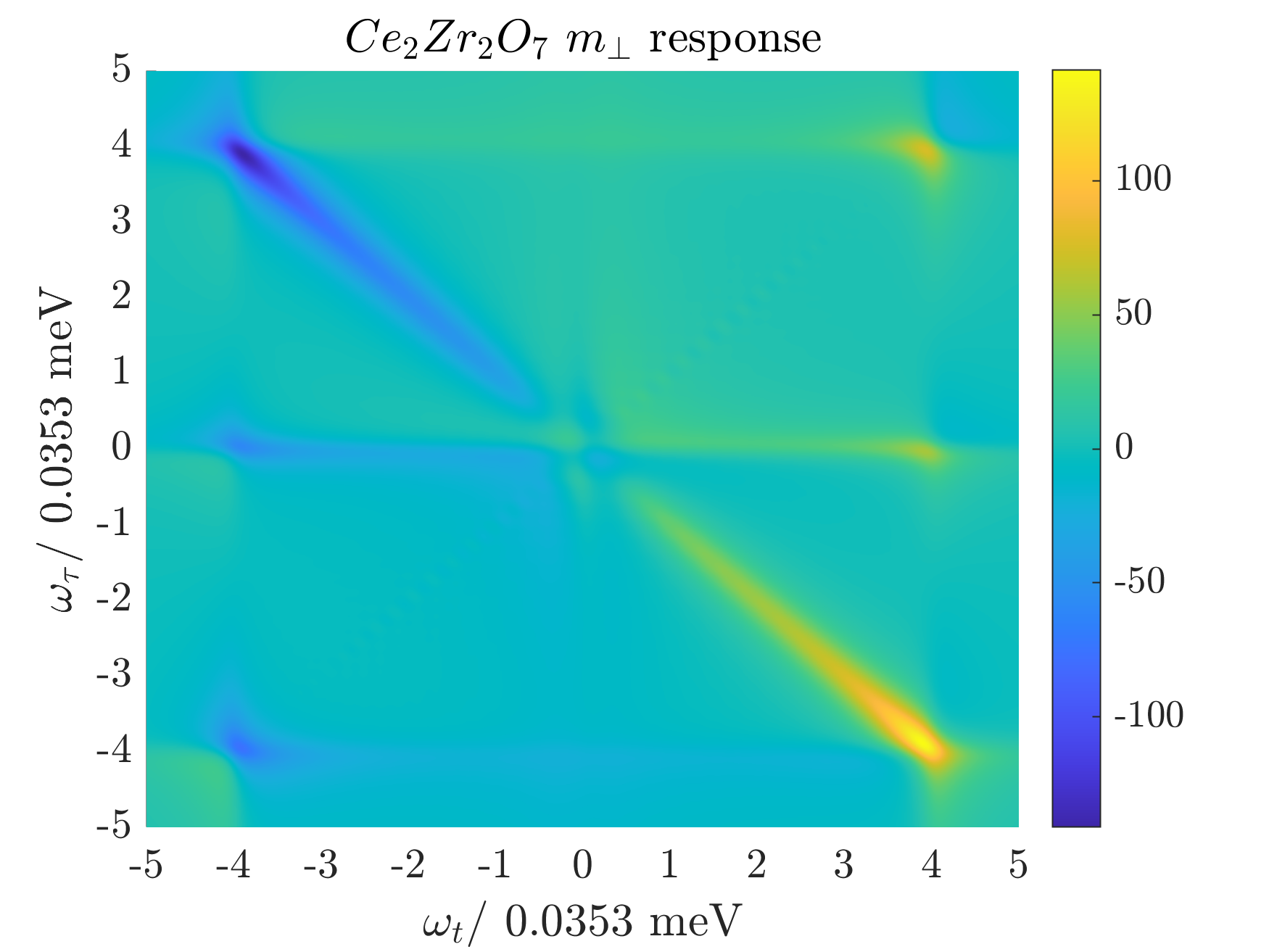}}
\subfigure[]{\includegraphics[scale=0.39]{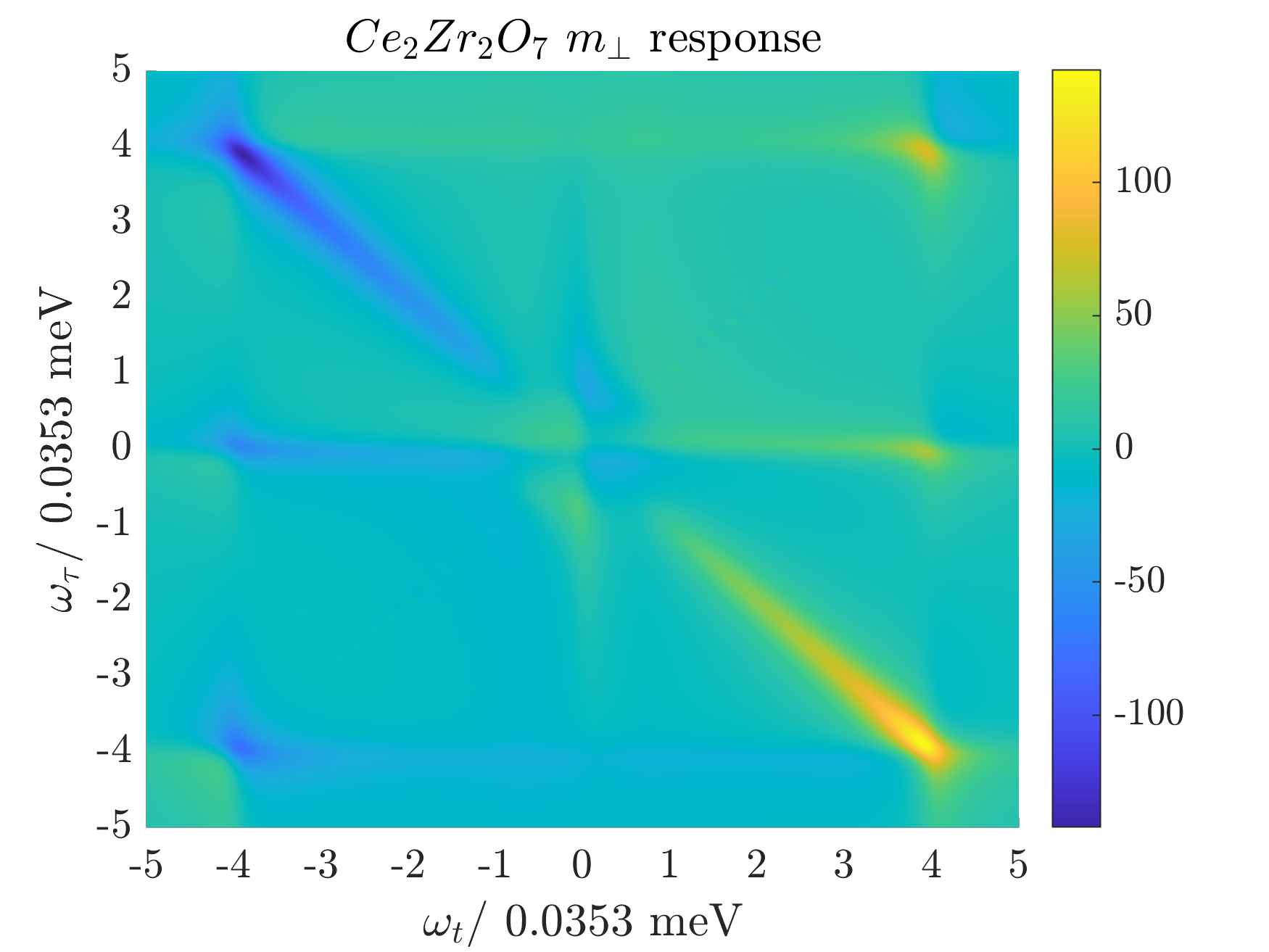}}
\caption{2DCS response of $\beta$ chains in Ce$_2$Zr$_2$O$_7$ to a [1$\bar{1}$0] polarized probe field, for varying values of the anisotropy $J_{\tilde{x}}-J_{\tilde{y}}$.
(a) $J_{\tilde{x}}-J_{\tilde{y}}=0.001\text{meV}$, as in the best fit parameters from \cite{smith22},
(b) $J_{\tilde{x}}-J_{\tilde{y}}=0.014\text{meV}$,
(c) $J_{\tilde{x}}-J_{\tilde{y}}=0.027\text{meV}$.
Calculations are made using Eq. (\ref{eq:Chi3_exact}), within the Hartree-Fock approximation on a system of 100 spins, along with large phenomenological depopulation and dephasing times $T_1=T_2=150\text{meV}^{-1}$. The energy gap is observed to grow as the $\beta$ chain Hamiltonian is taken further away from criticality, with the low energy edge of the rephasing streak along $\omega_t=-\omega_{\tau}$ moving away from the origin. The intensity of the rephasing streak closely tracks the spinon density of states, as shown in Fig. \ref{fig:Ce_mperp_DOS_cut}. The strength of the magnetisation response is in arbitrary units, and the ratio of the strength of the first field pulse to the second is $1:5$.}
\label{fig:Ce_mperp_Jx_vary}
\end{figure*}

In Fig. \ref{fig:Ce_mperp_Jx_vary}, we present plots of the response of the $\beta$ chains of Ce$_2$Zr$_2$O$_7$ to a [1$\bar{1}$0] probe field. Fig. \ref{fig:Ce_mperp_Jx_vary}(a) shows the response for the estimated model parameters \cite{smith22}, whilst (b) and (c) demonstrate how the response would vary with an increasing anisotropy $J_{x}-J_{y}$. The strong diagonal signatures in the lower right and upper left quadrants are the aforementioned rephasing signatures, which run from close to the origin out to $\omega_{t}=-\omega_{\tau}=|\tilde{J}_x+\tilde{J}_y|$. The increase in intensity that is observed at the upper end of the spectrum is due to the diverging density of states at the upper band edge, as the energy of a spinon pair $2\lambda_k$ is stationary at this maximum as a function of $k$.

The horizontal bar observed along the $\omega_{\tau}=0$ axis originates in the $\chi^{(3)}(t,t+\tau,t+\tau)$ part of the non-linear response. The two different time orderings contribute with different weights, with $\chi^{(3)}(t,t+\tau,t+\tau)$ contributing with a strength $A_0^2A_{\tau}$, and $\chi^{(3)}(t,t,t+\tau)$ with $A_0A_{\tau}^2$, as seen in Eq. (\ref{eq:Non_linear_response}). Thus, in principle, the contribution from  $\chi^{(3)}(t,t,t+\tau)$, which contains the rephasing term, can be favoured by tuning the pulse strengths \cite{hart23}, provided that the second order contribution also vanishes (as it does for the $\beta$ chains).

With increasing anisotropy $J_x-J_y$, we can see in the 2DCS response that the gap to excitations widens and the bandwidth of pair excitations narrows as we tune away from the critical point.

\begin{figure*}
\subfigure[]{\includegraphics[scale=0.39]{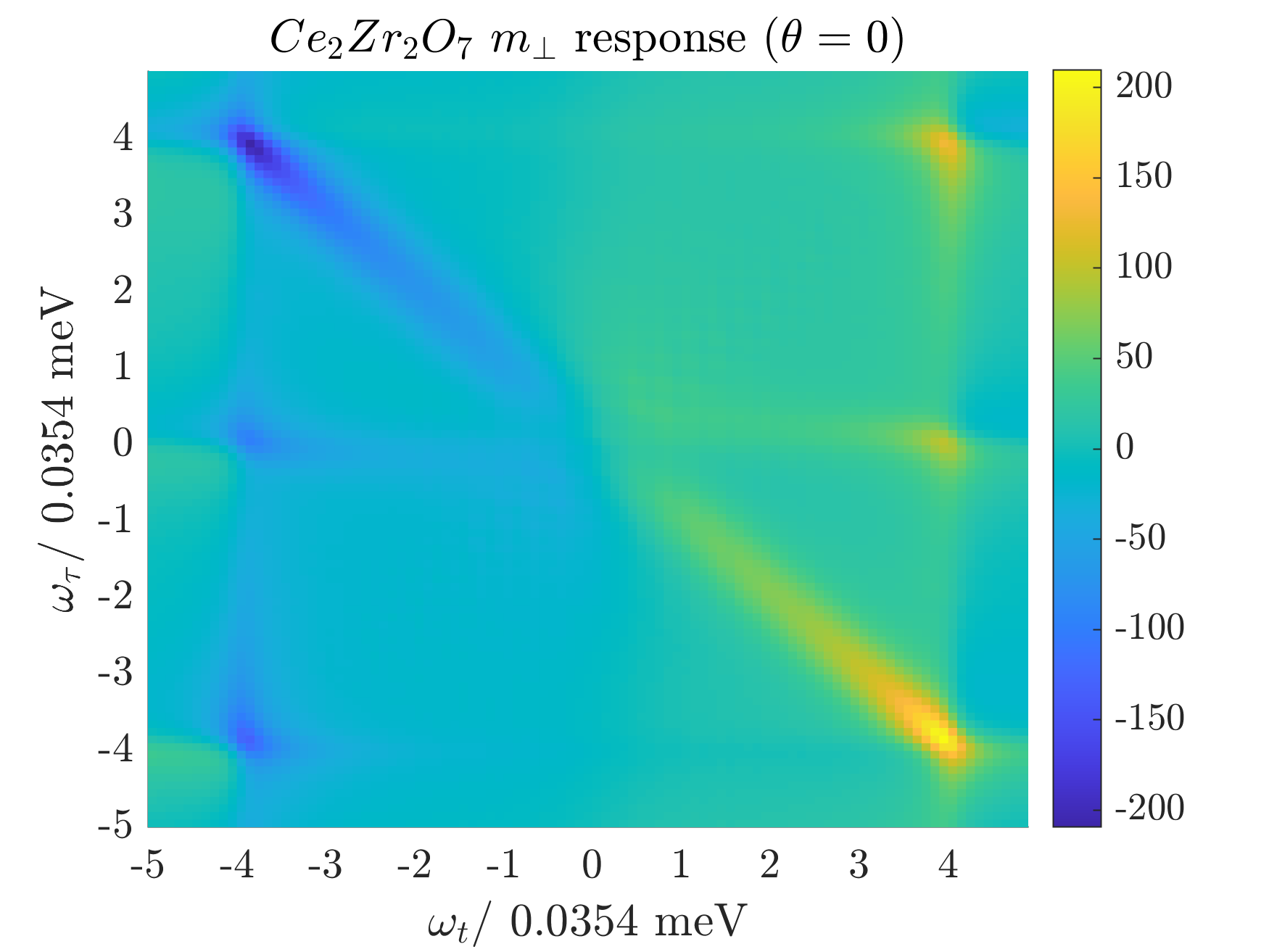}}
\subfigure[]{\includegraphics[scale=0.39]{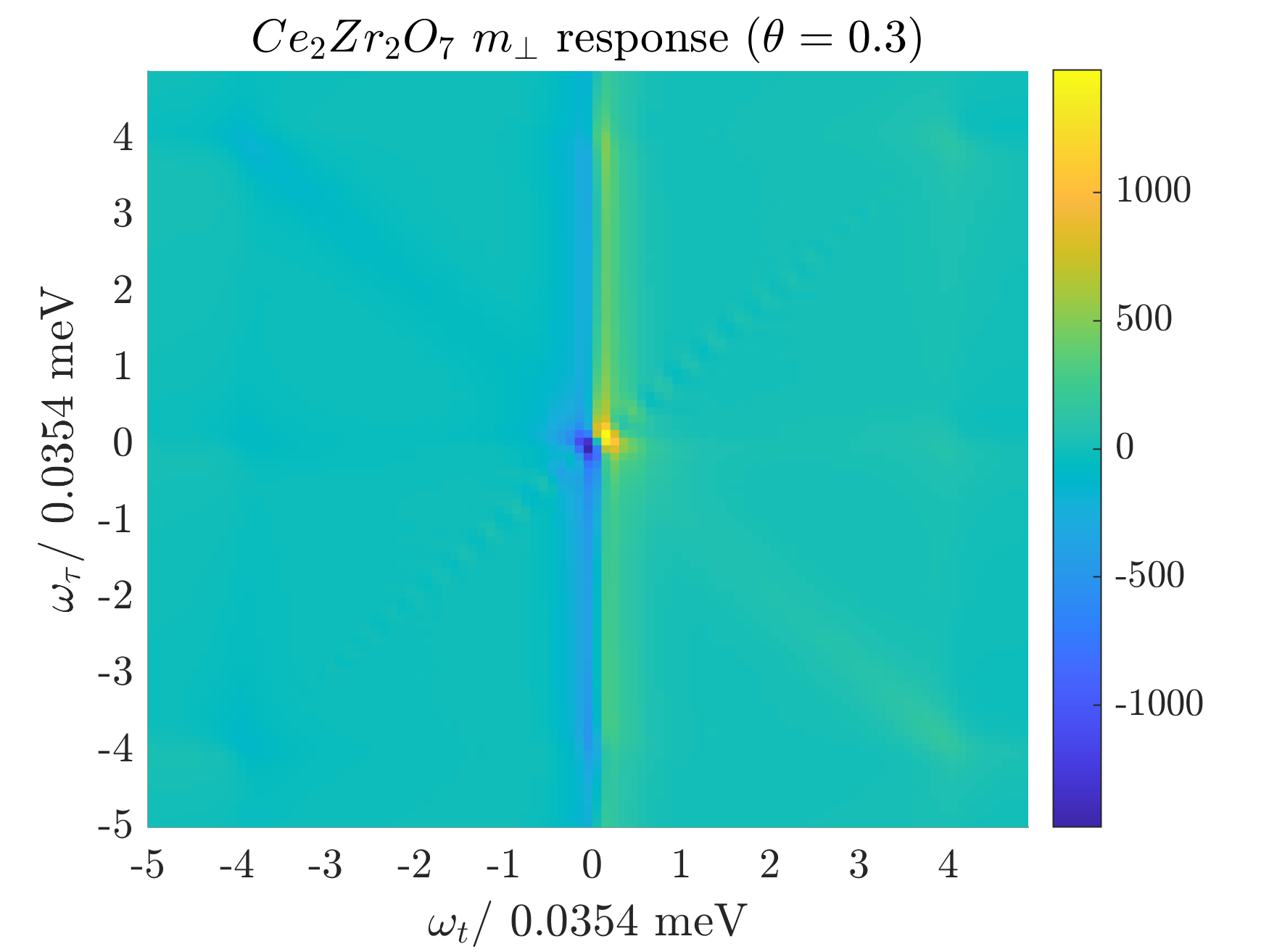}}
\subfigure[]{\includegraphics[scale=0.39]{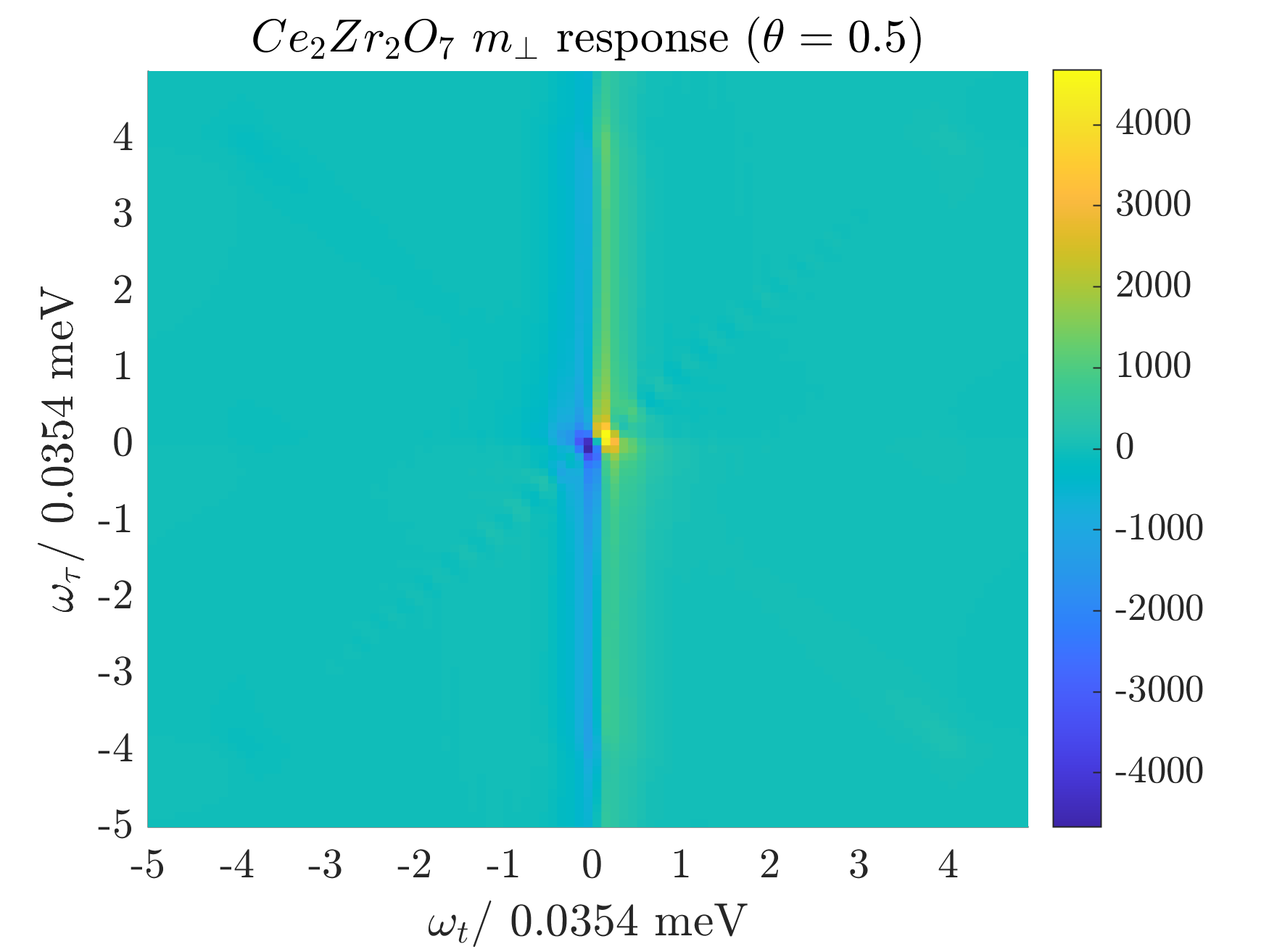}}
\caption{2DCS response of $\beta$ chains in Ce$_2$Zr$_2$O$_7$ to a [1$\bar{1}$0] polarized probe field, for increasing values of the dipolar-octupolar mixing angle $\theta$. Responses are generated using the matrix pfaffian method, and so the overall scale of plots in this figure differ from those generated without this expansion.
(a) $\theta=0$, as in the best fits in \cite{smith22},
(b) $\theta=0.3$,
(c) $\theta=0.5$. We observe that as $\theta$ is increased from zero, a large signal is introduced along the $\omega_t=0$ line, which dominates over the rephasing signal. A chain length of $L=40$ was used for these calculations, with $A_0:A_{\tau}=1:5$.
}

\label{fig:Ce_mperp_theta_vary}
\end{figure*}

In Fig. \ref{fig:Ce_mperp_theta_vary}, we again tune away from the predicted model parameters for Ce$_2$Zr$_2$O$_7$, but this time by introducing a finite mixing angle $\theta$. To calculate the non-linear response, we must make use of the matrix Pfaffian method discussed in Section \ref{subsec:lswt_and_ed}, which is somewhat more expensive computationally, and so we are limited to chain lengths of $L=40$, and a coarser frequency resolution. Following a similar procedure as that detailed in the supplementary material of \cite{wan19}, we impose open boundary conditions, and impose a cut off of 5 sites from the end of each chain in sums over sites to avoid the effects of edge states.

With finite mixing angle $\theta$ introduced to the Ce$_2$Zr$_2$O$_7$ model parameters, we observe that the response becomes dominated by a signal along $\omega_t=0$, whilst the rephasing contribution appears to remain approximately constant in amplitude with changing $\theta$. This is shown in Fig. \ref{fig:Ce_mperp_theta_vary}, for mixing angles of $\theta=0$, $\theta=0.3$, and $\theta=0.5$. As the response functions in Fig. \ref{fig:Ce_mperp_theta_vary} come only from the $\beta$ chains, whose Hamiltonian has no dependence on $\theta$, the effects for finite $\theta$ must originate in the changing matrix elements between the probe field and the pseudospin degrees of freedom.

\subsubsection{$[110]$ polarization: two-magnon excitations on $\alpha$ chains}
\label{subsubsec:czo_110}

If we instead choose to align our probe field and measure the magnetisation parallel to the [110] direction, we can extract the response of the $\alpha$ chains alone. The quantum phase transition on the $\alpha$ chains between a magnetically ordered phase and a field polarized phase is predicted occur at $|\tilde{J}_x-\tilde{J}_y|=2\tilde{h}$, which corresponds to a field of the order $H_c \approx 0.001$T for Ce$_2$Zr$_2$O$_7$. Thus for any moderately large field, the $\alpha$ chains are field polarized, and relatively far from a quantum phase transition. As these chains are in a field polarized phase, and not one hosting one-dimensional ferromagnetic/antiferromagnetic ordering like on the $\beta$ chains, the fundamental excitations are magnons, which are not fractionalized. 
However, for $\theta=0$, the probe field has no matrix element to excite single magnons in the dipole approximation, and thus we still see only responses from excitations of pairs of particles.
Thus the 2DCS response will still produce streak-like continua when probing the $\alpha$ chains, despite the absence of fractionalisation.

\begin{figure*}
\subfigure[]
{\includegraphics[scale=0.39]{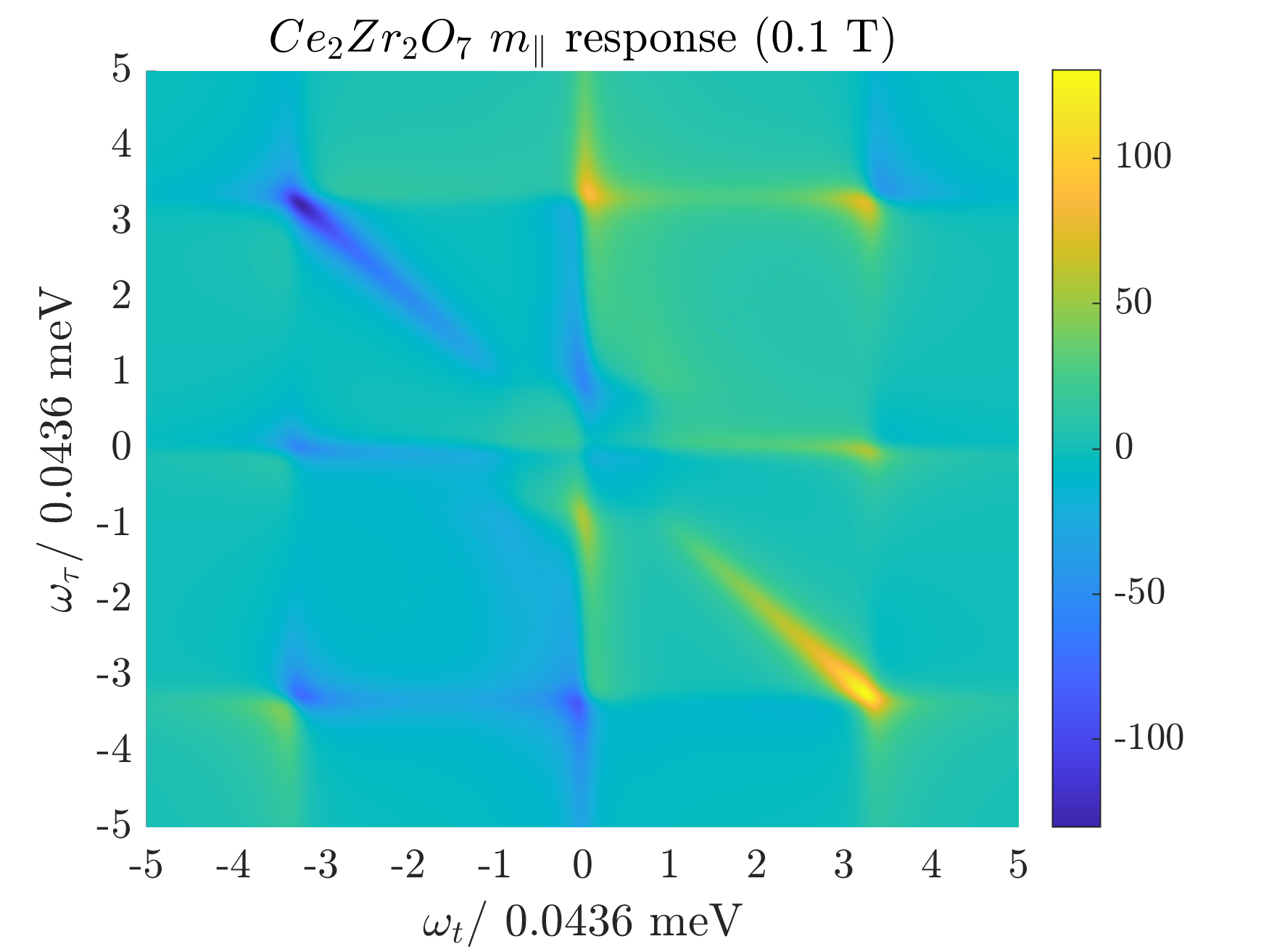}}
\subfigure[]
{\includegraphics[scale=0.39]{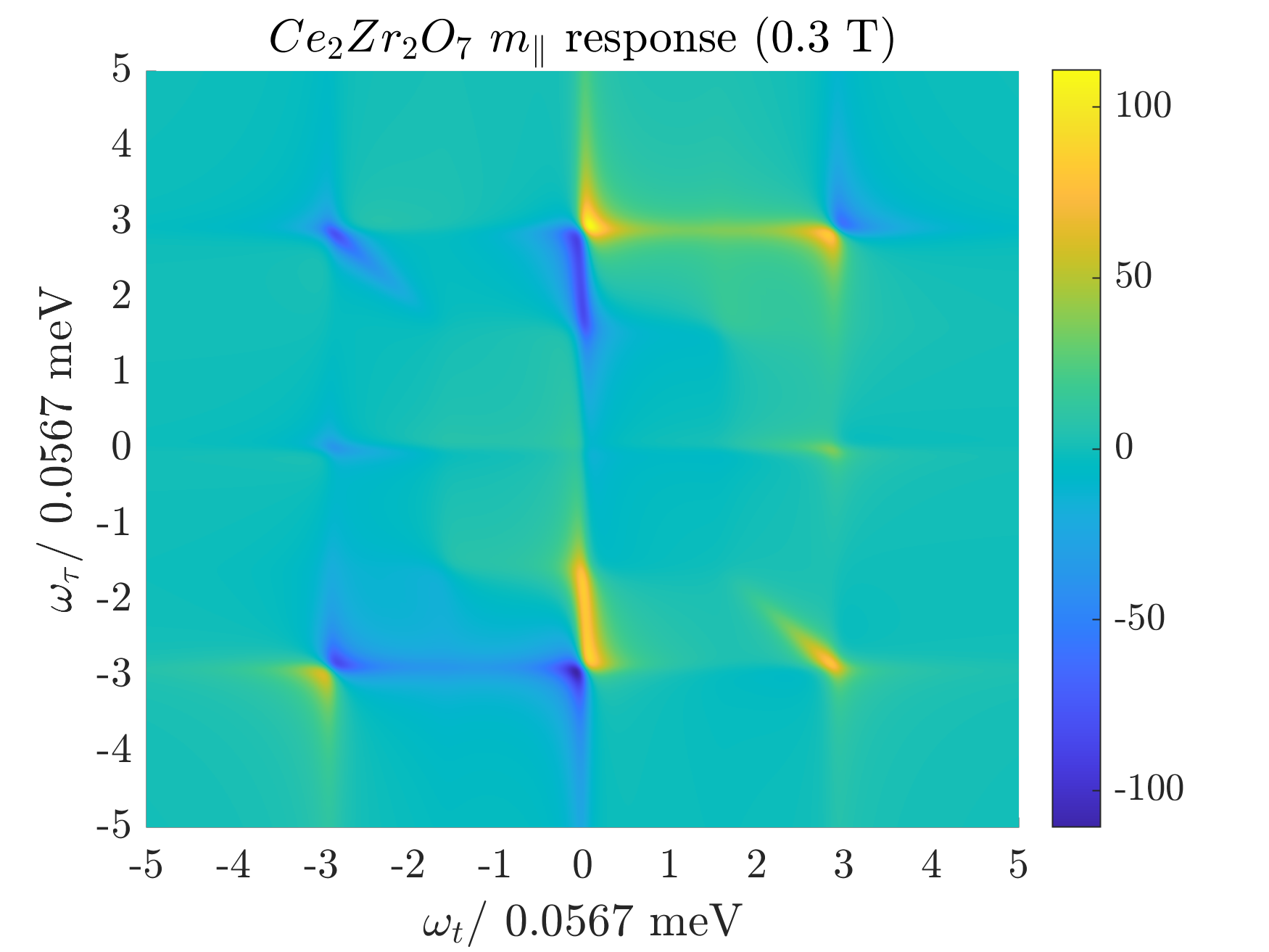}}
\subfigure[]
{\includegraphics[scale=0.39]{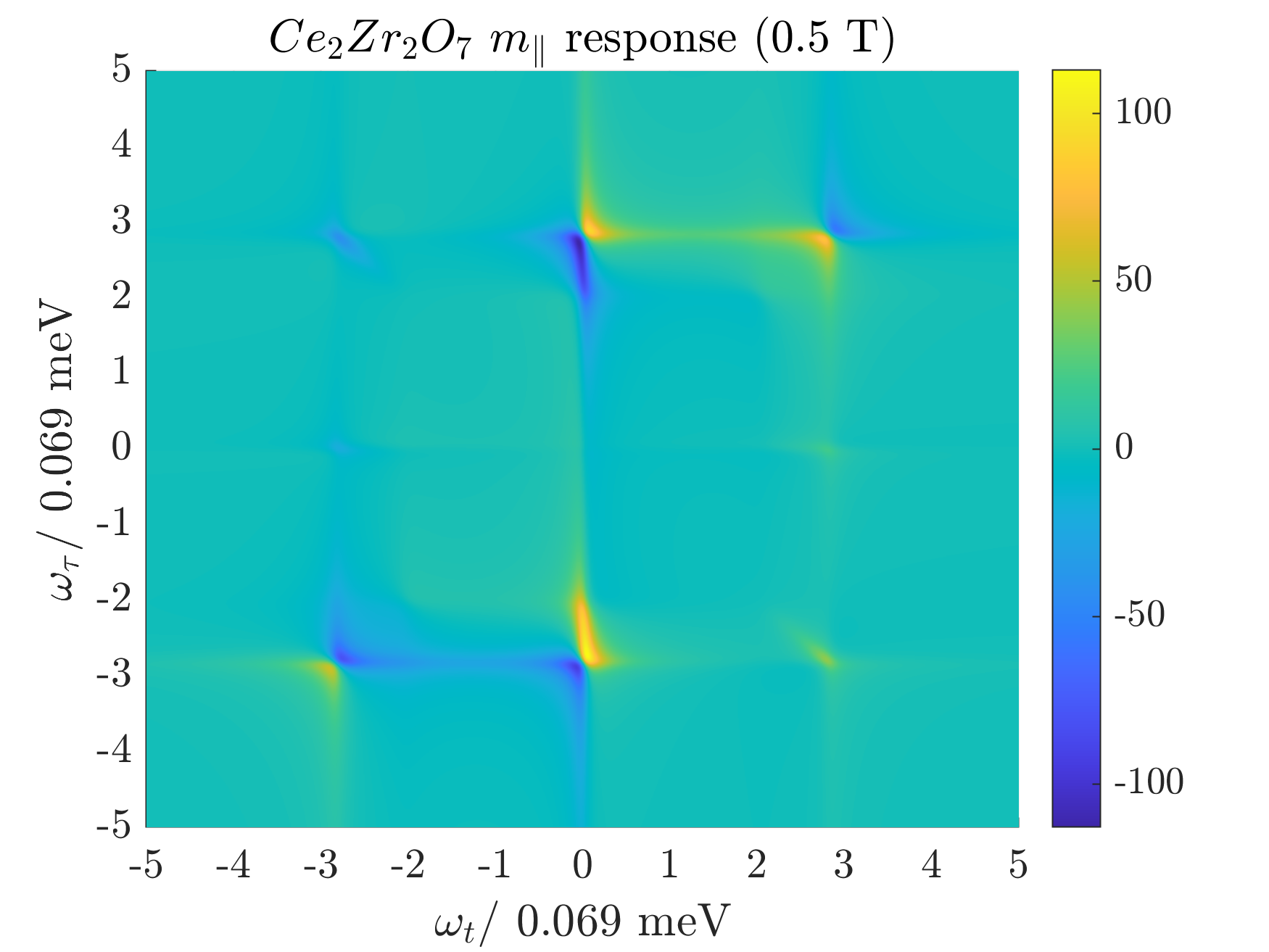}}
\caption{2DCS response of $\alpha$ chains in Ce$_2$Zr$_2$O$_7$ to a [110] probe field, for increasing strength of the external field $H$. Figures are produced using Eq. (\ref{eq:Chi3_exact}), within the Hartree-Fock approximation. In all panels, the experimentally estimated model parameters from \cite{smith22} are used.
Two-magnon scattering dominates the response, producing 
a streak-like continuum despite the absence of fractionalisation on the $\alpha$-chains.
It is observed that the lateral extent of the rephasing signature is greatly diminished in high fields, due to the energy cost of magnon pair creation increasingly dominating over the width of the 2-magnon continuum. For all figures, $L=100$, and the pulse ratio is $A_0:A_{\tau}=1:5$.}
\label{fig:Ce_mpara_h_vary}
\end{figure*}

In Fig. \ref{fig:Ce_mpara_h_vary}, we present the 2DCS response of the $\alpha$ chains of Ce$_2$Zr$_2$O$_7$ to a [110] field for increasing field strengths. We observe that as the field strength is increased from $0.1$T, to $0.3$T and $0.5$T that the two magnon bandwidth narrows considerably relative to the gap, with the 2DCS response at $0.5$T collapsing almost to single peaks as the energy scale for exciting two magnons ($h$) becomes much larger than the energy scale governing their dynamics ($J_{\mu}$).

\subsubsection{$[001]$ polarization: probing nearby criticality}
\label{subsubsec:czo_001}

Finally, we consider what the 2DCS response using a [001] polarized probe field can tell us about the excitations of Ce$_2$Zr$_2$O$_7$. Such a probe field couples to both $\alpha$ and $\beta$ chains, and does so uniformly across each chain, instead of alternating in sign from one site to the next. The relative $\pi$ phase shift in the momentum of the probe field to the [110] and [1$\bar{1}$0] probe fields (and to the external field $\mathbf{H}$) leads to qualitatively different 2DCS responses.

The response of the $\beta$ chains to the $[001]$ polarization is very different to the case of the $[1\bar{1}0]$ polarization: instead of being sensitive to the full spinon density of states, the measurement specifically picks out the lower band edge.
This polarization choice is then particularly favorable here as a measure of the proximate criticality on the $\beta$ chains.
In order to calculate the response of the $\beta$ chains in this setting, one can perform a basis transformation
and rotate the spin basis on every second site by $\pi$ around the $y$-axis.
This returns gives us back an alternating probe field (as for the $[1\bar{1}0]$ polarization), but with the signs of $J_x$ and $J_z$ reversed. Looking at Eqs. [(\ref{eq:XYh_sin_theta})-(\ref{eq:XYh_lambda})], we see that with this change $\sin\phi_k$ is very close to vanishing, and is only large near $k=\pi/2$, which now corresponds to the bottom band edge. Thus the $\beta$ chain response is a large spike of intensity close to the origin, which vanishes precisely at $J_x=J_y$.

On the $\alpha$ chains, the situation is now somewhat more complicated, as the chains are subject to an external field $H$ which alternates in sign, and a probe field which does not. For a suitable choice of basis, as discussed in Section \ref{subsec:noninteract}, the excitations on the $\alpha$ chains can be understood to exist in two bands, with the probe field creating pairs of particles of opposite bands, and driving momentum preserving transitions between bands.

\begin{figure*}
\subfigure[]
{\includegraphics[scale=0.39]{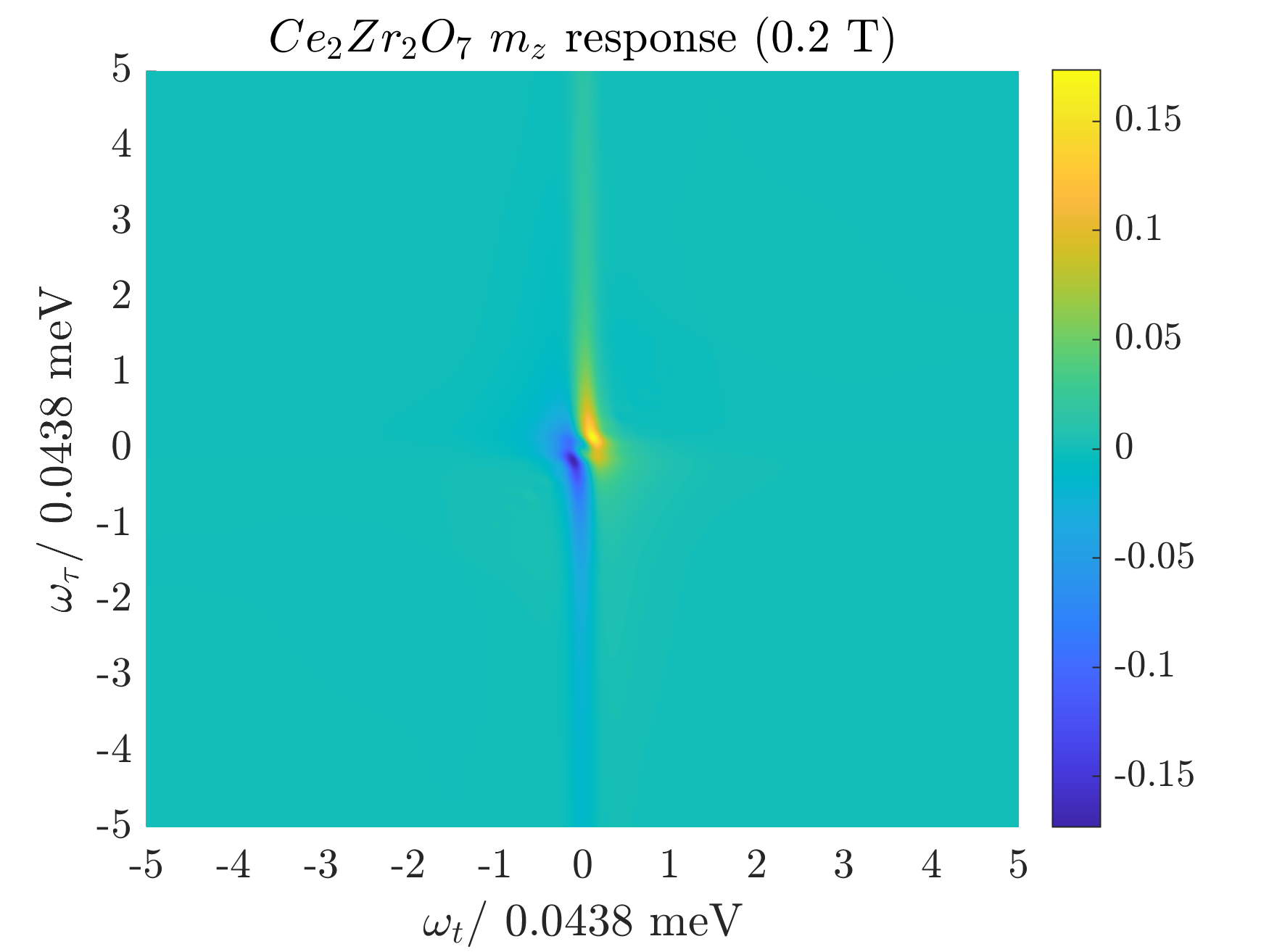}}
\subfigure[]
{\includegraphics[scale=0.39]{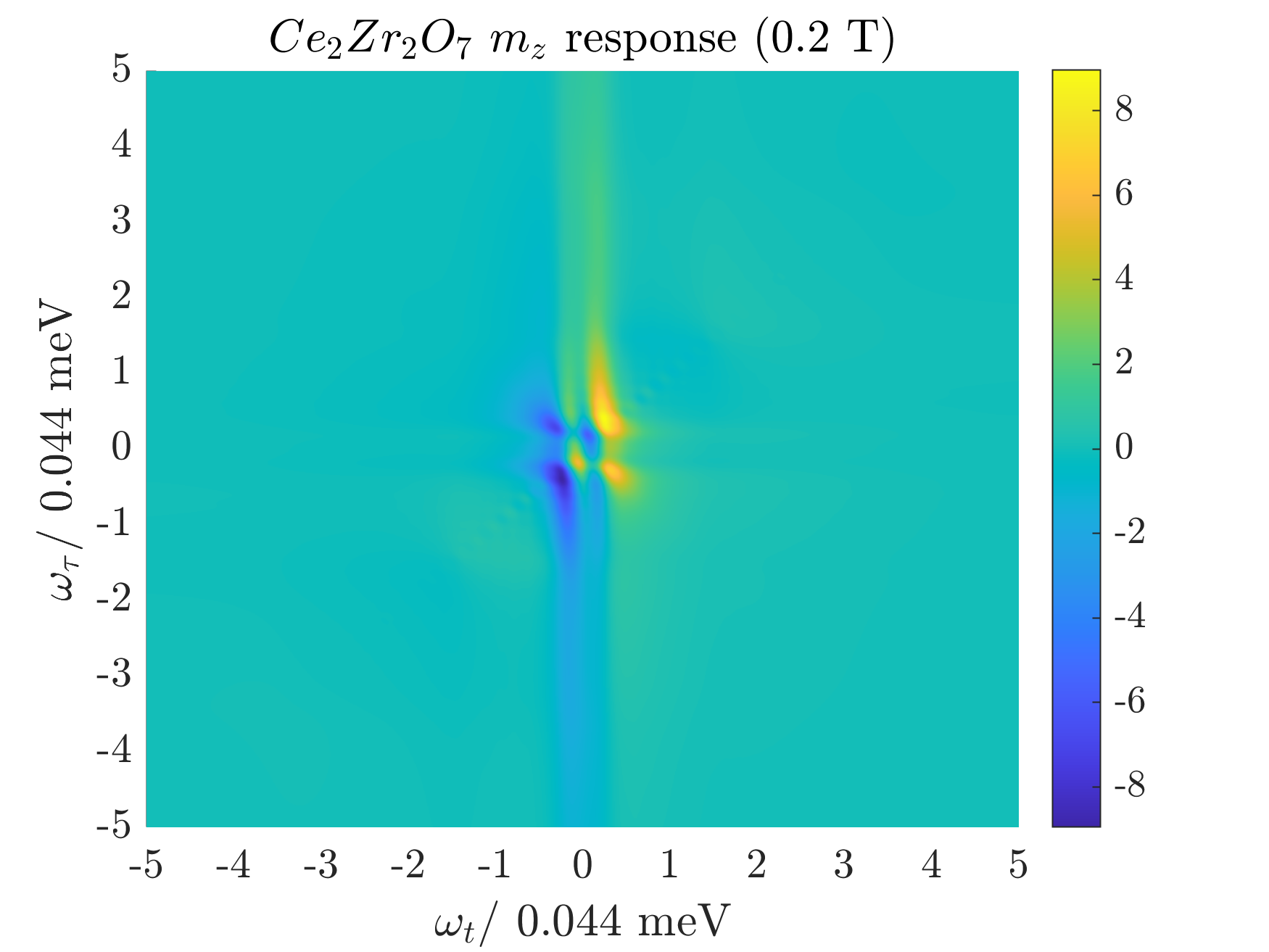}}
\subfigure[]
{\includegraphics[scale=0.39]{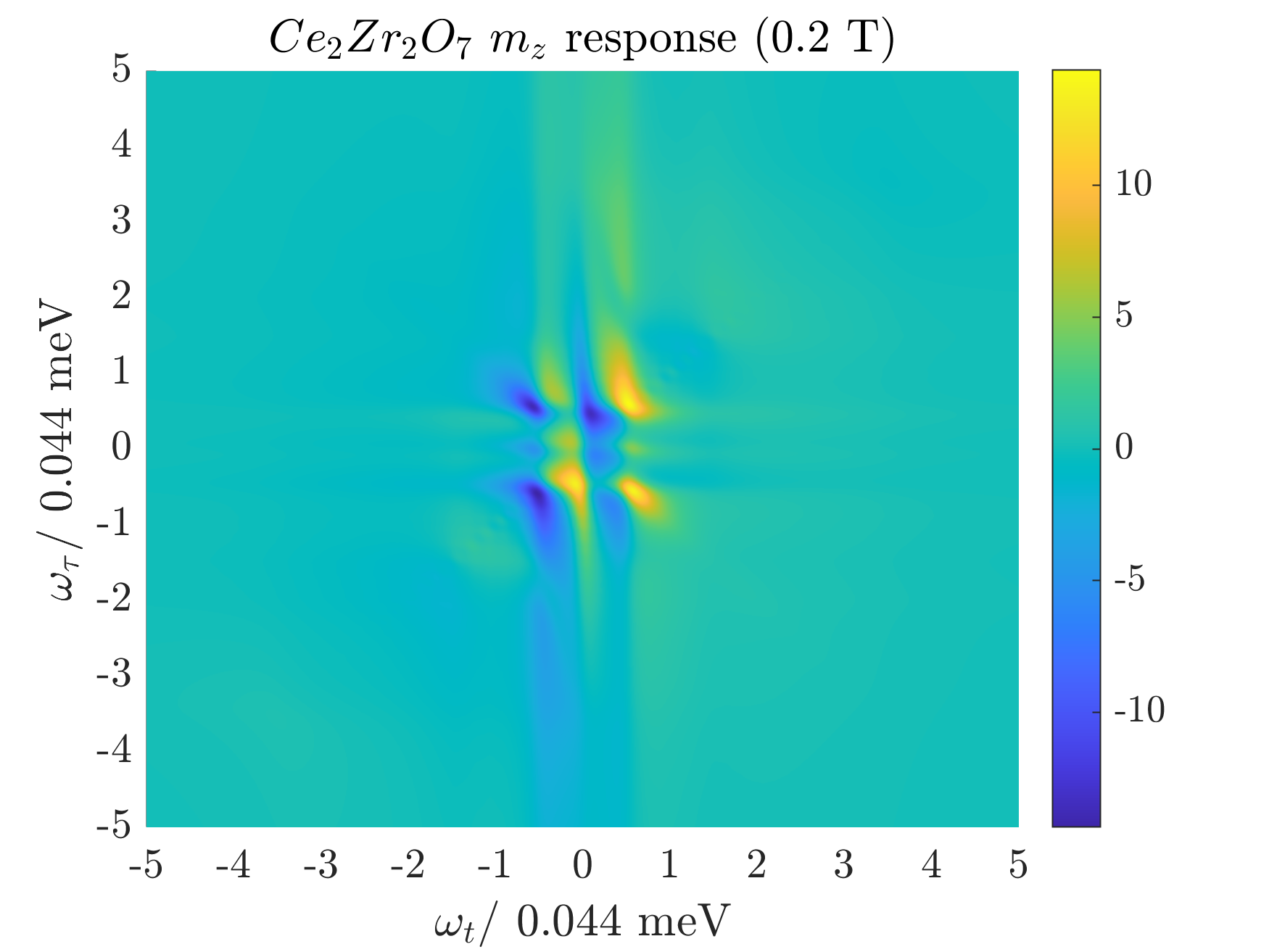}}
\caption{Total 2DCS response from $\alpha$ and $\beta$ chains in Ce$_2$Zr$_2$O$_7$ for a probe field in the [001] direction, calculated in the Hartree-Fock approximation,
while varying $J_{\tilde{x}}-J_{\tilde{y}}$. A field strength of $H=0.2$T is used in all panels. 
(a) $J_{\tilde{x}}-J_{\tilde{y}}=0.001 {\rm meV}$ as in the best fits from \cite{smith22},
(b) $J_{\tilde{x}}-J_{\tilde{y}}=0.014 {\rm meV}$,
(c) $J_{\tilde{x}}-J_{\tilde{y}}=0.027 {\rm meV}$.
The central peak is observed to resolve into the six distinct signatures produced by $\chi^{(3)}(t,t,t+\tau)$, which dominates for the pulse strength ratio chosen of $A_0:A_{\tau}=1:5$. As the anisotropy is increased, the $\beta$ chains are driven further from criticality, and the six features spread out from the center of the plot. At the isotropic point, the $\beta$ chain response collapses completely to $\omega_t=\omega_{\tau}=0$, and vanishes, leaving only the much weaker $\alpha$ response.}
\label{fig:Ce_mz_Jx_vary}
\end{figure*}

In Fig. \ref{fig:Ce_mz_Jx_vary}(a), the full 2DCS response for Ce$_2$Zr$_2$O$_7$ for a [001] probe field, calculated in the Hartree-Fock approximation, is presented for $H=0.2 \mathrm{T}$ and parameters taken from \cite{smith22}. A large central spike of intensity is observed, originating from the $\beta$ chain contribution to the full response. 
We find that the $\beta$ chain contribution is much stronger than that of the $\alpha$ chains, and there is therefore little variation in the full response with field strength.

For small increases in  $J_x-J_y$, as seen in Fig.\ref{fig:Ce_mz_Jx_vary}(b) and (c), the central peak is resolved into the six distinct signatures produced by $\chi^{(3)}(t,t,t+\tau)$. This is reflects both the widening of the gap to excitations, which moves each signature away from the origin, and $\sin\phi_k$ becoming large for values of $k$ further from $\pi/2$.

Approaching the critical point, these six peaks coalesce together, before disappearing entirely precisely at the isotropic point.

\begin{figure*}
\subfigure[]
{\includegraphics[scale=0.39]{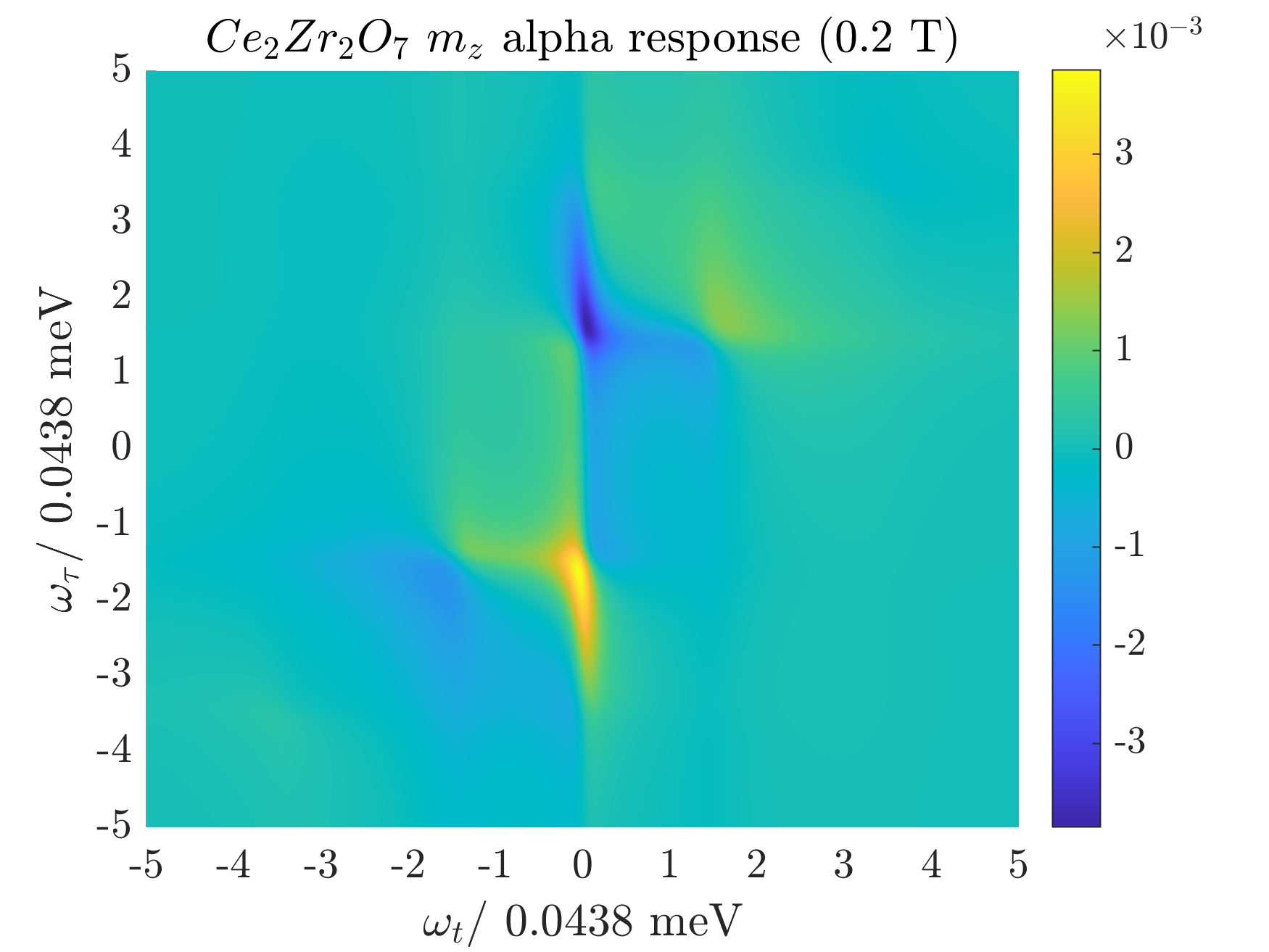}}
\subfigure[]
{\includegraphics[scale=0.39]{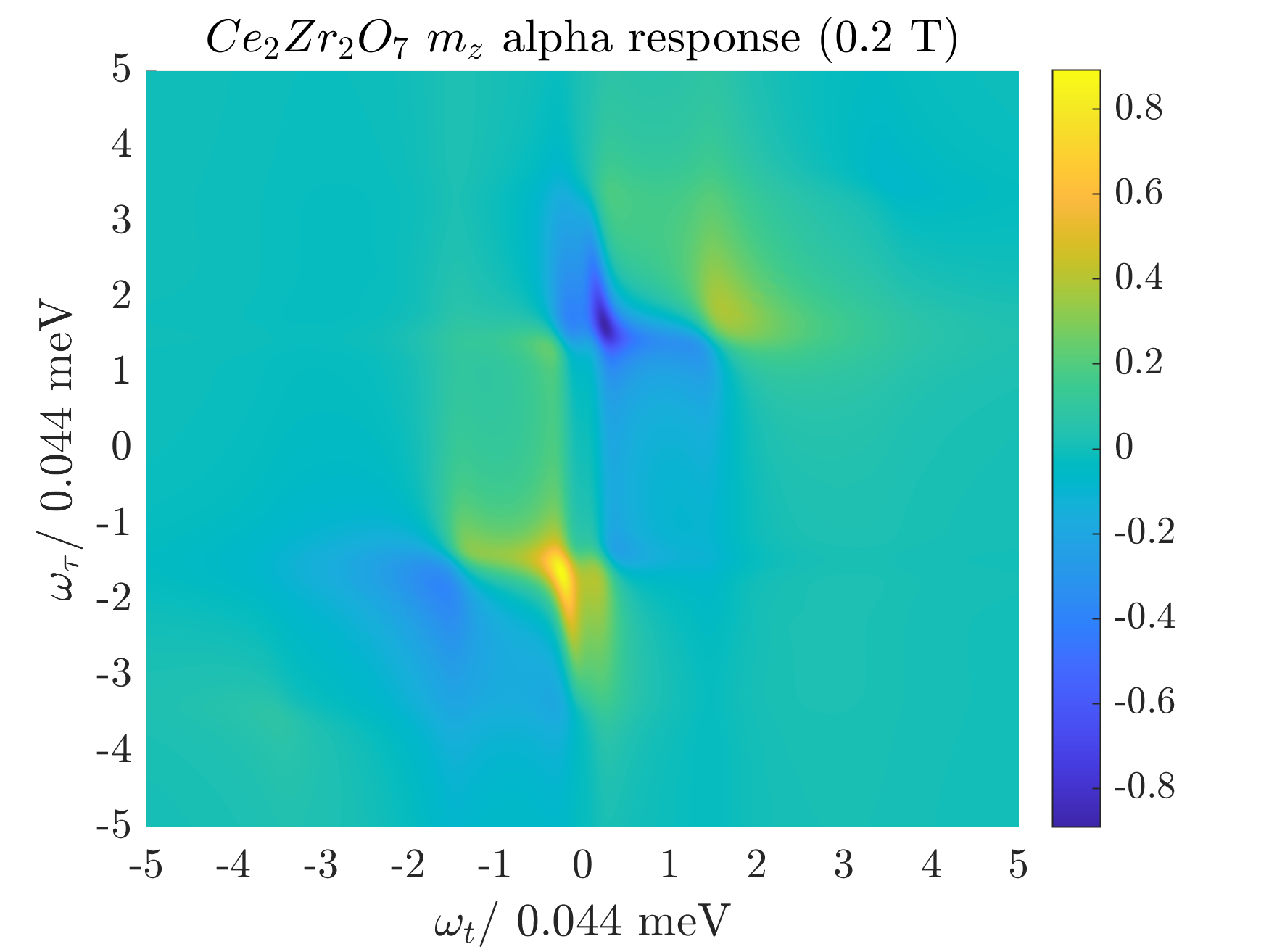}}
\subfigure[]
{\includegraphics[scale=0.39]{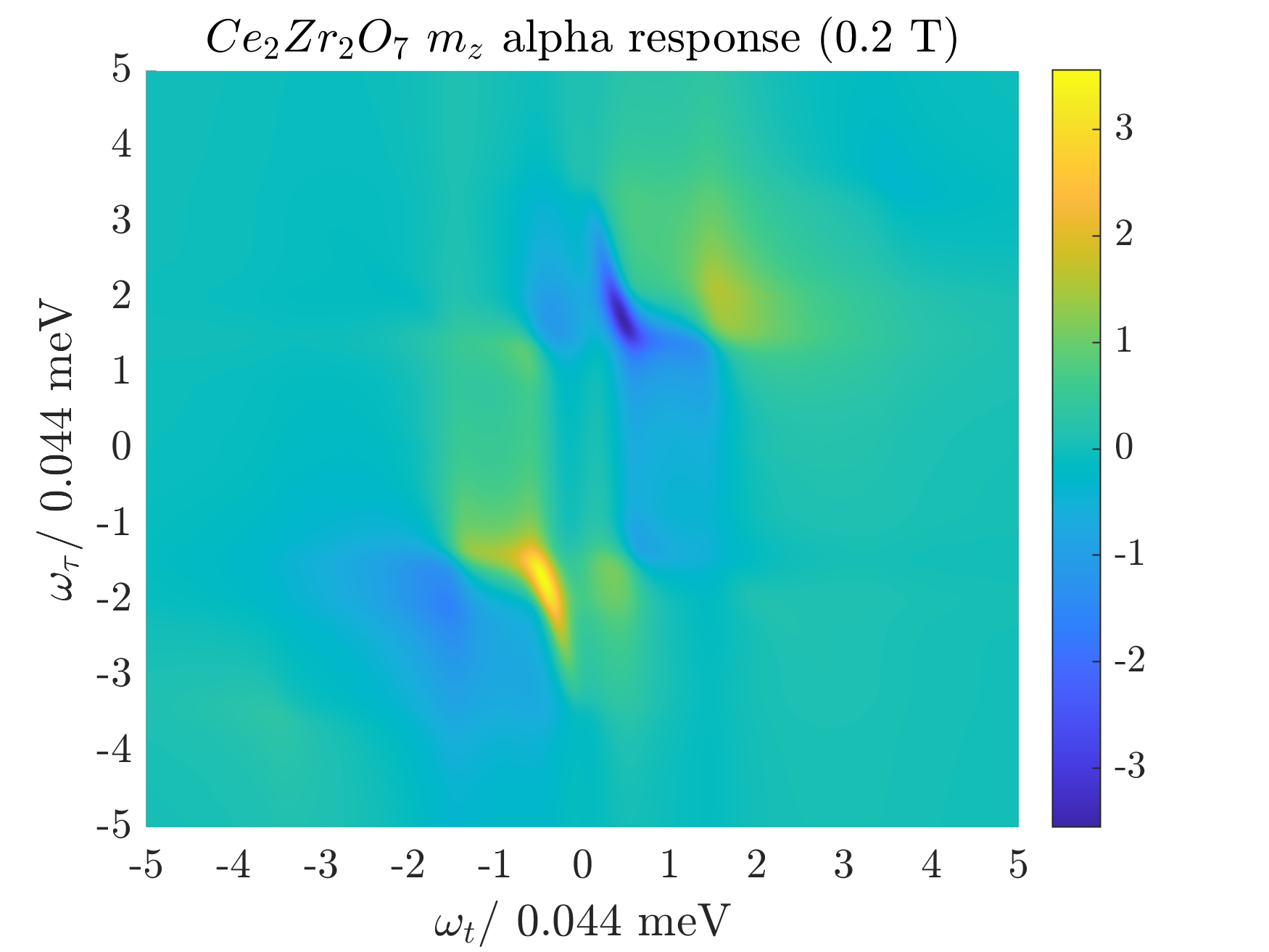}}
\caption{$\alpha$ chain contribution to the 2DCS response of Ce$_2$Zr$_2$O$_7$ to a [001] field, calculated in the Hartree-Fock approximation, while varying $J_{\tilde{x}}-J_{\tilde{y}}$. A field strength of $H=0.2$T is used in all panels. 
(a) $J_{\tilde{x}}-J_{\tilde{y}}=0.001 {\rm meV}$ as in the best fits from \cite{smith22}.
(b) $J_{\tilde{x}}-J_{\tilde{y}}=0.014 {\rm meV}$.
(c) $J_{\tilde{x}}-J_{\tilde{y}}=0.027 {\rm meV}$. 
We observe that the $\alpha$ chain response is roughly an order of magnitude weaker than the $\beta$ response. As the anisotropy is increased from left to right, we observe that the curvature in the signature close to the $\omega_t=0$ axis increases, indicating the contribution of inter-band magnon transitions driven by the probe field.}
\label{fig:Ce_mz_alpha_Jx_vary}
\end{figure*}

While the $\beta$ chain response vanishes at the isotropic point,  the $\alpha$ chain response will remain. We show the $\alpha$ chain response separately for varying values of the anisotropy in Fig. \ref{fig:Ce_mz_alpha_Jx_vary}. Here, the most notable features lie close to the $\omega_t=0$ axis. Referring to Eq. (\ref{eq:Chi3_exact_alternating}), this signature is produced by the term $A^{(2)}$, whose time dependence is $\sin(2\bar{\lambda_k}\tau)\cos(\Delta\lambda_kt)$, where $\Delta\lambda_k$ is the splitting between magnon bands, in the two-band picture described in Section \ref{subsec:noninteract}. The non-zero $\Delta\lambda_k$ shifts the response off of the $\omega_t=0$ axis, producing the slightly curved responses observed in Fig. \ref{fig:Ce_mz_alpha_Jx_vary}. This curvature could then be used to infer the energy changes in these inter-band transition processes, and thereby deduce the asymmetry in the magnon dispersion about $k=\pi/2$.

\subsection{Nd$_2$Zr$_2$O$_7$}
\label{subsec:nzo}
Calculating the two dimensional coherent spectroscopic response of Nd$_2$Zr$_2$O$_7$ within the CHAIN$_x$ phase is made more complicated by the presence of a non-zero mixing angle $\theta$. For the $\beta$ chains, we can still make calculations within the Hartree-Fock approximation, and use the matrix Pfaffian construction for  the expectation values involving ${S}^{\tilde{x}}$ operators now contributing to the response. For the $\alpha$ chains, we make predictions of the 2DCS response only in the high field limit, wherein we treat excitations about the field polarized state perturbatively.

For the $\beta$ chains, the Hartree-Fock re-scaled parameters are given by:
\begin{equation}
{J}^{\prime}_x = 0.099 \text{meV} \ {J}^{\prime}_y = -0.030 \text{meV} \nonumber
\end{equation}

\subsubsection{$[1{\bar1}0]$ polarization: absence of
spinon-rephasing signal for large $\theta$}
\label{subsubsec:nzo_1-10}

\begin{figure*}
\subfigure[]{\includegraphics[scale=0.38]{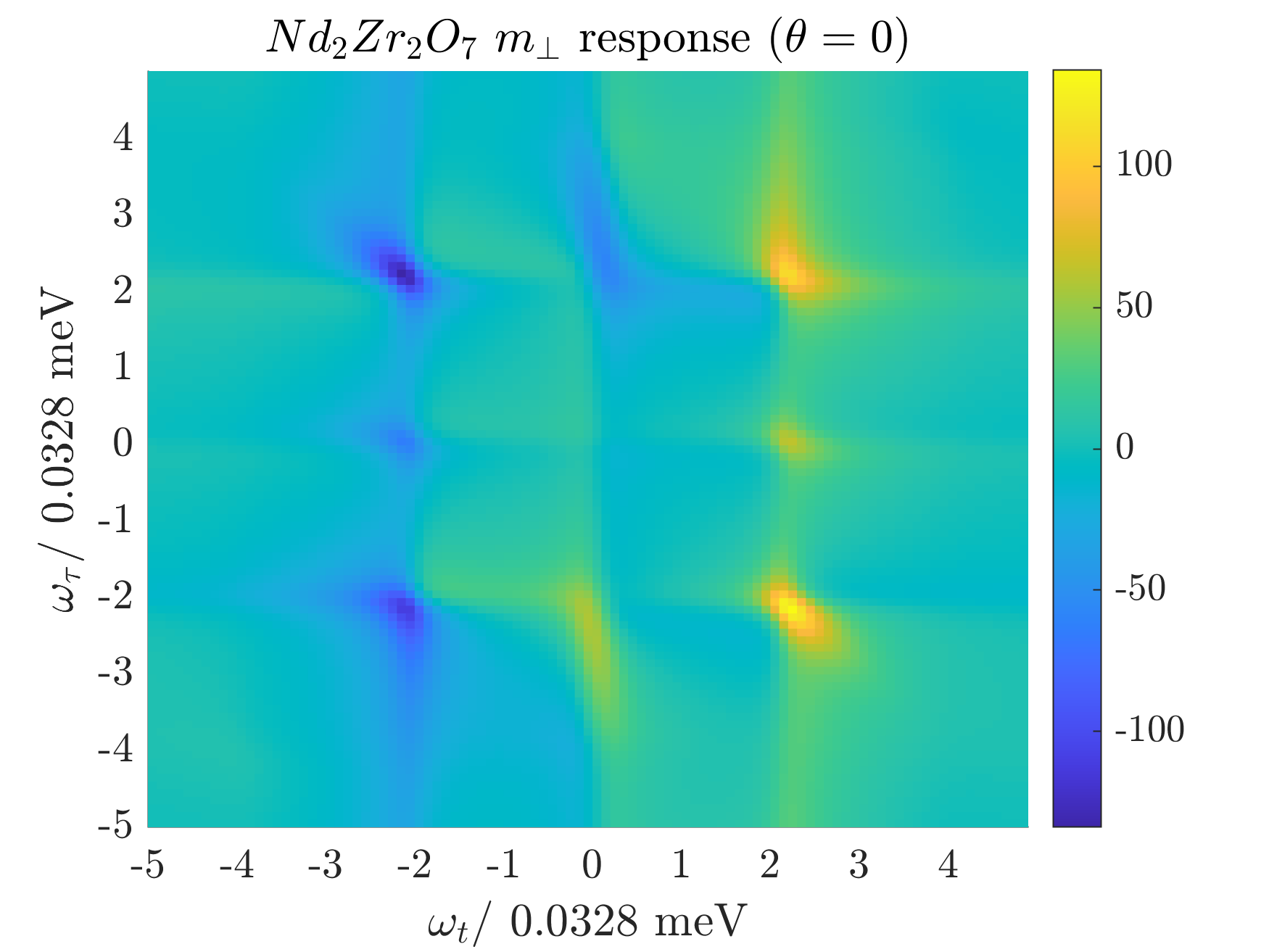}}
\subfigure[]{
\includegraphics[scale=0.38]{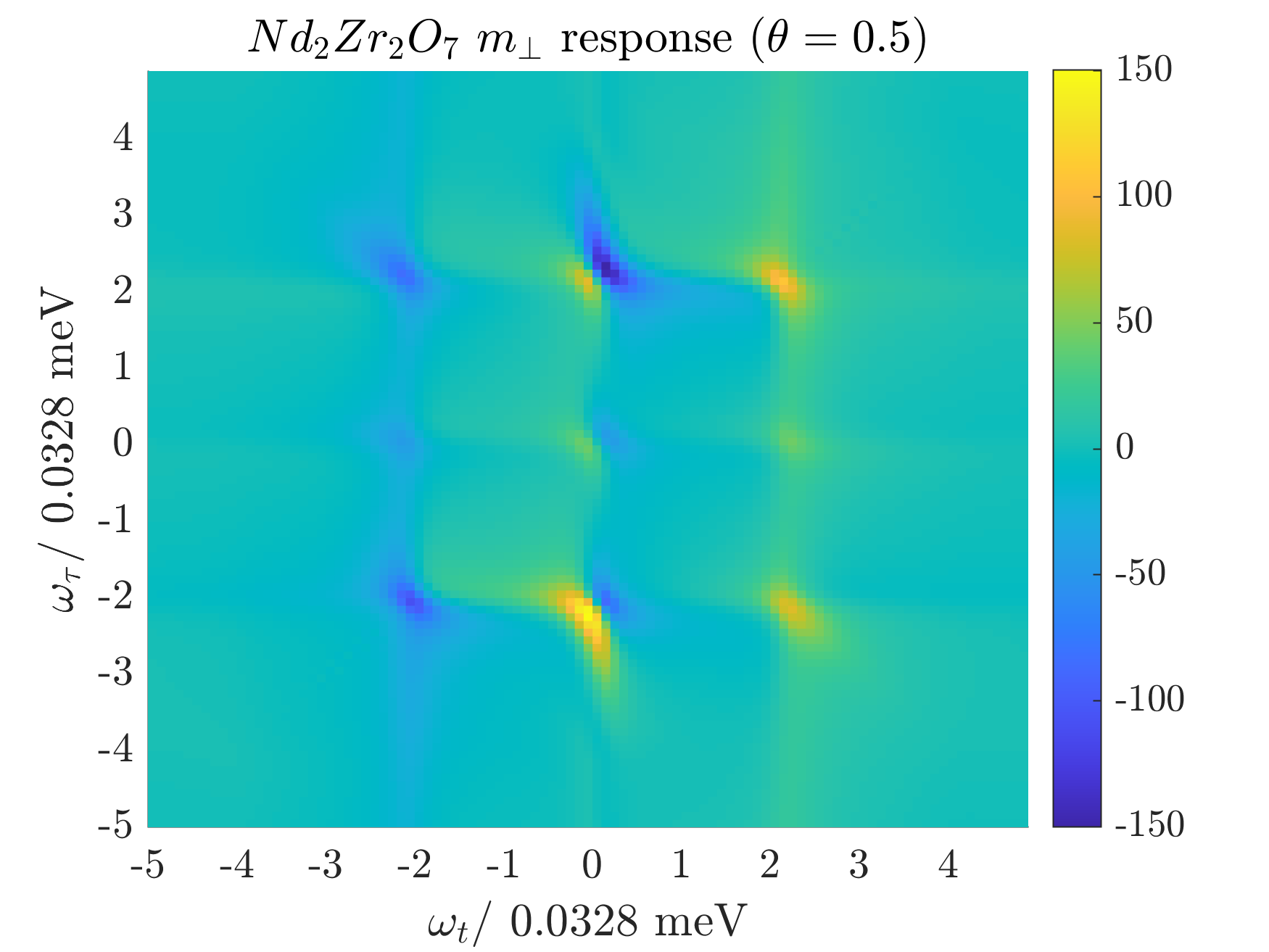}}
\subfigure[]{
\includegraphics[scale=0.38]{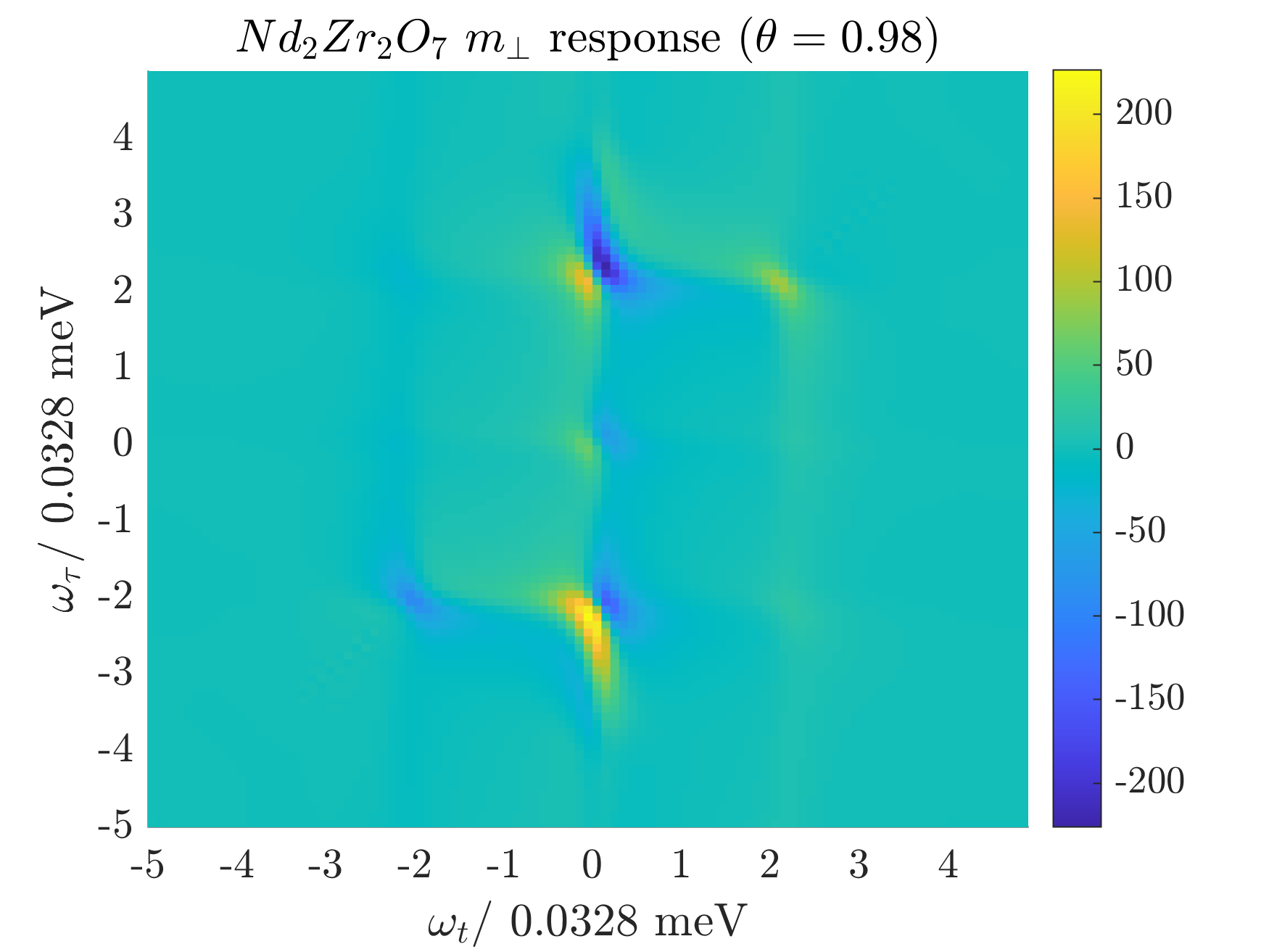}}
\caption{2DCS response of $\beta$ chains in Nd$_2$Zr$_2$O$_7$ to a [1$\bar{1}$0] field, for varying values of the parameter $\theta$. The other parameters take their reported values from \cite{xu19}. 
(a) $\theta=0$, (b) $\theta=0.5$ and (c) $\theta=0.98$  as found in \cite{xu19}. Calculations are made using the Hartree-Fock approximation, and by making use of the matrix Pfaffian method, which introduces an order one global rescaling of the responses due to approximations used in computations. Both the overall intensity of the response, and the relative intensity of different signatures vary with $\theta$, with a noticeable shift in relative intensity away from the rephasing signal, but no other qualitative changes to the response are observed for the $\beta$ chains over this range of dipolar-octupolar mixing angles. In all cases, $L=40$, and $A_0:A_{\tau}=1:5$.}

\label{fig:Nd_mperp_theta_vary}
\end{figure*}

In  Fig. \ref{fig:Nd_mperp_theta_vary}, we present several figures showing the 2DCS response of the $\beta$ chains in Nd$_2$Zr$_2$O$_7$ to a probe field in the [1$\bar{1}$0] direction, for the experimentally estimated model parameters, with varying $\theta$. 

We observe that at the reported value of $\theta=0.98$, the rephasing response is seemingly very weak, relative to the rest of the signal. Similar to the situation in Ce$_2$Zr$_2$O$_7$, we attribute this to the  strengthening of the $\omega_t=0$ and $\omega_t=\omega_{\tau}$ parts of the response for non-zero $\theta$. As seen in Fig. \ref{fig:Nd_mperp_theta_vary} (a) and (b), the rephasing contribution is similar in strength to the other parts of the signal for lower values of the field mixing angle.

As with Ce$_2$Zr$_2$O$_7$, it is only the strengths of different parts of the response that vary with $\theta$. The locations of response peaks are unchanged, which is to be expected, as the $\beta$ chain Hamiltonian is independent of $\theta$, and 
it is only the matrix elements with the probe field that are changing.

It is notable that values  $\theta$ not close to a multiple of $\pi/2$ seem to disfavour the rephasing part of the response, as this the most useful signature which cleanly separates out the contributions from
fractionalisation and finite lifetimes to the breadth of a continuum \cite{wan19}.
Thus whilst the general expectation is that 2DCS enables unambiguous measurements of spinon continua, the finite value of $\theta$ for Nd$_2$Zr$_2$O$_7$ seems to obscure this information somewhat.
This is different to the case of Ce$_2$Zr$_2$O$_7$, discussed above, where $\theta$ is small or vanishing and the rephasing signature is predicted to be clear (see Fig. \ref{fig:Ce_mperp_Jx_vary}).

\subsubsection{$[110]$ polarization: 1- and 2- magnon excitations}
\label{subsubsec:nzo_110}

If we instead probe the $\alpha$ chains of Nd$_2$Zr$_2$O$_7$ using a [110] field, we introduce for finite $\theta$ a coupling to the ${S}^{\tilde x}$ pseudo spins, or equivalently a non-zero value of $J_{xz}$, which allows us to probe singly created magnon excitations. As discussed in Section \ref{subsec:lswt_and_ed}, the ${S}^{\tilde x}$ terms in the Hamiltonian become highly non-local in Jordan-Wigner fermions, and so the leading order behaviour of the response is accessed in perturbation theory. The predicted response in the high field limit is given by Eq. (\ref{eq:Nd_high_h_Chi3}) and Eq. (\ref{eq:Nd_high_h_Chi2}).
In general, we predict that both one- and two- magnon excitations will be observable in this polarization,
with their relative strength in the signal being controlled by $\theta$. For the value of $\theta$ found in \cite{xu19} for Nd$_2$Zr$_2$O$_7$, the one-magnon contribution is found to dominate.

\begin{figure*}
\subfigure[]{\includegraphics[scale=0.38]{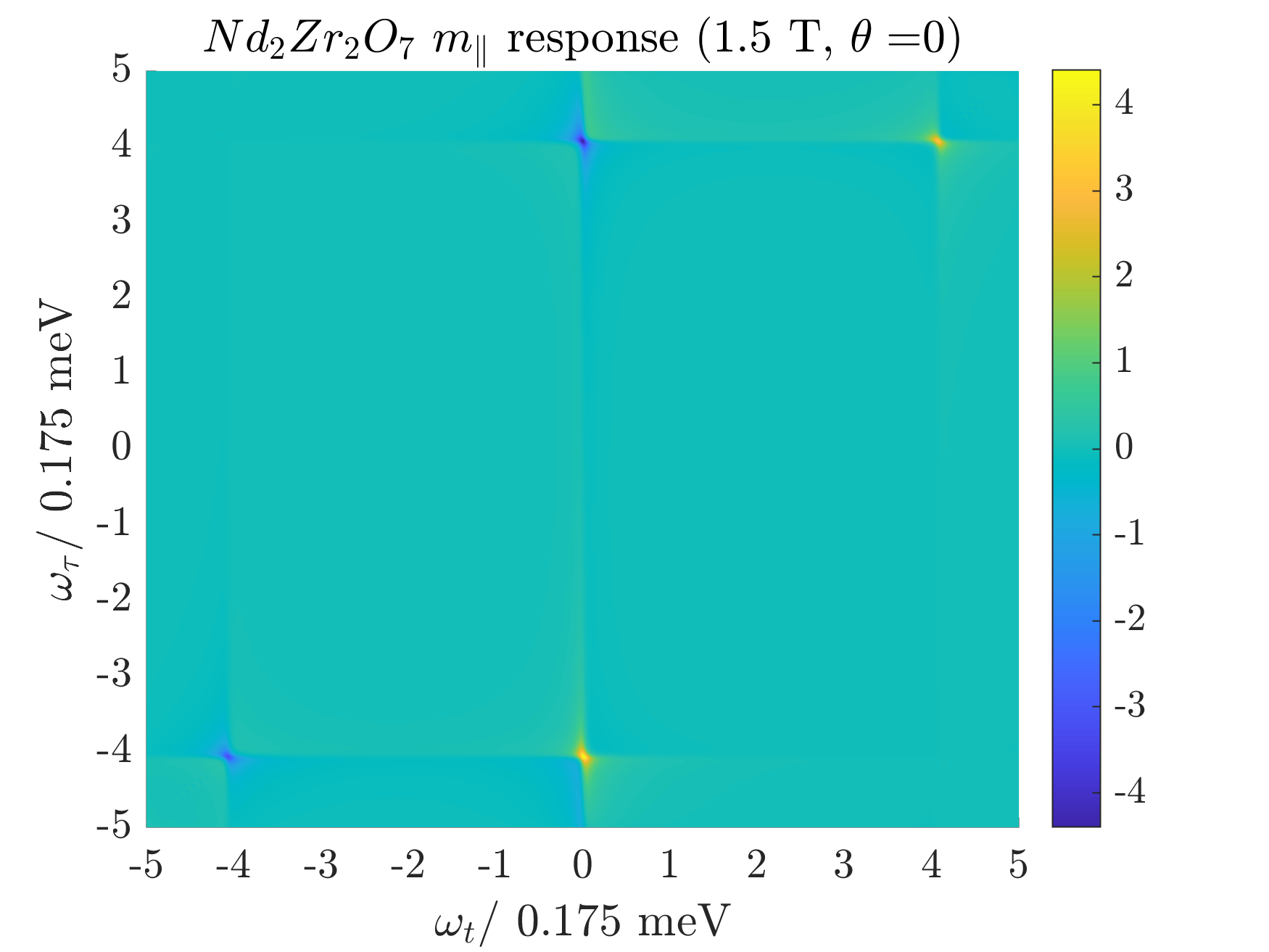}}
\subfigure[]{
\includegraphics[scale=0.38]{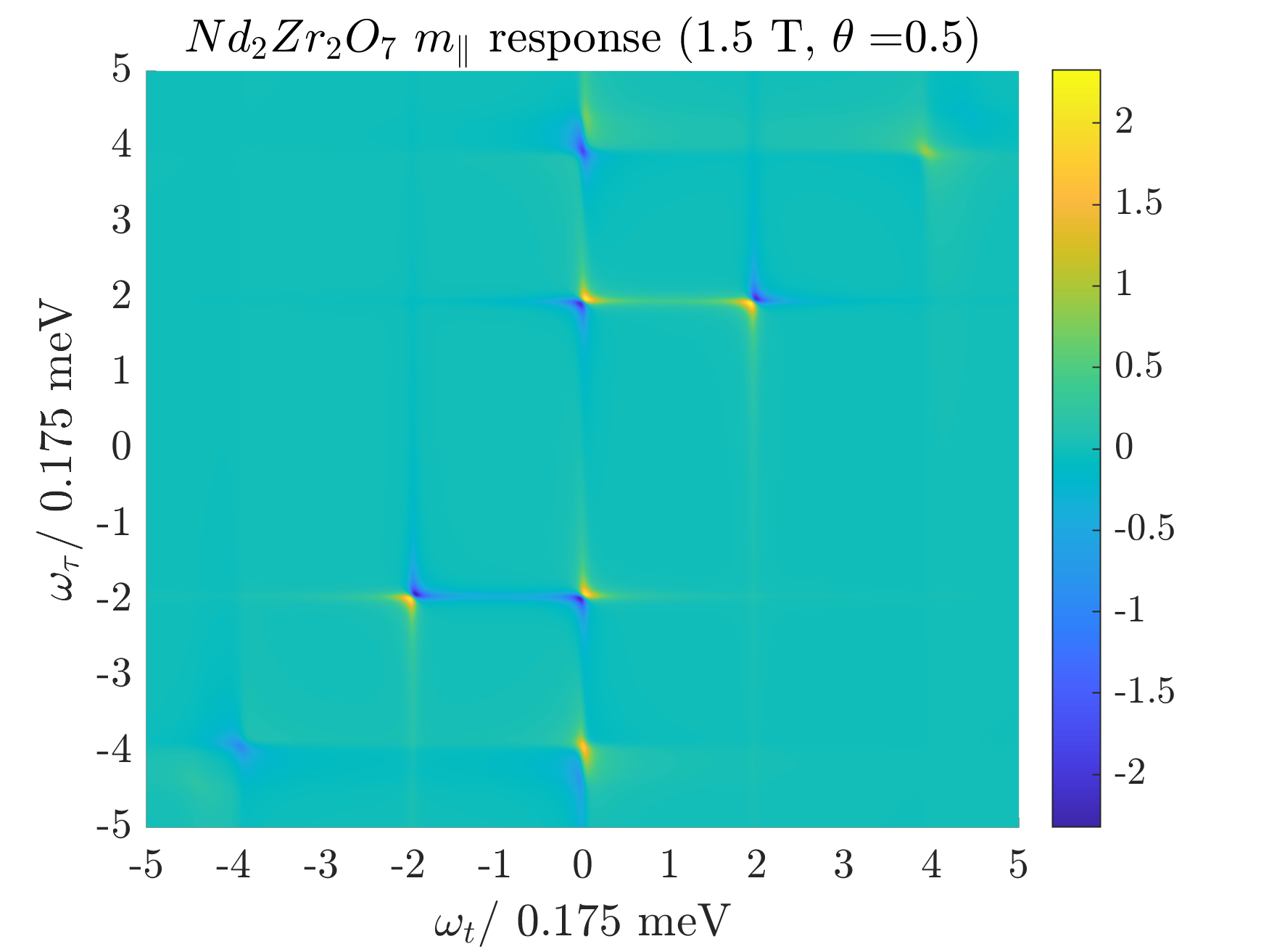}}
\subfigure[]{
\includegraphics[scale=0.38]{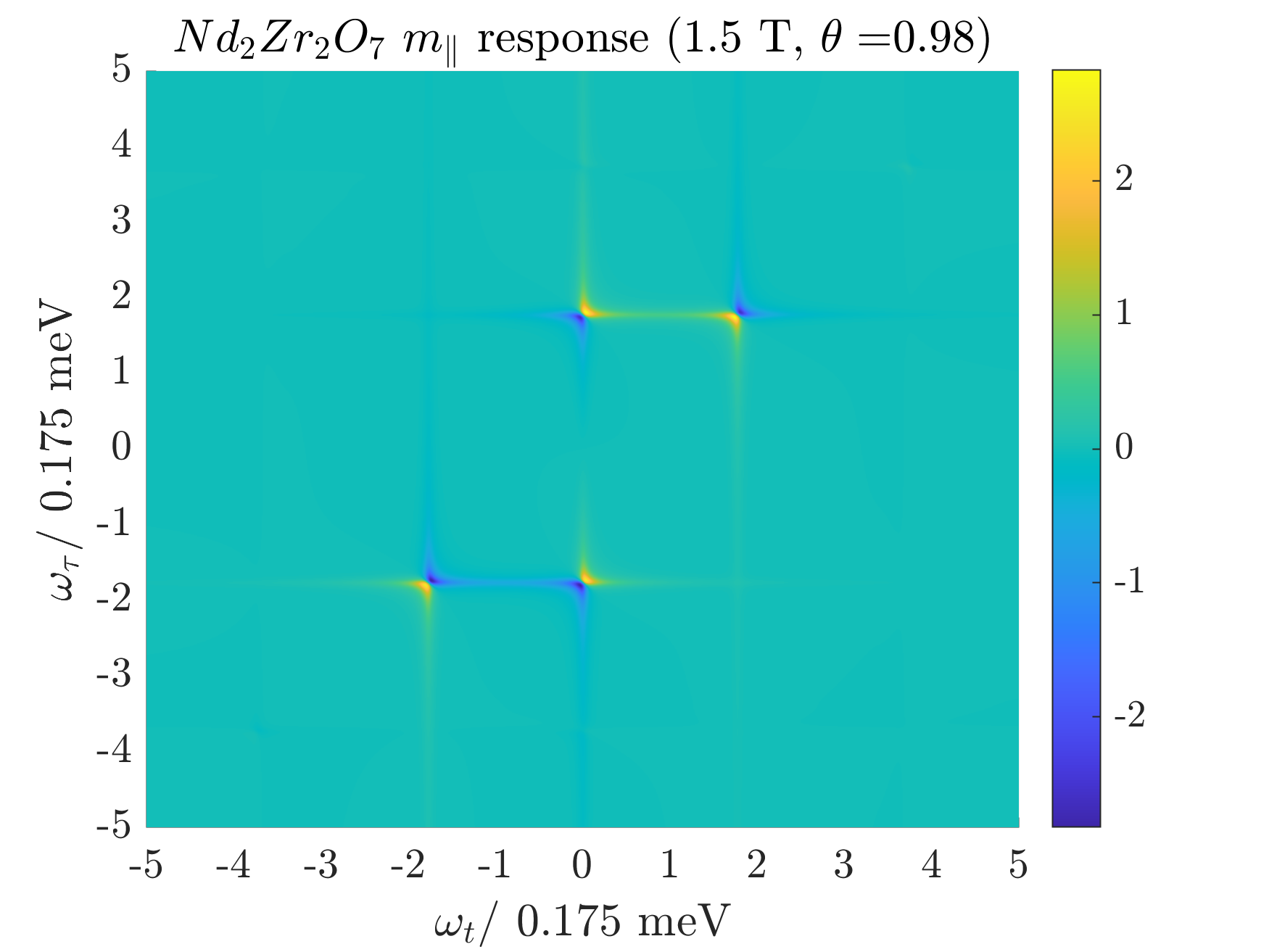}}
\caption{2DCS response of $\alpha$ chains in Nd$_2$Zr$_2$O$_7$ to a [110] probe field, as calculated using perturbation theory in the high field (large $h$) limit, with varying mixing angle $\theta$. Parameters apart from $\theta$ are set to their values from \cite{xu19}.(a) $\theta=0.0$, (b) $\theta=0.5$, (c)  $\theta=0.98$ as in \cite{xu19}. 
We observe that $\theta$ controls the relative visibility of the one- and two- magnon contributions to the response.
For $\theta=0.5$ both contributions are observable, leading to two distinct sets of peaks. The lower energy, single magnon, peaks are also visible at $\theta=0.98$. At $\theta=0$, only the two magnon response is visible. For all figures, $L=40$, and the pulse ratio is $A_0:A_{\tau}=1:5$.}
\label{fig:Nd_alpha_pert}
\end{figure*}

In Fig. \ref{fig:Nd_alpha_pert}, the $\alpha$ chain response in the perturbative limit is plotted for varying values of $\theta$.  Fig. \ref{fig:Nd_alpha_pert}(a) shows the response for the experimentally estimated parameters, wherein four response peaks are observed. In the high field limit of the exactly soluble model, the rephasing response decays as $(J_{\mu}/h)^4$, whilst other signatures fall away more slowly with $(J_{\mu}/h)^2$. We similarly see here that, keeping terms only up to the second power in $J_{\mu}/h$, no rephasing-like signature is observed.

The four peaks that can be seen occur at energies roughly equal to $h$, the energy cost of a single spin flip, indicating that these are indeed single magnon excitations. As $\theta$ is tuned away from its reported value to $0.5$ in the middle panel of Fig. \ref{fig:Nd_alpha_pert}, the amplitude of the two magnon signature increases, and both sets of peaks can be seen simultaneously, and are distinct from one another. Tuning $\theta$ to zero completely, we observe that the single magnon peaks vanish, leaving only the higher energy two magnon signatures.

The response from the two magnon sector is not visible at $\theta=0.98$ because the value of $J_{xx}-J_{yy}$ is much smaller than $J_{xz}$ for the reported model parameters, and so the probe field has a much greater amplitude to excite magnons one at a time. $J_{xx}-J_{yy}$, and hence the two magnon response, vanishes completely at $\theta = \pm 1.07(2)$ and $\theta = \pm 2.07(0)$ radians. Tuning $\theta$ towards $0$, $J_{xx}-J_{yy}$ increases whilst $J_{xz}$ decreases with $\sin(2\theta)$. This leads to the observed cross-over between a dominant single magnon response and a dominant two magnon response.

\begin{figure*}
\subfigure[]{\includegraphics[scale=0.38]{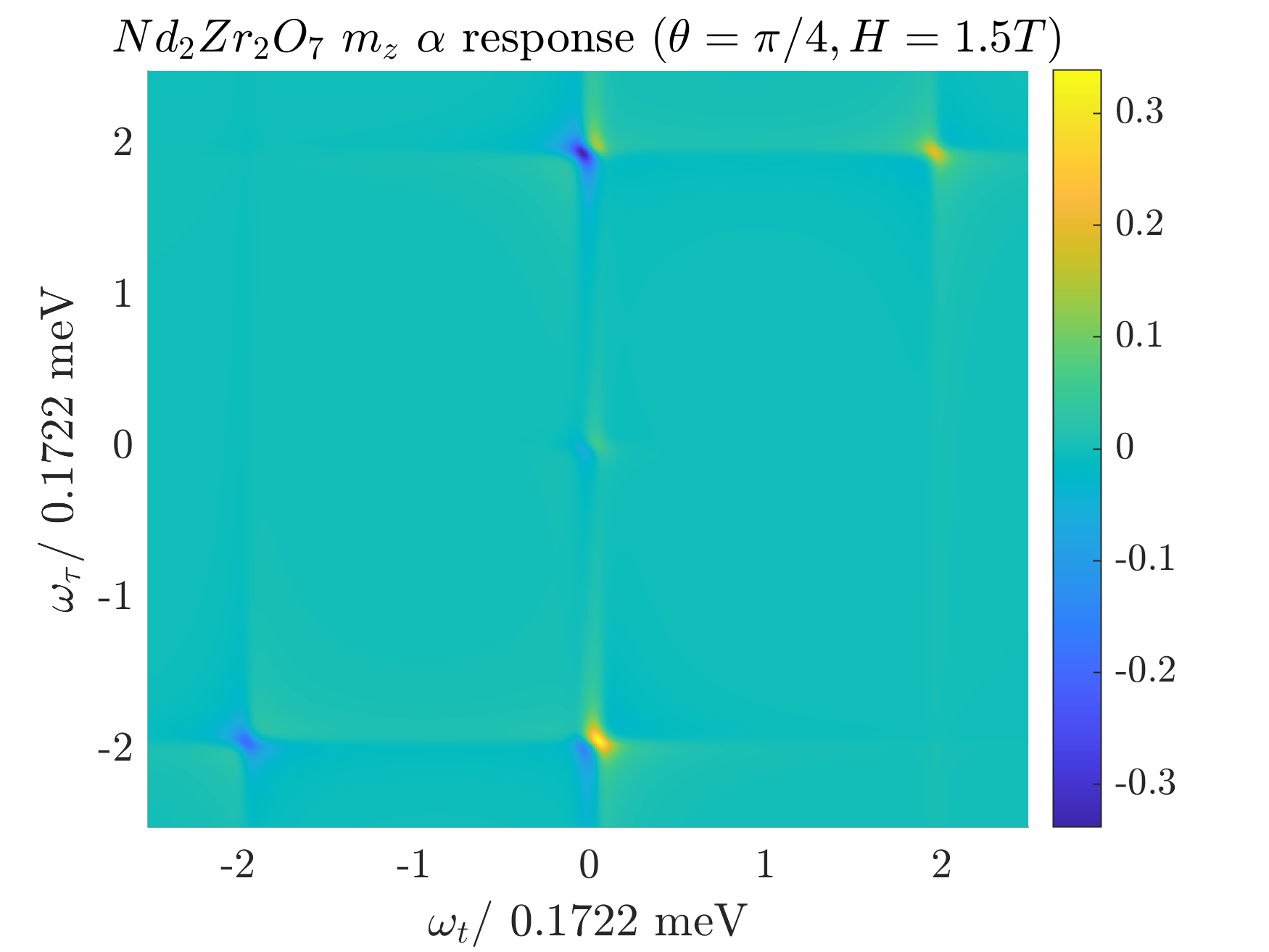}}
\subfigure[]{
\includegraphics[scale=0.38]{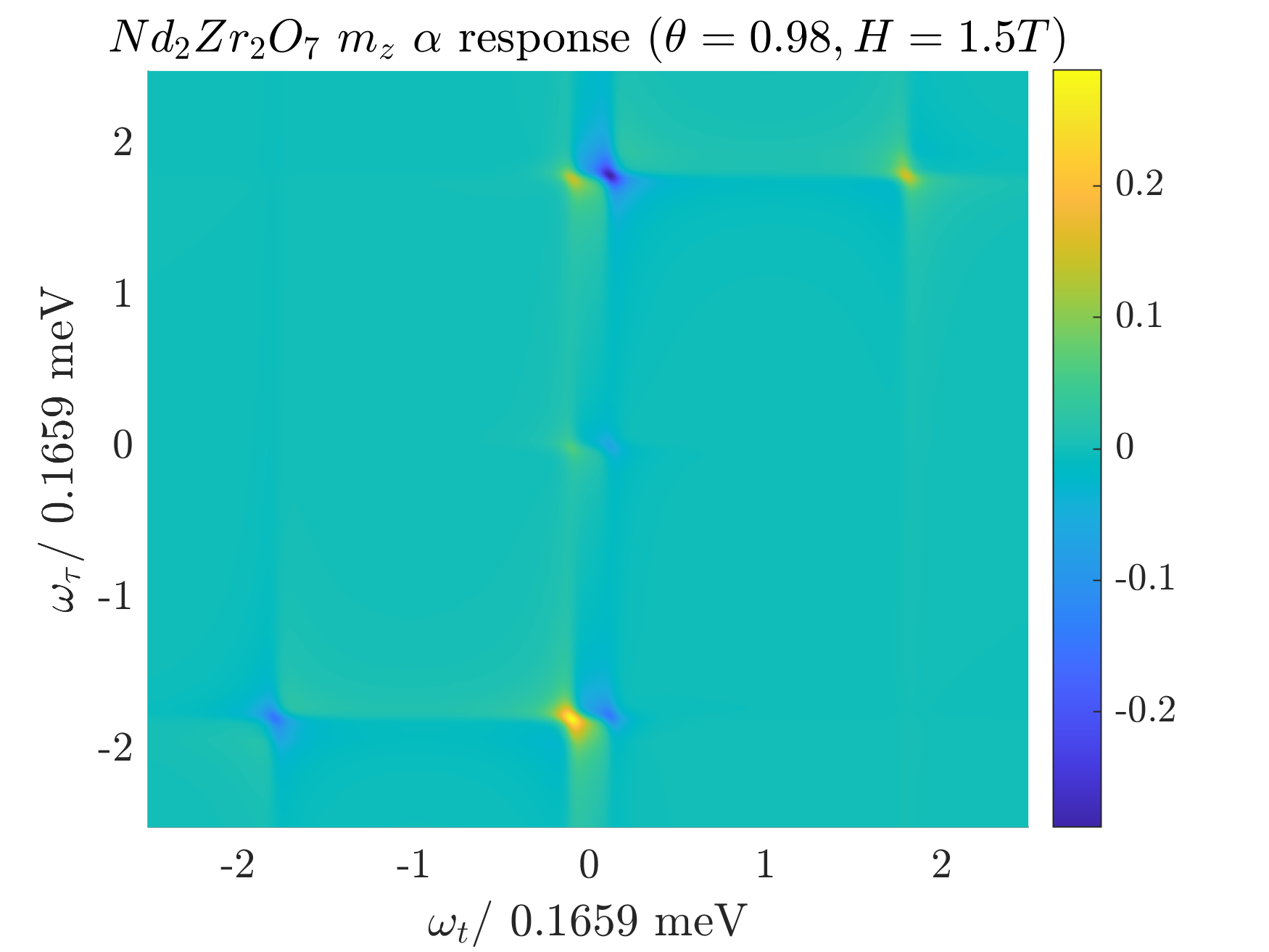}}
\subfigure[]{
\includegraphics[scale=0.38]{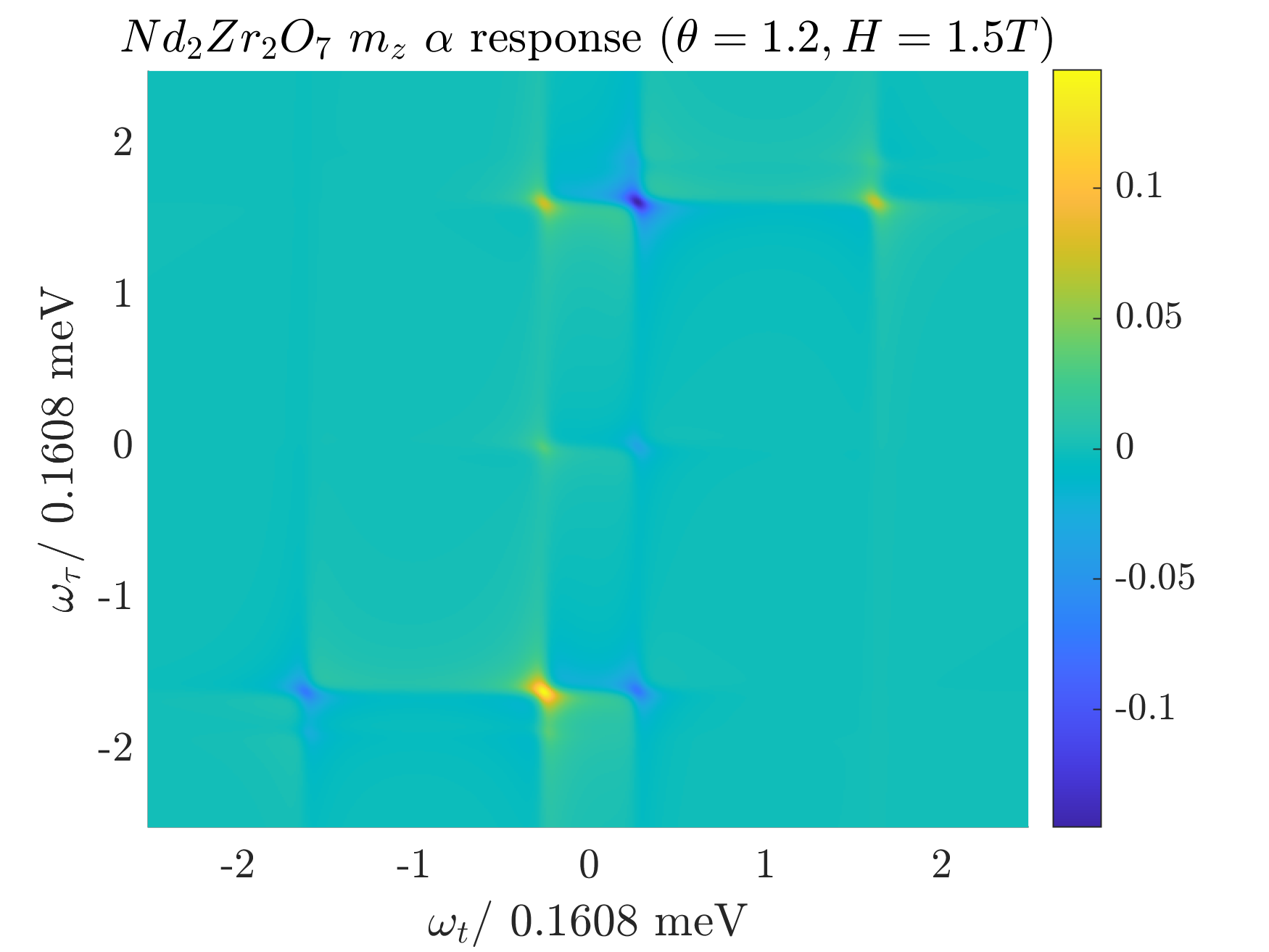}}
\caption{2DCS response of only the $\alpha$ chains in Nd$_2$Zr$_2$O$_7$ to a [001] probe field, as calculated using perturbation theory in the high field limit, with varying mixing angle $\theta$. Parameters apart from $\theta$ are set to their values from \cite{xu19}.(a) $\theta=\pi/4$, (b)  $\theta=0.98$ as in \cite{xu19} and (c) $\theta=1.2$. We observe that as $\theta$ is increased over this range of values, the two peaks on the $\omega_t=0$ line separate into two signals, and a third set of peaks is generated close to the origin. This splitting is due to transitions between $k=0$ and $k=\pi$ magnon states induced by the probe field, which has an effective momentum of $\pi$ in the basis in which the field polarized ground state is uniformly ordered.
}

\label{fig:Nd_mz_alpha_contribution}
\end{figure*}

\begin{figure*}
\subfigure[]{
\includegraphics[scale=0.38]{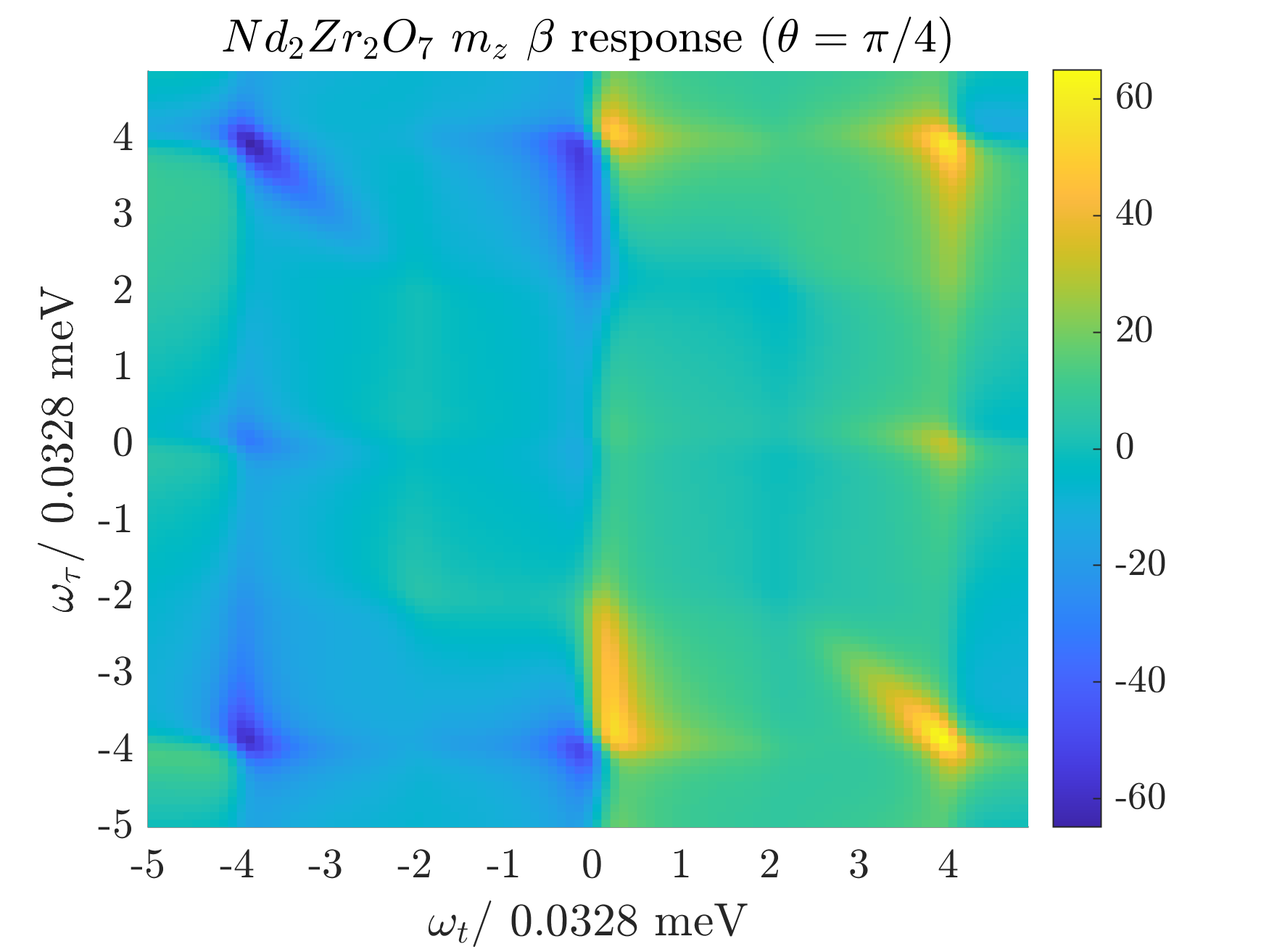}}
\subfigure[]{
\includegraphics[scale=0.38]{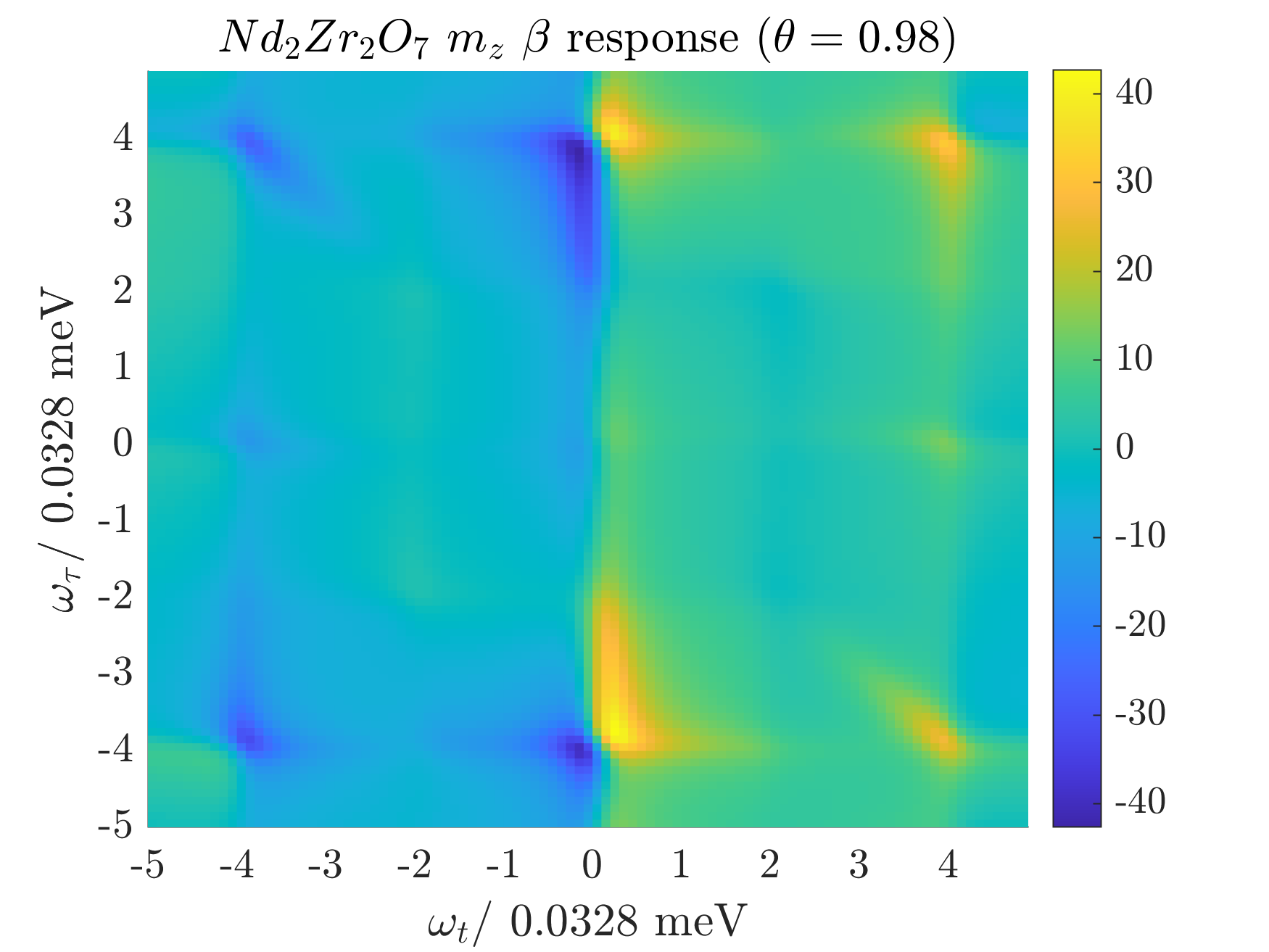}}
\subfigure[]{
\includegraphics[scale=0.38]{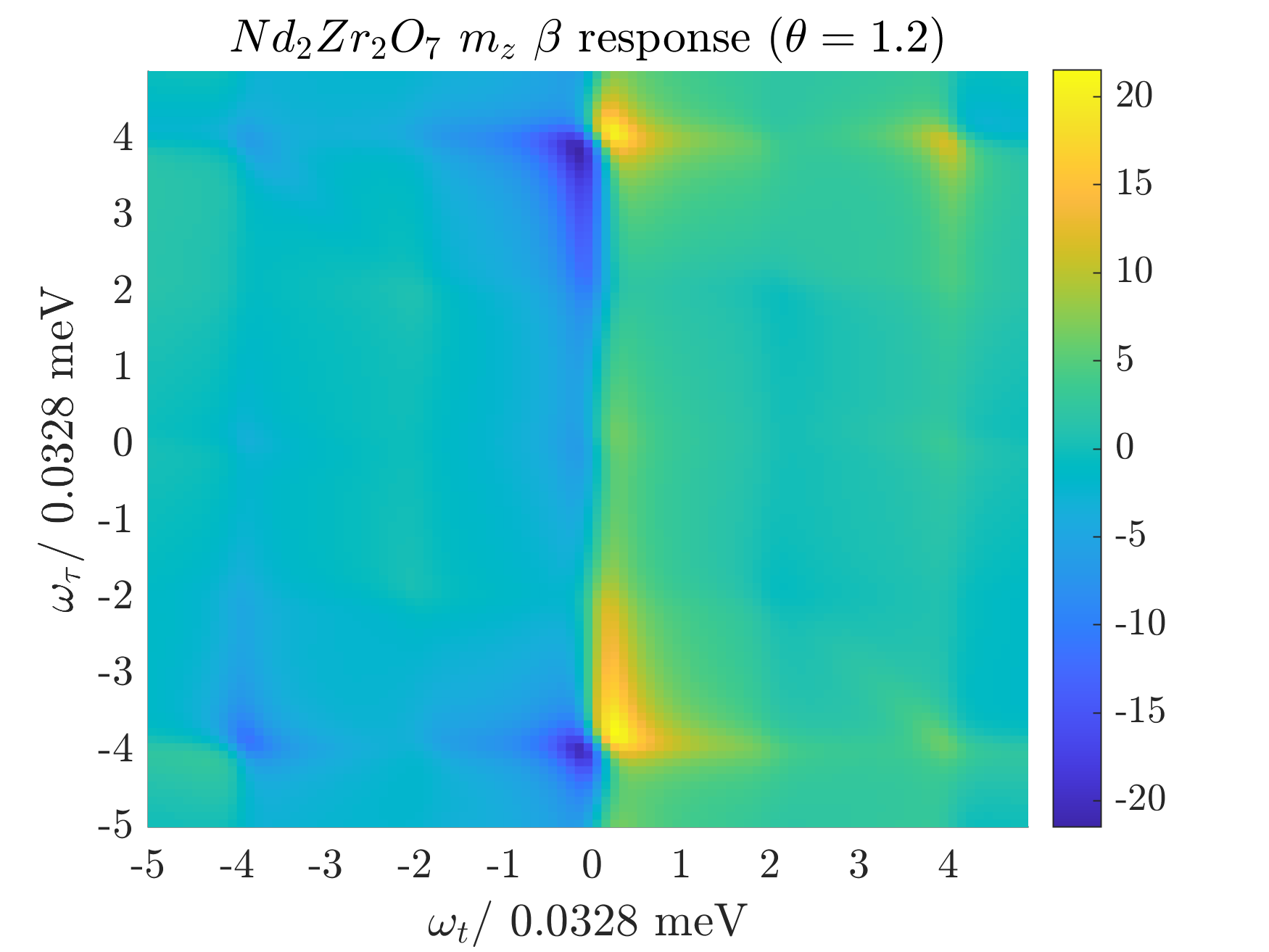}}
\caption{2DCS response of just the $\beta$ chains in Nd$_2$Zr$_2$O$_7$ to a [001] field, for varying values of the parameter $\theta$. Parameters other than $\theta$ take their reported values \cite{xu19}. In (a) $\theta=\pi/4$, in (b) $\theta=0.98$, the best fit value from \cite{xu19}, and in (c) $\theta=1.2$. Plots are generated in the Hartree Fock approximation, making use of the matrix Pfaffian expansion, which introduces a global rescaling of the response due to approximations made in computations. Much as in Fig. \ref{fig:Nd_mperp_theta_vary}, only variation in the relative intensities of different parts of the response is observed as $\theta$ is changed. Notably, the rephasing signal is not so heavily suppressed at $\theta=0.98$ as it was for the [1$\bar{1}$0] probe field, as seen in (b), though the $\omega_t=0$ signal again dominates if $\theta$ is further increased, as seen in (c). In all cases, $L=40$, and $A_0:A_{\tau}=1:5$.}
\label{fig:Nd_mz_theta_vary}
\end{figure*}

\subsubsection{$[001]$ polarization: observing rephasing at large $\theta$}
\label{subsubsec:nzo_001}

Subjecting Nd$_2$Zr$_2$O$_7$ to a [001] probe produces responses from both $\alpha$ and $\beta$ chains. The contribution to the response from the $\beta$ chains is qualitatively very similar as for the $[1\bar{1}0]$ polarization, whilst the $\alpha$ chain response shows signatures of probe field induced transitions between $k=0$ and $k=\pi$ states.
The $\alpha$ chains are again analysed in a high field perturbative limit, whilst the $\beta$ chain response is calculated using the matrix Pfaffian expansion in the Hartree-Fock approximation. As differing methods are used for each type of chain, the responses for each are presented separately.

First, as discussed in Section \ref{subsec:lswt_and_ed}, we find that the $\alpha$ chain response comes only from single magnon excitations to second order in perturbation theory. Fig. \ref{fig:Nd_mz_alpha_contribution} presents the predicted response for values of $\theta$ varied around that given in \cite{xu19}. Response peaks occur at the excitation energies of $k=0$ and $k=\pi$ magnons, as well as at the transition energy $\Delta\lambda=\lambda_0-\lambda_{\pi}$. As in Section \ref{subsubsec:czo_001}, this can be understood in a two band picture with the doubling of the unit cell coming from the different position dependencies of the static external magnetic field and the probe fields. It is observed that the two peaks on the line $\omega_t=0$ each split in two, and move apart perpendicular to this axis a distance $\Delta\lambda$. Also, a further set of peaks of opposite sign appear near the origin, which vanish if $\Delta\lambda=0$.

As the mixing angle $\theta$ is varied, the difference between the $k=0$ and $k=\pi$ magnon energies changes also, and we find that as the mixing angle is increased from $\pi/4$, through $\theta=0.98$, to $\theta=1.2$, the increasing value of $\Delta\lambda$ can be observed in the increasing separation of peaks either side of the $\omega_{t}=0$ axis.

The contribution from the $\beta$ chains is presented in Fig. \ref{fig:Nd_mz_theta_vary}, as calculated using the matrix Pfaffian construction. The response is calculated for various values of $\theta$, tuning the matrix elements between the pseudospins and the probe field. The most noticeable difference seen in the [001] $\beta$ chain response when compared the [1$\bar{1}$0] response, is that at the reported mixing angle of $\theta=0.98$ \cite{xu19} the rephasing signal is far more visible, and is not completely dominated by other signatures in the response. Examining Fig. \ref{fig:Nd_mz_theta_vary}, one can see that the relative intensities do still evolve with $\theta$ for this polarization choice, with the relative amplitude of the rephasing signature being higher in Fig. \ref{fig:Nd_mz_theta_vary} (a) with $\theta=\pi/4$, and lower in Fig. \ref{fig:Nd_mz_theta_vary} (c) for $\theta=1.2$, than at $\theta=0.98$. We note that the total intensity of the response from the $\beta$ chains using a [001] probe is somewhat weaker than for the [1$\bar{1}$0] polarization.

These results again show that, whilst a large non-zero $\theta$ may obscure the rephasing response for the $\beta$ chains, it is not absent entirely. Further, as it is only the overall intensity that is modulated with $\theta$, this response can still be used to probe properties of fractionalized spinon excitations on the $\beta$ chains. They also suggest the $[001]$ polarization as more favourable than [1$\bar{1}$0] for the  excitations of the $\beta$ chains in Nd$_2$Zr$_2$O$_7$.

\newpage

\section{Conclusion}
\label{sec:summary}

In this article we have explored how the polarization of magnetic probe fields in two dimensional coherent spectroscopy influences the measured response, and we have demonstrated how this polarization can be used as a powerful control parameter. By analysing the detailed coupling as a function of polarisation, we can infer  information ranging from the detailed nature of the microscopic degrees of freedom all the way to the determination, and isolation, of different types of excitations in an exotic magnet.

We have focused on the materials Ce$_2$Zr$_2$O$_7$ and Nd$_2$Zr$_2$O$_7$, both fascinating examples of frustrated quantum magnets presenting field induced Chain phases of effectively one dimensional character \cite{xu18, smith23}. By varying the probe field polarization, the 2DCS response from fractionalised, unpolarised $\beta$ chains can be isolated from the response of non-fractionalised, polarised $\alpha$ chains. 

The strong dependence of the 2DCS response on the coupling between the probe field and a system's degrees of freedom is also highlighted in how the dipolar-octupolar mixing angle $\theta$ influences the response, with this mixing angle being observed to control the relative intensities of one- and two- magnon signatures.

In the particular case of Ce$_2$Zr$_2$O$_7$, $\theta=0$ and we observe that the response to a $[1\bar{1}0]$ probe, which accesses the $\beta$ chains, is qualitatively similar to the non-fractionalised response of the $\alpha$ chains to a $[110]$ field, as the magnetic dipole operator is only able to create magnons in pairs. In Nd$_2$Zr$_2$O$_7$, $\theta$ is non-zero, and single magnon excitations dominate the $[110]$ response. We see also that large values of $\theta$ can somewhat suppress the relative amplitude of the rephasing streak along the $\omega_t = - \omega_{\tau}$ line.

A different polarisation choice can be used to probe another aspect of the physics of Ce$_2$Zr$_2$O$_7$. The chain phase of this material is believed to be close to a critical point, and  we find that the [001] polarization serves as a particularly sensitive measure of the proximity to criticality.

We hope that the predictions presented here will inspire
experiments to test them, exploring the quantum
excitations of field induced chain phases.
This may be challenging on a technical level, since the relatively weak exchange interactions in these systems ($\sim0.1 {\rm meV}$) mean that both excellent frequency
resolution and very low temperatures would be required.
Nevertheless, there is no reason of principle as to why these obstacles cannot be overcome and we expect that as 2DCS becomes a more widely used tool in the study of quantum magnetism, the approach will be increasingly optimised to meet these challenges.

\section*{Acknowledgements}
We thank Benedikt Placke for useful
discussions.
This work was in part supported by the Deutsche
Forschungsgemeinschaft under grants SFB 1143 (project-id 247310070) and the cluster of excellence ct.qmat (EXC 2147, project-id 390858490).

\appendix

\section{Details of Matrix Pfaffian method for $\beta$ chains with non-zero $\theta$} \label{sec:Pfaffian_Appendix}

In this appendix, we shall outline some further details of the matrix Pfaffian method used to obtain third order susceptibilities for $\beta$ chains for non zero mixing angle $\theta$. This method is also outlined in the supplementary material of \cite{wan19}, and the first part of this appendix will broadly review the construction outlined there. Following that, we will detail some further technical points and simplifications made to improve the computation time required for these calculations.

As discussed in Section \ref{subsec:lswt_and_ed}, we can express the Hartree-Fock re-scaled Hamiltonian of the $\beta$ chains in terms of Majorana modes $\alpha_j$ and $\beta_j$. The Hamiltonian is written in terms of the vectors $\alpha=(\alpha_1,\alpha_2,...)$ and $\beta=(\beta_1,\beta_2,...)$ and the matrix $\bar{J}$ as:

\begin{equation}
H=\frac{i}{4}\left(\alpha^{T}\bar{J}^T\beta-\beta^T\bar{J}\alpha\right).
\end{equation}
We chose to impose open boundary conditions, which gives the matrix $\bar{J}$ the following form:

\begin{equation}
\bar{J}=
\begin{pmatrix}
0 & -2J_x & & & \\
-2J_y & 0 & -2J_x & & \\
& & \ddots & \ddots & \\
& & -2J_y & 0 & -2Jx \\
& & & -2J_y & 0\\
\end{pmatrix}.
\end{equation}

If a singular-value decomposition of $\bar{J}$ is then performed, the Hamiltonian can be diagonalized. Writing $J = U\Lambda V^T$, $\gamma = V^T \alpha$ and $\delta = U^T \beta$, we arrive at
\begin{equation}
H = \frac{i}{4} (\gamma^T \Lambda \delta - \delta^T \Lambda \gamma) .\label{eq:SVD_Hamiltonian}
\end{equation}
If $\lambda_n$ are the singular values of $J$, we can then rewrite Eq. (\ref{eq:SVD_Hamiltonian}) as
\begin{equation}
H = \frac{i}{4} \sum_n \lambda_n (\gamma_n \delta_n - \delta_n \gamma_n).
\end{equation}
The Hamiltonian is now diagonal in terms of the $\gamma$ and $\delta$ Majorana modes, whose time evolution in the Heisenberg picture can now be easily evaluated.

The aim of this construction is to evaluate expectation values of the form $\langle \sigma^x_j(s_1)\sigma^x_l(s_2)\sigma^x_m(s_3)\sigma^x_n(s_4)\rangle$. In terms of Majoranas, each $\sigma^x_j(s_i)$ is a long string of operators:
\begin{equation}
    \sigma^x_j(s_i)=\left.\alpha_1i\beta_1\alpha_2i\beta_2...\alpha_{j-1}i\beta_{j-1}\alpha_j\right|_{t=s_i} \ .
\end{equation}
Each four-spin expectation value thus includes a total of $2(j+l+m+n)-4$ Majorana operators. Fortunately, the Hamiltonian is quadratic in Majorana modes, and so we can make use of Wick's theorem to decompose these strings into sums over products of bilinear terms. To perform this expansion, we label each operator appearing in $\langle \sigma^x_j(s_1)\sigma^x_l(s_2)\sigma^x_m(s_3)\sigma^x_n(s_4)\rangle$ from left to right with an index $\mu$ which runs from 1 to $2(j+l+m+n)-4$. We then denote the two-operator expectation value between the mode with index $\mu$ and that with $\nu$ by $A_{\mu,\nu}$. In performing the Wick expansion, we must sum over all possible pairings of operators. Each term will be a product of $(j+l+m+n)-2$ $A_{\mu,\nu}$ factors, multiplied by a sign determined by the Majorana anticommutation relations. We can tabulate all of these possible pairings by considering all permutations on a list of size $N=2(j+l+m+n)-4$, $S_N$. For each permutation, the pairs are found by grouping neighbouring elements of the list (entries $S(1)$ and $S(2)$ are paired, then $S(3)$ and $S(4)$ and so on). 

As formulated, $S_N$ over counts the number of distinct terms in the Wick expansion. First, we have a factor of $(N/2)!$ that arises from re-orderings of the $A_{\mu,\nu}$ factors. Second, the full permutation group will include terms where the order of $\mu$ and $\nu$ in a given bi-linear $A_{\mu,\nu}$ are flipped. Only the ordering $\mu<\nu$ appears in the Wick expansion, and so we for now restrict the set of allowed permutations to only those where this constraint is satisfied on all $A_{\mu,\nu}$. The resulting expression is given by:
\begin{equation}
\frac{1}{(N/2)!} \sum_{\substack{S \in S_N \\ \mu<\nu}} \text{sgn}(S)\prod_i^{(N/2)} A_{S(2i-1),S(2i)} .
\end{equation}
The factor of $\text{sgn}(S)$ correctly gives the Majorana exchange factor for each term in the Wick expansion.

We can now massage this expression into the form of a matrix Pfaffian by defining $A_{\mu,\nu}=-A_{\nu,\mu}$, to absorb the change in $\text{sgn}(S)$, and then relaxing the $\mu<\nu$ constraint. This over-counts the Wick expansion by a factor of $2^{(N/2)}$, and so dividing through by this factor, one obtains:
\begin{equation}
\frac{1}{2^{(N/2)}(N/2)!} \sum_{S \in S_N }\text{sgn}(S)\prod_i^{(N/2)} A_{S(2i-1),S(2i)} = \text{Pf}(A) .
\end{equation}
We then recognise the resulting expression as the definition of the matrix Pfaffian of $M$, a skew-symmetric matrix whose elements in the upper triangle are bilinear expectation values of the form $\langle O_{\mu}O_{\nu}\rangle$, with $O_{\mu}$ the $\mu^{\text{th}}$ Majorana operator appearing in $\langle \sigma^x_j(s_1)\sigma^x_l(s_2)\sigma^x_m(s_3)\sigma^x_n(s_4)\rangle$.

Thus the problem is reduced to evaluating all required Majorana bilinears. These will be of the form $\langle \alpha_{m}(s_i)\alpha_{n}(s_j)\rangle$, $\langle \alpha_{m}(s_i) i\beta_{n}(s_j)\rangle$, $\langle i\beta_{m}(s_i) \alpha_{n}(s_j)\rangle$, or $\langle i\beta_{m}(s_i)i\beta_{n}(s_j)\rangle$. Using the relationships $\gamma = V^T \alpha$ and $\delta = U^T \beta$, these in turn can be expressed in terms of two-operator $\gamma$ and $\delta$ expectation values. The only such expectation values which are non-zero are between Majorana modes with the same index $n$, which are given by:
\begin{align}
\langle \gamma_n(s_1)\gamma_n(s_2)\rangle =& \cos(\lambda_n (s_1-s_2)) -i\sin(\lambda_n (s_1-s_2)), \\
\langle i\delta_n(s_1)i\delta_n(s_2)\rangle =& -\cos(\lambda_n (s_1-s_2)) +i\sin(\lambda_n (s_1-s_2)), \\
\langle \gamma_n(s_1)i\delta_n(s_2)\rangle =& i\sin(\lambda_n (s_1-s_2)) - \cos(\lambda_n (s_1-s_2)), \\
\langle i\delta_n(s_1)\gamma_n(s_2)\rangle =& -i\sin(\lambda_n (s_1-s_2)) + \cos(\lambda_n (s_1-s_2)).
\end{align}

Following this general recipe, each of the required expectation values for non-zero mixing angle susceptibilities can be constructed.

We will now briefly discuss some further simplifications made in performing this calculation to improve performance. First, as also discussed in \cite{wan19}, open boundary conditions were found to produce results that more closely agreed with the exactly solvable limit at $\theta=0$, provided the contribution from spins close to the end of the chain are excluded to remove the influence of edge states. This is accomplished by applying a sinusoidal windowing function to the magnetisation operator:
\begin{equation}
M = \sum_{j=R+1}^{L-R} (\cos\theta S^{\tilde{z}}_j+\sin\theta S^{\tilde{x}}_j)\sin\left(\frac{\pi(j-R)}{L+1-2R}\right)
\end{equation}
A sinusoidal window was used in place of a uniform window in an attempt to reduce artifacts caused by a hard cut off. In our calculations, $L=40$ and $R=5$ were used. 

In performing the $\theta \neq 0$ calculation for the $\beta$ chains, one must evaluate $\mathcal{O}(L^4)$ matrix Pfaffians, of sizes up to $(8L-4) \times (8L-4)$. This is a relatively expensive computation, and so we have employed two further simplifications to reduce the number of terms required. Most importantly, we find that, although open boundary conditions have been employed, we can make use of approximate translational invariance near the centre of the chain to fix one of the four spin indices appearing in the expectation values, reducing the number of Pfaffians required by a factor of $L$. Comparing with exact results, this simplification does not seem to introduce significant error. We further half the required number of Pfaffians by making use of mirror symmetry.

Finite broadening Lorentzian broadening is introduced to the response by introducing a uniform exponential time decay factor of $\exp(-\Gamma t)$ to the non-linear response. $\Gamma$ is chosen to be $0.0067$ meV in all cases.

\bibliography{biblio.bib}

\end{document}